\documentclass{aa}

\usepackage{graphicx}
\usepackage[colorlinks=true,linkcolor=blue,citecolor=blue, urlcolor=blue]{hyperref}
\usepackage{pifont}
\usepackage{sidecap}
\usepackage{txfonts}
\usepackage{lipsum}
\usepackage{tikz}
\usetikzlibrary{shapes.geometric, arrows}
\newcommand*\samethanks[1][\value{footnote}]{\footnotemark[#1]}

\defcitealias{2023martin}{MS23}
\defcitealias{2023andrae}{A23}

\begin{document}

   \title{The Pristine survey}
   \subtitle{XXVIII. Journey to the Galactic outskirts: Mapping the outer halo red giant stars down to the very metal-poor end}
   \titlerunning{Journey to the Galactic outskirts}

   \author{Akshara~Viswanathan\inst{\ref{kapteyn},\ref{uvic}}\fnmsep\thanks{Both authors contributed equally\\ \email{astroakshara97@gmail.com;\\Amanda.Bystrom@ed.ac.uk}}
          \and
          Amanda~Byström
          \inst{\ref{edin},\ref{kapteyn}}\fnmsep\samethanks[1]
          \and
          Else~Starkenburg\inst{\ref{kapteyn}}
          \and
          Anne~Foppen\inst{\ref{kapteyn}}
          \and
          Jill~Straat\inst{\ref{kapteyn}}
          \and
          Martin~Montelius\inst{\ref{kapteyn}}
          \and
          Federico~Sestito\inst{\ref{uvic}}
          \and
          Kim~A.~Venn\inst{\ref{uvic}}
          \and
          Camila~Navarrete\inst{\ref{nice}}
          \and
          Tadafumi~Matsuno\inst{\ref{heid}}
          \and
          Nicolas~F.~Martin\inst{\ref{stra},\ref{mpia}}
          \and
          Guillaume~F.~Thomas\inst{\ref{iac},\ref{laguna}}
          \and
          Anke~Ardern-Arentsen\inst{\ref{cam}}
          \and
          Giuseppina~Battaglia\inst{\ref{iac},\ref{laguna}}
          \and
          Morgan~Fouesneau\inst{\ref{heid}}
          \and
          Julio~Navarro\inst{\ref{uvic}}
          \and
          Sara~Vitali\inst{\ref{chile}}
          }

   \institute{Kapteyn Astronomical Institute, University of Groningen, Landleven 12, 9747 AD Groningen, The Netherlands\label{kapteyn}
   \and
   Dept. of Physics and Astronomy, University of Victoria, P.O. Box 3055, STN CSC, Victoria BC V8W 3P6, Canada\label{uvic}\and
   Institute for Astronomy, University of Edinburgh, Royal Observatory, Blackford Hill, Edinburgh EH9 3HJ, UK\label{edin}\and
   Universit\'e C\^ote d'Azur, Observatoire de la C\^ote d'Azur, CNRS, Laboratoire Lagrange, Nice, France\label{nice}\and
   Astronomisches Rechen-Institut, Zentrum f\"ur Astronomie der Universit\"at Heidelberg, M\"onchhofstra{\ss}e 12-14, 69120 Heidelberg, Germany\label{heid}\and
   Universit\'e de Strasbourg, CNRS, Observatoire astronomique de Strasbourg, UMR 7550, F-67000 Strasbourg, France\label{stra}\and
   Max-Planck-Institut f\"ur Astronomie, K\"onigstuhl 17, D-69117 Heidelberg, Germany\label{mpia}\and
   Instituto de Astrof\'isica de Canarias, Calle V\'ia L\'actea s/n, 38206 La Laguna, Santa Cruz de Tenerife, Spain\label{iac}\and
   Universidad de La Laguna, Avda. Astrof\'isico Francisco S\'anchez, 38205 La Laguna, Santa Cruz de Tenerife, Spain\label{laguna}\and
   Institute of Astronomy, University of Cambridge, Madingley Road, Cambridge CB3 0HA, UK\label{cam}\and
   N\'ucleo de Astronom\'ia, Facultad de Ingenier\'ia y Ciencias Universidad Diego Portales, Ej\'ercito 441, Santiago, Chile\label{chile}
    }

   \date{Received 31 August 2024/ Accepted 5 October 2025}

  \abstract
   {The outer Galactic halo remains relatively unexplored, particularly regarding its metallicity distribution, merger debris, and the population of very and extremely metal-poor ([Fe/H] < --2.5) stars.}
   {Using photometric metallicities from the Pristine survey data release 1 (PDR1) and Pristine-\textit{Gaia} synthetic (PGS) catalogue and \textit{Gaia} DR3 astrometry, we constructed well-characterised samples of bright (G < 17.6) red giant branch (RGB) stars in the outer halo. With accurate distances, these samples enable studies of the halo’s metallicity distribution, accreted debris, and very metal-poor (VMP) substructures beyond 40 kpc.}
   {We selected giants by excluding stars with reliable \textit{Gaia} parallaxes in brightness ranges where dwarfs are measurable. Purity and completeness were validated against the Pristine spectroscopic training set. Distances were derived using BaSTI isochrone fitting combined with Pristine metallicity estimates.}
   {The photometric distances reach $\sim$100 kpc (PDR1) and $\sim$70 kpc (PGS), with typical uncertainties of 12\% and scatter up to 20-40\% compared to parallax- and StarHorse-based distances. The PDR1 sample provides a nearly unbiased metallicity–distance view, while the PGS sample offers an all-sky map, especially at the very metal-poor end. Using PDR1-giants, we traced the halo metallicity distribution function out to 101 kpc, fitting a three-component Gaussian mixture model. The most metal-poor component becomes increasingly dominant with distance, as beyond 50 kpc, 40–50\% of the stars are very metal-poor ([Fe/H] < --2.0). With added radial velocities, we identified metallicity trends in integrals-of-motion space and investigated accreted debris. The PGS sample reveals substructures, including the Pisces Plume, where 41 VMP stars are linked to the Magellanic stream.}
   {We publish two RGB catalogues: PDR1-giants (180,314 stars, with 10,096 very metal-poor candidates and 2,096 beyond 40 kpc) and PGS-giants (2,420,898 stars, with 75,679 very metal-poor candidates and 267 beyond 40 kpc). These catalogues represent extensive resources for future outer halo studies.}

   \keywords{Galaxy: stellar content - Galaxy: halo - Galaxy: kinematics and dynamics - Galaxy: structure - Galaxy: formation - Methods: data analysis}

   \maketitle
%
\section{Introduction}\label{1}
The stellar halo provides a unique insight into the assembly history of our Galaxy. Due to the long dynamical timescales, the halo has not undergone complete phase mixing. Therefore, measuring the orbital and chemical properties of halo stars can help reconstruct significant events in the Galaxy's history. This is one of the main goals of the field of Galactic archaeology.

Early theories about the formation of the stellar halo considered both `dissipative' \citep{1962eggen} and `dissipationless' \citep{1978searle} formation channels. These are now referred to as `in situ' and `accretion' (or  `ex situ') channels. In the in situ scenario, halo stars are born within the Galaxy and are later dynamically heated to halo-like orbits \citep{2015cooper,2017bonaca,2018koppelman}. Conversely, the accretion scenario, supported by hierarchical assembly in a cold dark matter cosmology, suggests that the halo was partially formed by the tidal disruption of smaller dwarf galaxies \citep{1999helmi,2005bullock,2006abadi}.
In theory, the combined orbital and chemical properties of stars should provide a powerful method for understanding the origin of the halo. Such data can help differentiate between in situ and accreted stars based on distance, metallicity, and other factors. A key objective of Galactic archaeology is to identify the number of significant events contributing to the accreted halo, estimate the progenitor masses and orbital properties, and ultimately reconstruct the stellar halo's formation. This research field has a rich history of using chemical and orbital properties to study the halo's origins \citep[e.g.][]{1991ryan,2000chiba,2004venn,2007carollo,2017bonaca,2018helmi,2018belokurov,2018koppelman,2018malhan,2019koppelman,2019myeong,2019iorio,2020naidu,2020yuan,2022wang,2023ibata}.

Significant progress has been made in understanding the origin and nature of the Galactic halo, yet several key questions remain open. Recent work has shed light on the relative contributions of in situ and ex situ components, but the full extent of the in situ (hot thick disc) population and its dominance across the halo mass spectrum is still debated \citep{2018helmi,2020belokurov,2023lee,2023khoperskovb}. Investigating the radial and vertical extent of this component is essential for constraining the Galaxy’s accretion history. On the ex situ side, it remains uncertain whether the outer halo is primarily built from a few massive merger events or a multitude of lower-mass, ultra-faint galaxies \citep[e.g.][]{2015deason,2023deason,2023khoperskov}. The radial variation of metallicity—i.e. whether it transitions to a more metal-poor spherical halo beyond 40 kpc—has also been proposed but not yet conclusively established \citep[e.g.][]{dietz2020,Liu2022,2025medina}. Whether such trends are driven by specific merger events or reflect a smooth component from disrupted globular clusters is also still under discussion. Furthermore, evidence of apocentric pile-ups from ancient mergers is now emerging \citep[e.g.][]{2021balbinot,2022perottoni,2024zhu}, though the completeness of our current samples may be limited by selection effects. In particular, colour cuts and the use of metallicity-sensitive standard candles may bias our view of the outer halo as being exclusively metal poor \citep[e.g.][]{2019conroyb,2020youakim,2021bonifacio}. By probing the distant metal-poor halo using individual stars, we are effectively resolving the remnants of galaxies with stellar masses in the range M$_*$ = 10$^6$–10$^7$ M$_\odot$, allowing us to study the accreted high-redshift galaxy population in unprecedented detail, which is currently beyond the mass ranges that can be studied using direct observation of high redshift galaxies.

Pre-\textit{Gaia}, attempts to directly explore the distant halo faced the challenge of targeting rare far away stars without the advantage of \textit{Gaia} parallaxes to exclude nearby contaminants, and therefore colour-cuts or low metallicity cuts had to be implemented. The main problem with these methods is that the selection and its consequences have not been explored fully to understand the advantages and limitations.
Fortunately, the observational landscape is rapidly improving on multiple fronts. \textit{Gaia} has measured proper motions and parallaxes for over 1.8 billion stars down to G$\sim$22 \citep{2023gaia}. While the majority of halo stars are too distant for precise parallax measurements and too faint for \textit{Gaia}'s radial velocity measurements, these new data are transforming our understanding of the stellar halo close to the Sun \citep[e.g.][]{2022lovdal,2022ruizlara,2023dodd}.
Additionally, very metal-poor (VMP; [Fe/H] < --2.0) stars are considered `fossils' of the early Universe, preserving its primordial chemical composition. Likely among the oldest stars, possibly even the first-generation \citep[population III,][]{2005beers,2015frebel,2018hansen,2018starkenburg} stars, VMP stars are key tracers of the Milky Way's formation and have greatly advanced our understanding of its assembly history \citep[see e.g.][]{1970bond,1981bond,1985beers,1992beers,2002christelieb,2010nissen,2020yuan}.

To date, nearly all observational work on the stellar halo has used tracers biased by metallicity. These biases arise because halo stars are rare and generally more metal poor than disc stars, combined with the need to efficiently use spectroscopic resources \citep{2009starkenburg,2013an,2017battaglia}. Before \textit{Gaia}, the most efficient method to distinguish the halo from disc stars was by selecting stars with low metallicities. This bias can occur at two stages: when selecting targets for spectroscopic follow-up and when identifying halo stars from the final sample. For instance, the SDSS calibration stars used by \citet{2007carollo,2010carollo} to study the stellar halo were selected based on their blue colours. The SDSS SEGUE sample of K giants, used to study the halo at great distances \citep{2015xue,2016das}, was selected for spectroscopic follow-up using colour-cuts favouring low metallicities. Photometric metallicities of F/G turnoff stars are another common method for studying the stellar halo, but such samples also favour lower metallicity stars due to their construction based on colour-cuts \citep{2008ivezic,2011sesar,2017zuo}. Additionally, rarer populations such as RR Lyrae and blue horizontal branch stars have been extensively used due to their status as standard candles \citep{2011deason,2017cohen,2017hernitschek,2017sesar,2018thomas,2019starkenburg,2019iorio}. However, these populations also preferentially trace metal-poor stars. The biases incurred by using these populations to study the halo are difficult to overcome without near-perfect knowledge of the underlying population and what fractions of stars were included in the sample. These different observational methods have led to sometimes conflicting conclusions about the chemical-orbital structure of the stellar halo. 

To study the metallicity structure of the distant Galactic halo, it is essential to construct a sample in which the effects of selection, distance uncertainties, and metallicity biases are well understood and corrected. Tip of the red giant branch (RGB) stars serve as bright, luminous tracers that offer significantly better proper motion precision—tangential velocity uncertainties of $\Delta_v \leq 30$ km/s at 80 kpc—compared to RR Lyrae or blue horizontal branch (BHB) stars \citep{2023chandraa}. Thus, the RBG stars enable us to probe the halo at greater distances with improved kinematics in a magnitude-limited sample.
However, this kinematic advantage comes with a trade-off in distance precision, as RGB distance estimates depend sensitively on metallicity. To address this, we based our selection on photometric metallicity catalogues from the Pristine survey and \textit{Gaia} XP-based narrow-band CaHK magnitudes, which provide complementary coverage in depth and sky area. In this work, we use these catalogues to construct a sample of RGB stars suitable for distant halo studies, carefully tracking and correcting for selection biases. Our approach allows us to probe the outer halo’s metallicity distribution and search for very metal-poor substructures, including features linked to past merger events and interactions with the Large and Small Magellanic Clouds (respectively, LMC and SMC). 

The rest of the paper is divided into five sections. In Section \ref{2} we describe the input photometric metallicity catalogues, the spectroscopic training sample of the Pristine survey on which our RGB selection pipeline was tested, and the quality cuts used in the input photometric metallicity catalogues. In Section \ref{3} we describe the full RGB selection pipeline solely based on photometry and parallax without the need for good distances, radial velocity, or other astrometric parameters; the photometric distance calculation using an isochrone fitting code; and the final purity and completeness of our RGB selection pipeline, using spectroscopic surface gravity determined independently from the \textit{Gaia} XP spectra. Section \ref{4} presents some of our main results, including a description of the two catalogues of giants, a bias-corrected view of the metallicity distribution functions out to large distances, a 6D subset using literature spectroscopic radial velocities and their orbital properties, associating them with several known accretion events, an all-sky view of the Milky Way outer Galactic halo in the VMP end and its implications. Finally, in Section \ref{6} we present the main results, conclusions, and outlook in a broad context.

\section{Data}\label{2} 

In this section, we describe the input photometric metallicity catalogues from the Pristine survey's first data release (PDR1) and the \textit{Gaia} XP spectra ran on the Pristine survey model, Pristine-\textit{Gaia} synthetic (PGS), which were used to construct the two RGB catalogues that probe the outer Galactic halo. We also describe the quality cuts applied to these input catalogues before running them on our RGB selection pipeline. We also briefly describe the Pristine survey's training sample on which we run our selection pipeline to test the purity and completeness of our selection methods.

\subsection{Pristine Data Release 1 catalogue of photometric metallicities}\label{2.1}

The Pristine survey observes the northern sky using the MegaCam wide-field imager located on the Canada France Hawaii Telescope at Mauna Kea. 
This survey utilises a narrow-band filter centred on the calcium (Ca) II H$\&$K lines (CaHK) in the near UV at 3968.5 and 3933.7 \AA, which are highly sensitive to metallicity. 
When combined with SDSS g, r, and i broad-band filters or \textit{Gaia}'s BP-RP broad-bands, this narrow-band filter has been proven to provide reliable estimates of stellar metallicity  \citep{2017starkenburg,2023martin}.
This pre-selection method is particularly effective at identifying very and extremely metal-poor stars 
\citep[VMP, \text{[Fe/H]}<-2.0, and EMP, \text{[Fe/H]}<-3.0,][]{2017youakim,2019aguado,2024viswanathanb}, 
and picks up ultra metal-poor stars (UMP, [Fe/H]<-4) \citep[the remnants of first stars,][]{2018starkenburg}.

The Pristine photometry, in conjunction with SDSS broad-band photometry, has paved the way for subsequent medium and high-resolution spectroscopic studies. These investigations specifically target stars with the lowest metallicity estimates derived from Pristine CaHK observations. The outcomes of these spectroscopic follow-up efforts have been successful, as evidenced by various works such as \cite{2017caffau,2018starkenburg,2019bonifacio,2020caffau,2020venn,2021kielty,2021lardo,2022lucchesi,2023caffau,2023lombardo}.
The combination of metallicity-sensitive medium/narrow-band photometry with broadband photometry has resulted in significant samples of VMP and EMP star candidates. 

The new version of the Pristine survey now uses \textit{Gaia} broad-bands instead of SDSS broad-bands to infer these photometric metallicities down to [Fe/H]$\sim$-4.0. 
PDR1 was made public by \citet[][hereafter \citetalias{2023martin}]{2023martin} for every star in the Pristine survey that has a \textit{Gaia} XP spectrum released by the newest \textit{Gaia} data release 3 down to magnitudes of G$\sim$17.6 \citep{2023gaia,2023deangeli}.
The high accuracy of \textit{Gaia} XP data allowed for the reprocessing and recalibration of the entire Pristine CaHK dataset, consisting of approximately 11,500 images taken since 2015. This update expanded the survey to cover over 6,500 square degrees. The enhanced photometric catalogue now achieves a precision of 13 millimagnitudes (mmag), a significant improvement from the initial precision of 40 mmag. 
In the updated methodology, the Pristine approach for determining the photometric metallicity of stars from CaHK and broadband magnitudes now relies exclusively on \textit{Gaia} broadband data (G, G$_{BP}$, G$_{RP}$) rather than SDSS data. An iterative method for extinction correction has been implemented, incorporating corrections on both \textit{Gaia} broadband magnitudes and CaHK synthetic (or Pristine CaHK) narrowband magnitudes, considering the star’s photometric temperature and metallicity. 
Variable stars are a significant source of contamination in photometrically selected low-metallicity catalogues, as their varying brightness can mimic the colours of metal-poor stars \citep{2017starkenburg,2023lombardo}. 
To mitigate the effects of photometric variability, which can lead to inaccurate metallicities, a variability model based on the photometric uncertainties of the 1.8 billion \textit{Gaia} sources were also included as described in section 5 of \citetalias{2023martin}.

Spectroscopic analyses of red giant stars have confirmed that approximately 38\% of the extremely metal-poor star candidates have [Fe/H] values below -3.0, considering quality flags during target selection, which makes these catalogues some of the most successful ones at finding extremely metal-poor stars \citep{2024viswanathanb}. However, it is crucial to consider potential variability in these stars and to implement rigorous photometric quality cuts to ensure accurate characterization of their metallicities \citep{2023lombardo}. Thanks to \textit{Gaia} DR3, several quality cuts have been improved and recent spectroscopic follow-ups have had a much larger success at finding very and extremely metal-poor stars \citep{2024viswanathan,2024viswanathanb}. 
We used PDR1, to create our first sample of red giants in the outer halo. 

\subsection{Pristine-\textit{Gaia} synthetic catalogue of photometric metallicities}\label{2.2}

The detailed creation of the PGS catalogue is described in \citetalias{2023martin}, and we only provide a brief summary here. Using the latest \textit{Gaia} data release \citep[][DR3]{2022gaia}, spectrophotometric XP information \citep{2023deangelinew} was used to construct an extensive catalogue of synthetic CaHK magnitudes, mimicking the narrow-band photometry used in the Pristine survey \citep{2023montegriffo}. 
Additionally, several recent studies have released metallicity estimate catalogues based on these \textit{Gaia} XP spectra \citep{2023andrae,2023zhang,2024xylakis}.

Using both Pristine CaHK magnitudes and XP-based synthetic CaHK magnitudes within the Pristine model, two catalogues of photometric metallicities for reliable stars were made public: the PGS catalogue and the PDR1 catalogue of photometric metallicities, which includes stars common to both Pristine and the XP catalogue from \textit{Gaia} DR3. The latter, serving as the first data release of the Pristine survey, provides deeper data with better signal-to-noise (S/N) ratios for stars in common.
These catalogues enable the construction of reliable samples of metal-poor stars, with a particular focus on V/EMP stars. The PGS catalogue offers photometric metallicities over a large portion of the sky, while the PDR1 catalogue, limited to the Pristine survey’s footprint, delivers high-quality metallicities and extends to significantly fainter stars.
We used this PGS catalogue to create our second sample of red giants in the outer halo.

\subsection{Quality cuts on PDR1 and PGS catalogues}\label{2.3}

To construct a pure and complete sample of RGB stars out to large distances, we use the PDR1 and PGS catalogues of photometric metallicities. It is important to note that we use the photometric metallicities inferred using the giants subsample of the training sample in the rest of this paper, i.e. the metallicities inferred for each star in Pristine if it were a giant. We used the following quality cuts on these input catalogues to end up with reliable photometric metallicities and to allow for an efficient selection of giants based on photometry and astrometry, most of which follow the suggestions from \citetalias{2023martin}:

\begin{itemize}
    \item   The photometric metallicity uncertainty is less than 0.5 dex (0.5*(\texttt{FeH\_CaHKsyn\_84th} - \texttt{FeH\_CaHKsyn\_16th})<0.5 dex, or 0.5*(\texttt{FeH\_Pristine\_84th} - \texttt{FeH\_Pristine\_16th})<0.5 dex).
    \item The 84th percentile value of the probability distribution function (PDF) of the photometric metallicity is greater than -3.999 (\texttt{FeH\_CaHKsyn\_84th}>-3.999, or \texttt{FeH\_Pristine\_84th}>-3.999).
    \item The percentage of Monte Carlo iterations used to determine [Fe/H] uncertainties inside the grid are greater than 80\% (\texttt{mcfrac\_CaHKsyn}>0.8, or \texttt{mcfrac\_Pristine}>0.8).
    \item The CASU photometric data reduction flag (\texttt{merged\_CASU\_flag} = -1 (or -2), that denotes whether a source is very likely (or likely) point-sources - this is applicable only for sources with PDR1 observations).
    \item The extinction on the B-V magnitude is less than 0.5 (\texttt{E(B-V)}<0.5).
    \item The photometric quality cut is defined as |{$\text{C}^*$}|<{3$\sigma_{^\text{C}*}$}. \texttt{Cstar} is the \textit{Gaia} DR3 corrected flux excess, C$^*$, as defined in equation 6 of \citealt{2021riello}, and \texttt{Cstar\_1sigma} is the normalised standard deviation of C$^*$ for the G magnitude of this source, $\sigma_{^\text{C}*}$, as defined in equation 18 of \citealt{2021riello} (abs(\texttt{Cstar})<\texttt{3*Cstar\_1sigma}).
    \item The probability of a star being a variable star to be less than 30\% \texttt{Pvar}<0.3.
    \item The Gaia astrometric quality cut (\textit{Gaia}'s renormalised unit weight error \texttt{RUWE} < 1.4).
\end{itemize}

\subsection{Spectroscopic training sample used by the Pristine survey model}\label{2.4}

The Pristine survey's model to derive photometric metallicities from CaHK narrow-band magnitudes is limited to FGK stars because, for hotter stars, the CaHK absorption lines are too weak to serve as reliable metallicity indicators. On the cooler end of the spectrum, very cool M stars and cool K giant have prominent molecular bands that significantly lower the level of the pseudo-continuum in the relevant wavelength range, making it challenging to measure the CaHK absorption features. 
Therefore, \citetalias{2023martin} restricts their analysis to stars with 0.5 < (G$_{BP,0}$ - G$_{RP,0}$) < 1.5, covering evolutionary stages from the upper main sequence and turn-off to the tip of the RGB for an old, VMP stellar population. This colour interval corresponds to a temperature range of approximately 3900 < T$_{\rm eff}$ < 7000 K.
The colour cut is also necessary due to the lack of VMP stars in the training sample in the colour ranges beyond this.

Following the methodology described in \citet{2017starkenburg}, \citetalias{2023martin} use a training sample of 66, 000 stars to map the de-reddened (CaHK, G, G$_{BP}$, G$_{RP}$) colour space onto photometric metallicities. 
A major component of this sample consists of SDSS/SEGUE stars \citep{2009yanny,2013smee} within the Pristine footprint, with an average signal-to-noise ratio per pixel greater than 25 over the 400–800 nm wavelength range. Additionally, the SDSS pipeline must provide log \textit{g} values, adopted T$_{\rm eff}$ < 7000 K, radial velocity uncertainty < 10 km/s, and adopted spectroscopic metallicity [Fe/H] with an uncertainty < 0.2 dex. 
For our outer halo studies, red giant stars are necessary, which are less numerous in the training sample. Therefore, \citetalias{2023martin} complements the SDSS sample with APOGEE DR17 giants.
To cover the rare very, extremely and ultra metal-poor stars as much as possible in the training sample and to provide reliable photometric metallicities in the VMP end, the training sample is supplemented by stars from the Pristine survey's spectroscopic follow-up programs \citep{2017youakim,2019aguado,2020venn,2021kielty,2021lardo,2022lucchesi}, VMP stars from the third data release of the LAMOST survey \citep{2018li} corrected for spurious stars in the low-temperature range as described by \citet{2020sestito} using the latest LAMOST DR8 catalogue, the PASTEL sample \citep{2016soubiran} as used by \citet{2022huang}, and high-resolution observations of the Boötes I dwarf galaxy \citep{2013gilmore,2016frebel}. 

In order to have a uniform sample of stars with homogeneously analysed atmospheric parameters to test our RGB selection pipeline, we use only the SEGUE subset of the training sample and excluding the other subsamples of the training set defined by \citetalias{2023martin} in Section \ref{3}.

\section{Methods}\label{3} 

The pipeline for constructing the final catalogue was tested on the training sample described in section \ref{2.4}. 
Because this sample contains spectroscopic log \textit{g} values, we can reliably divide the stellar sample into dwarfs and giants by their log \textit{g} measurements. 
We choose to define dwarfs as all stars with log \textit{g} > 3.5 and giants as all stars with log \textit{g} < 3.5. 
This division allows us to compute the purity and completeness of giants of every new cut we apply to the catalogue, and we can choose to apply the cuts that maximise both. 
We define purity as the number of stars with log \textit{g} < 3.5 divided by the total number of stars after our selections and completeness as the number of giants after our selections divided by the total number of giants in the training sample (stars with log \textit{g} < 3.5).
We apply the colour cut 0.5 < G$_{BP,0}$ - G$_{RP,0}$ < 1.5 to the training sample to match the colour range over which Pristine photometric metallicities are assigned \citepalias{2023martin}. 
It is important to note that the spectroscopic training sample based on SEGUE is not fully representative of our input PDR1/PGS catalogues of photometric metallicities. 
This is because of different on-sky coverage, magnitude range and target selection effects.
Therefore, the purity and completeness derived from them do not necessarily trace the purity and completeness of our final catalogue of RGB stars. However, the training sample allows us to efficiently test our method, which in turn allows us to better understand the consequences of our selection pipeline.

In the subsections \ref{3.1}, \ref{3.2} and \ref{3.3} we introduce the motivation for the two cuts we apply to the catalogue.
In subsection \ref{3.4} we explain how distances to the stars are derived.
The main advantage of our method is that we select RGB stars using only parallax\footnote{Throughout this work, when we refer to parallax, we mean the corrected parallax. The parallax ($\pi$) is corrected for its individual zero-point offset ($\pi_{\rm offset}$) using the \texttt{gaiadr3\_zeropoint} python module, following the procedure outlined in \citet{2021lindegreen}.} and photometry, without the need for good distances, radial velocities, or atmospheric parameters.

\subsection{Parallax-based colour-absolute magnitude diagram cut}\label{3.1}

For FGK stars, the main contamination in the selection of giants is dwarfs in the same colour range. An obvious difference between giant and dwarf stars is where they are located in the colour-absolute magnitude diagram (CaMD), so if we can create an CaMD of the sample, we can use the location of stars in it to separate them.
Because all of our stars have \textit{Gaia} parallaxes, we can invert these parallaxes to get stellar distances, and use the apparent magnitudes to get absolute magnitudes using the distance modulus equation, to produce an CaMD.
However, inverting parallaxes to get distances can only be done for good quality parallaxes, \texttt{parallax\_over\_error>5}, as recommended by the Gaia consortium \citep{2018bailer-jones}. 
In practice this means that this cut, which we refer to as the parallax-based CaMD cut, is only applied to the subset of stars that have `good enough' parallaxes \footnote{The `good enough' parallax is defined as $\pi$>0 and $f$ = $\Delta\pi$/$\pi$ $\in$ (0, 1)}. 
We use inverted parallax simply as a means to remove nearby dwarfs with good enough parallax, and therefore our quality cut on parallax errors can be less strict than what is recommended when using the parallax to obtain distances. 
For all stars with bad parallaxes, there is no equivalent way of applying these cuts.
The following part of the method is thus only applied to the good enough parallax subset of the catalogue.

We define good (or well-defined) parallax as being neither negative nor zero, as both imply a non-physical distance, as well as the fractional parallax uncertainty being less than some value $f$ = $\Delta \pi$/$\pi$, with $f \in (0, 1)$. 
It is important to note that we do not use the inverted parallaxes to get distances for these stars, but simply use them to remove dwarfs that have good enough parallaxes (see subsection \ref{3.4} for the description of photometric distance inference).

To compute the purity and completeness for different values of $f$, we needed to first perform the parallax-based CaMD cut, so we needed to define a division between dwarfs and giants. 
We visually inspected the CaMD of the training sample for $f \leq 0.1$, i.e. we only looked at stars with very good parallaxes (uncertainties less than 10\%), and we constructed the following piece-wise division between dwarfs and giants:

\begin{equation}\label{eq:dwarfgiantsdivision}
\begin{aligned}
    M_G = 0.9 \text{ } \&& \text{ }  G_{BP,0}-G_{RP,0} < 0.8, \\  
    M_G = 6.25(G_{BP,0}-G_{RP,0}) - 1.7 \text{ } \&& \text{ } 0.8 < G_{BP,0}-G_{RP,0} < 0.952, \\
    M_G = 4.25 \text{ } \&& \text{ } 0.952 < G_{BP,0}-G_{RP,0}.
\end{aligned}
\end{equation}

\noindent All the stars that fall below these lines are assumed to be dwarfs. We then tested different values of $f$ using the above distinction between dwarfs and giants.

\begin{figure}
    \centering
   \includegraphics[width=\hsize]{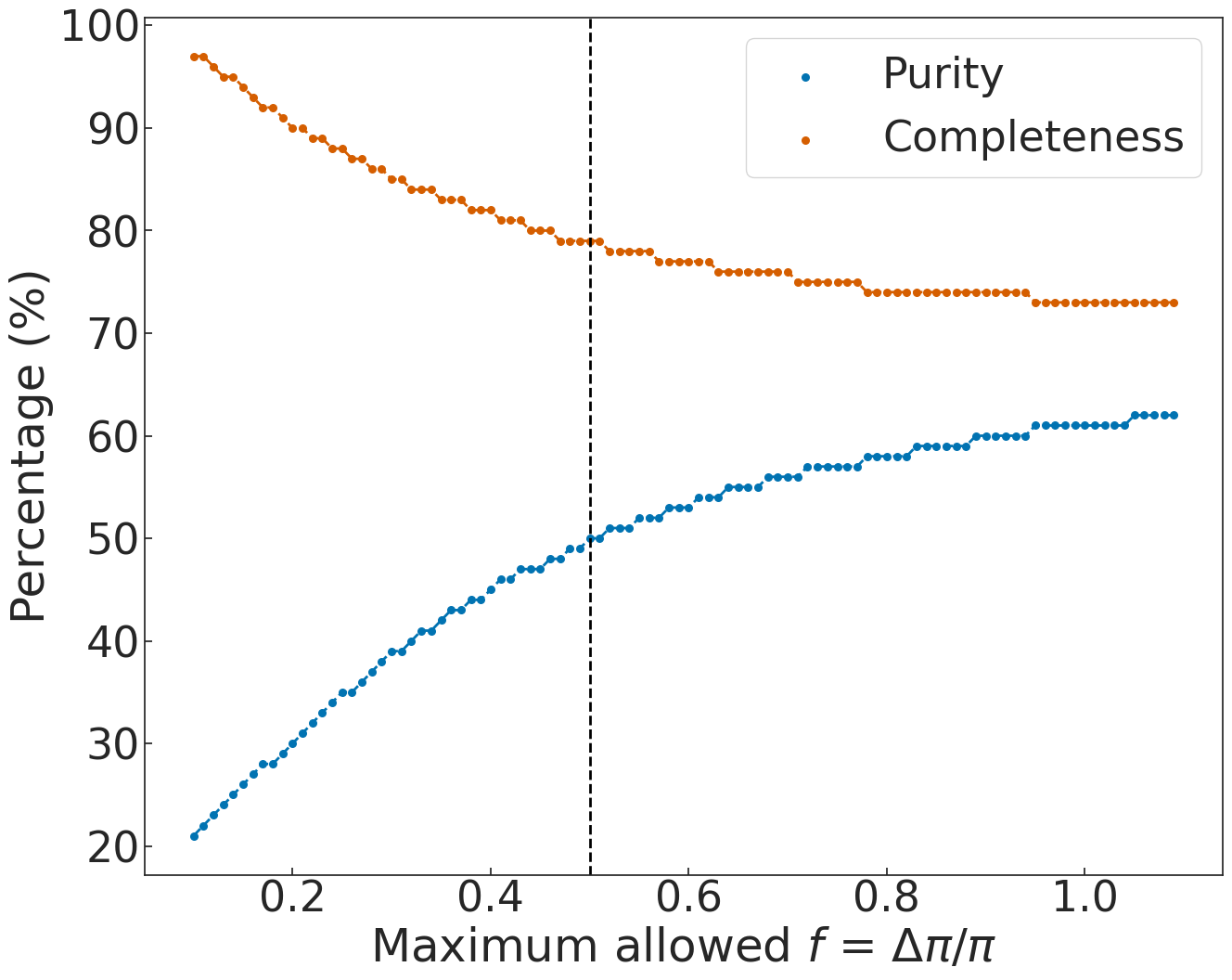} 
      \caption{
      Purity (blue) and completeness (orange) for different parallax-based CaMD cuts (along with bad parallax pool of giants) in the training sample. The choice of 50\% ($f$=0.5) is justified by the plateau reached in purity and completeness, with a good balance between the two values. The discreteness of the plot is due to the small size of the training sample (N=53666) used to test our cuts. All values are given in percentages. 
              }
         \label{fig:puritycompletenessfvalues}
\end{figure}

To identify a value of the fractional parallax uncertainty $f$ that maximises both purity and completeness of the parallax-based CaMD cut, we needed to compute the purity and completeness for several different values of $f$. 
We required that $f$ cannot be negative or larger than 100 \%, meaning that we only consider values of $f \in (0, 1)$. 
The purity and completeness for this range in $f$ is seen in Figure \ref{fig:puritycompletenessfvalues}. 
This figure shows that as $f$ increases, the purity monotonically increases (more dwarfs are correctly removed), and completeness monotonically decreases (more giants are incorrectly removed). 
It is important to note that the choice of \( f \) affects both the removal of dwarfs and the selection of giants. An \( f = 0.1 \) cut means that stars with positive parallaxes and \( f < 0.1 \), located above the division line in equation \ref{eq:dwarfgiantsdivision}, are considered giants. Additionally, all stars with \( f > 0.1 \), including those with negative parallaxes, are also selected as potential giant candidates.
This is the reason why finding a good balance in the choice of $f$ is important.
From Figure \ref{fig:puritycompletenessfvalues}, we see that as $f$ increases, the purity increases. 
There is an inherent effect on the distance probed and the quality of the parallax. 
Therefore, with increasing  $f$, we are more likely to select stars at larger distances, i.e. brighter giants, on a fixed CaMD.
With increasing $f$, we are also more likely to have a cleaner selection of giants based on their `bad' parallax.
From Figure \ref{fig:puritycompletenessfvalues}, we also see that as $f$ increases, completeness decreases. 
This is because, with increasing $f$, we are more complete for the `good' parallax giants but less complete for the `bad' parallax giants and the two effects together make the completeness go down at larger $f$.

For our RGB selection, we choose to use $f = 0.5$, as it falls approximately where the increase in purity and decrease in completeness have a lower gradient and start to reach a plateau without having to compromise on the parallax uncertainty too much.
The chosen 50\% uncertainty on parallax is much higher than the \textit{Gaia}-recommended 20\%. However, we stress that we only use parallax as a proxy to remove dwarfs with good enough parallax and not assign distances based on the inverted parallax.  
We only need good enough parallaxes to have a reliable dwarf-giant division.
If we use \citet{2021bailer-jones} photogeometric distances instead of inverted parallax distances, the purity and completeness results remain unchanged.
After this cut is applied to the data, the purity is 50\% and completeness is 79\%.

\begin{figure}
    \centering
   \includegraphics[width=\hsize]{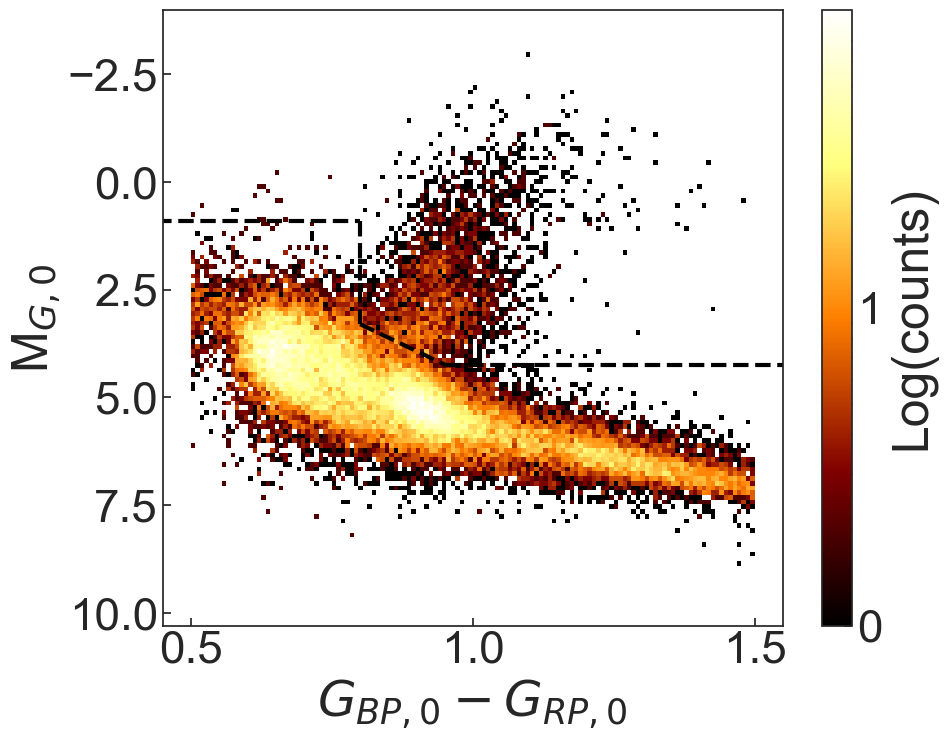} 
      \caption{Colour-absolute magnitude diagram of the training sample, with absolute magnitudes computed using \textit{Gaia} parallaxes, $\pi$, with the conditions that $\pi > 0''$ and that the fractional parallax uncertainty $f = 0.5$. 
              }
         \label{fig:ACMDtrainingsamplegiantdwarfcut}
\end{figure}

Figure \ref{fig:ACMDtrainingsamplegiantdwarfcut} shows the CaMD of the training sample (SEGUE), and the division between dwarfs and giants defined in equation \ref{eq:dwarfgiantsdivision}; by visual inspection, we see that it works for our choice of $f$$\leq$0.5 (with the giant sequence being distinctly visible), and it is the definition we used to construct the final RGB catalogues.

\subsection{Magnitude cut}\label{3.2}

The previous section describes a method to clean the catalogue of dwarfs when good enough parallaxes are available.
After this method had been applied, the sample consisted of giants with good enough parallaxes and dwarfs and giants with bad parallaxes.
\textit{Gaia} DR3 astrometry only probes a volume of 2 kpc for dwarfs with good enough parallax \citep{2023viswanathan}.
We expect that all dwarfs with $G > 17.6$ are too far away to have good enough parallaxes. 
We introduce a cut where all stars fainter than $G = 17.6$ are removed, to remove these bad parallax dwarfs.
Although there are some giants at these magnitudes, meaning that this cut reduces completeness, dwarfs with bad parallax dominate the sample at these faint magnitudes (as dwarfs are 100 times more common than giants in stellar evolution) in our training sample, and we cannot distinguish both without the use of atmospheric parameters. 
The choice of the exact value of 17.6 in \textit{Gaia} G magnitude (which is also the magnitude limit of the PGS catalogue because it is based on \textit{Gaia} XP spectra that has a magnitude limit of about 17.6) is justified by looking at the histogram of \textit{Gaia} G magnitudes for the dwarf stars with good enough parallax as described in detail in Appendix \ref{A}.
In combination with the CaMD cut, this cut increases purity to 65\% and decreases completeness to 64\%.

\begin{figure}
    \centering
   \includegraphics[width=\hsize]{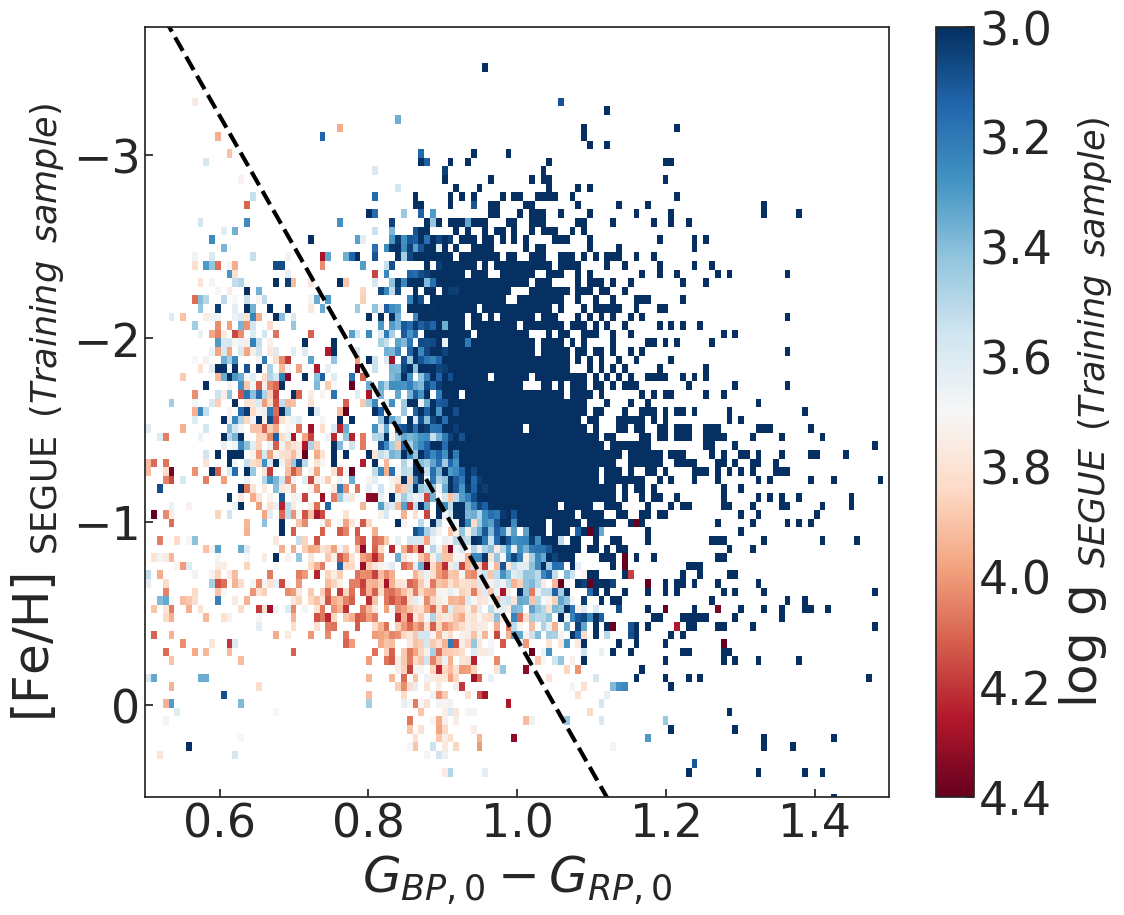} 
      \caption{Training sample (CaMD cut on good enough parallax giants and bad parallax giants selected after the magnitude cut) colour-coded by log \textit{g} and the colour-metallicity cut in equation \eqref{eq:colourmetallicitycut}, shown as a black line. Any stars that lie under this line are removed. The cut is designed to remove mainly subgiant stars probed unevenly at different metallicity ranges, and the colour-coding shows that most stars underneath this line has log \textit{g} roughly greater than 3.5 (not a giant). The cut preserves the RGB (the blue region) and cuts away the SG/MSTO (the red region)}
         \label{fig:colormetallicitycut}
\end{figure}

\begin{figure*}
    \includegraphics[width=0.9\textwidth]{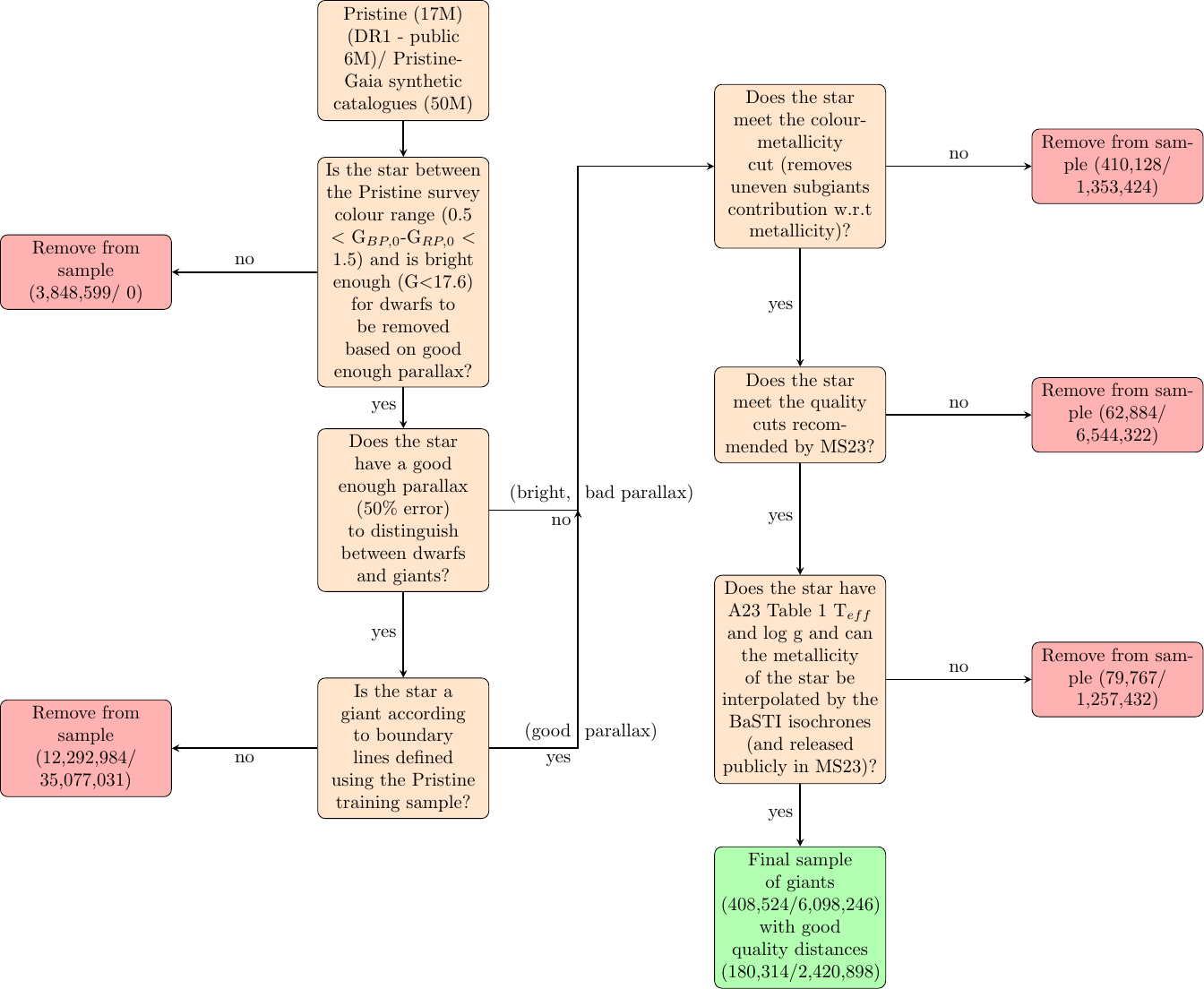}
    \caption{Flowchart of the steps involved in creating the two RGB star catalogues using the parent samples from PDR1 and PGS catalogues. The number of stars removed at each selection step for the two input catalogues are shown in the red boxes, the method and cuts used to select RGB stars on every step is shown in orange boxes, and the final sample with counts are shown in the green box.}
    \label{fig:flowchart}
\end{figure*}

\subsection{Colour-metallicity cut}\label{3.3} 

The fraction of dwarfs to giants changes as a function of both colour and metallicity: as the metallicity decreases, the colour at which the subgiant branch turns into the giant branch becomes bluer (and the absolute magnitude decreases).
We thus introduced a cut that removes the subgiants based on their metallicities and colours.
Using five MIST isochrones \citep{2016choi,2016dotter} with [Fe/H] from –0.25 to –2.25 dex (in 0.5 dex steps), we identify the sub-giant branch turning point as a function of Gaia BP–RP colour and metallicity. A linear fit is applied to these points using \texttt{scipy}, which is used to cut away sub-giant branch stars. This linear function is\footnote{This range in metallicity reflects the metallicity distribution of the sample. Note that the Pristine survey metallicities go down to --4 dex, but for halo-like ages around 11 Gyr the subgiant branch turn-off starts to behave in unexpected ways for metallicities below --2.5 dex \citep[see][]{2016dotter}, and we do not use isochrones below these metallicities for the interpolation.}

\begin{equation}\label{eq:colourmetallicitycut}
    \text{G}_{BP,0} - \text{G}_{RP,0} = 0.14 \times \text{[Fe/H]} + 1.05,
\end{equation}

\noindent where any stars with a colour bluer (smaller) than this are cut away. 
The training sample in metallicity versus colour view, colour-coded by the training sample's log \textit{g} is shown in Figure \ref{fig:colormetallicitycut}. We can see that the subgiant stars with log \textit{g} > 3.5 are removed (in red) using this colour-metallicity cut, retaining the RGB stars (in blue).
There exists a trend in surface gravity with respect to metallicity for when the stars have a sub-giant branch, which is what we clearly see on the right side of the dashed black line in Figure \ref{fig:colormetallicitycut}. Therefore, the blue overdensity of stars at Gaia BP-RP colour of 0.6 are relatively metal-poor sub-giant branch stars with smaller surface gravity.
If we change the model isochrone to PARSEC or BaSTI, the impact of the final colour-metallicity cut on the input catalogues is less than 1\%.
We also refrain from using the training sample to decide the colour-metallicity cut, as the training sample is not fully representative of our input catalogues, and is very small in size after the previous selections (N=7882). Therefore, we only want to use the training sample to study the consequences of our selections.
This cut, together with the CaMD cut and the magnitude cut, increases purity to 90\% and decreases completeness to 58\%.

\begin{figure*}
\centering
\sidecaption
    \includegraphics[width=6cm]{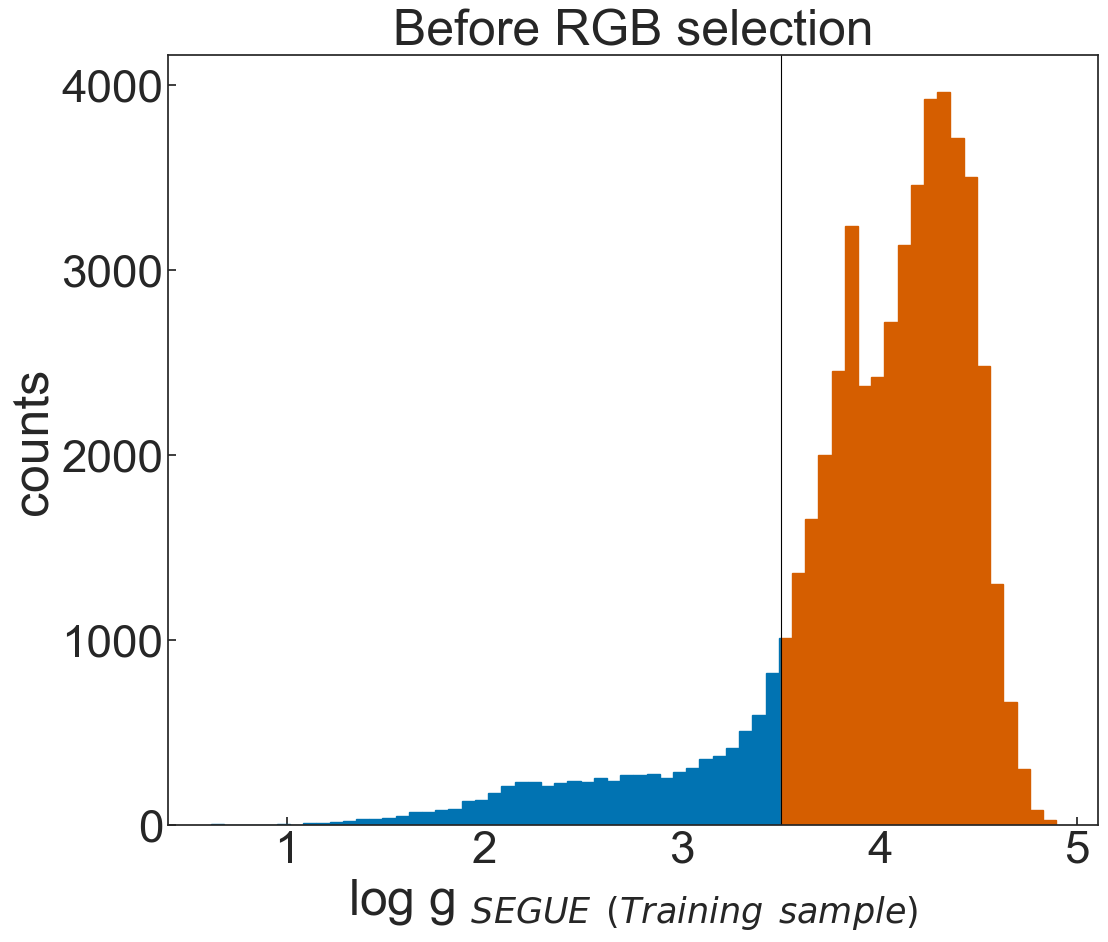}    \includegraphics[width=6cm]{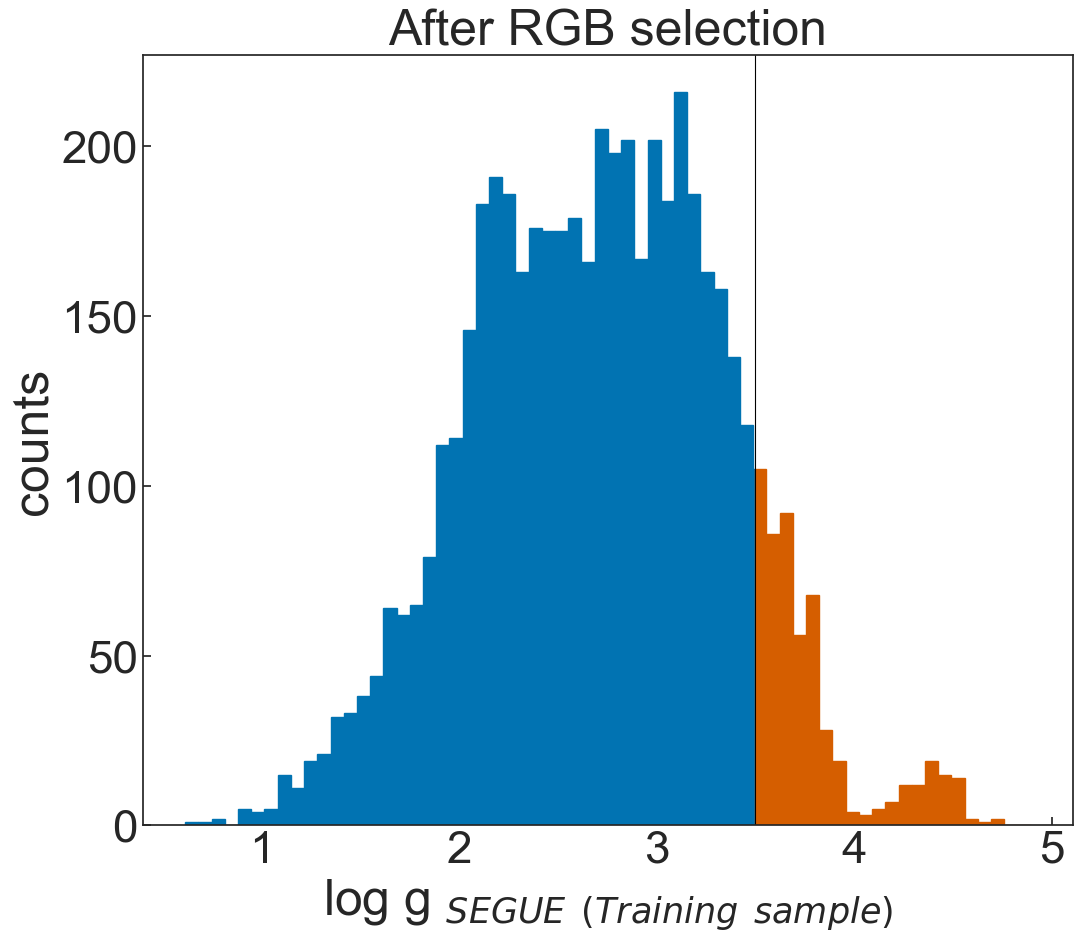}
\caption{Distribution of the training sample's surface gravities before and after the catalogue pipeline (described in Figure \ref{fig:flowchart}) is applied. The vertical line at log \textit{g}=3.5 shows the separation used to validate the giants sample selection. The overdensity near log \textit{g}$\sim$4.5 (orange region, right panel) arises from contamination by metal-rich turn-off and sub-giant stars in SEGUE, where their surface gravities overlap with the lower part of the RGB. Note that the log \textit{g}<3.5 cut is an approximate validation threshold; a metallicity-dependent log \textit{g} cut provides a more refined RGB selection.} 
\label{fig:loggdistribution} 
\end{figure*}

Our full RGB selection pipeline is shown as a flowchart in Figure \ref{fig:flowchart}. The log \textit{g} distribution of the training sample before and after the RGB selection pipeline has been applied is shown in Figure \ref{fig:loggdistribution}, which shows the efficiency of our RGB selection on the training sample.

\subsection{Photometric distance derivation}\label{3.4} 

To map the Galactic outer halo, we need reliable distances out to $\sim$100 kpc. This cannot be achieved using \textit{Gaia} parallaxes as these reach only up to 5-10 kpc for giants. This means we need to use assumptions such as isochrone models to derive photometric distances to probe the Galactic outer halo. These isochrones are heavily dependent on the assumed metallicities, especially on the RGB. Given that our input catalogues are based on PDR1 and PGS catalogues with reliable photometric metallicities down to -3.5, we can use them as input metallicities to derive photometric distances. We tried our method on several different isochrone models such as MIST \citep{2016dotter,2016choi}, PARSEC \citep{2012bressan,2020pastorelli}, and BaSTI isochrones \citep{2018hidalgo}. We find the standard deviation of our photometric distances with inverted-parallax distances for the subsample that has good parallax \footnote{good parallax in this subsection refers to stars with $\pi$>0 and $f$ = $\Delta\pi$/$\pi$ $\leq$ 0.05 or 0.1 or 0.2, depending on what is specified} ($f$ $\leq$ 0.1) is minimum for BaSTI isochrones. Additionally, the difference between BaSTI and PARSEC is much smaller than the difference between BaSTI and MIST. This can be attributed to the fact that MIST isochrones do not have $\alpha$-enhancement implemented whereas BaSTI and PARSEC do. In the remainder of this work, we use BaSTI isochrones because of its agreement with the shape and slope of the isochrone in the RGB using \textit{Gaia} XP/\textit{Gaia} BP-RP colours and the fact that it is available down to lower metallicities ([Fe/H]$\sim$-3.2) compared to PARSEC ([Fe/H]$\sim$-2.2). We use BaSTI isochrones at a fixed age of 10 Gyr, in the $\alpha$-enhanced version \citep{2004pietrinferni}, with a Reimers mass loss parameter of 0.3, assuming a Kroupa initial mass fraction \citep{2001kroupa}, a fraction of unresolved binaries of 30\%, and a minimum mass ratio for binaries of 0.1, between metallicities of -3.2 and -0.08. We use an effective temperature range of 3000 to 7000 K, \textit{Gaia} BP - RP extinction corrected colour between 0.5 and 1.5, and a log \textit{g} range (-0.5 to 4.2) just enough to infer photometric distances for RGB stars and remove outliers in the process.
The log \textit{g} in BaSTI model isochrones are calculated using the Stefan-Boltzman equation.
We only select the RGB part of the stellar model isochrone and do not fit for any other stellar populations. 

\begin{figure}
\centering
Pristine DR1 (PDR1) giants
  \includegraphics[width=0.5\textwidth]{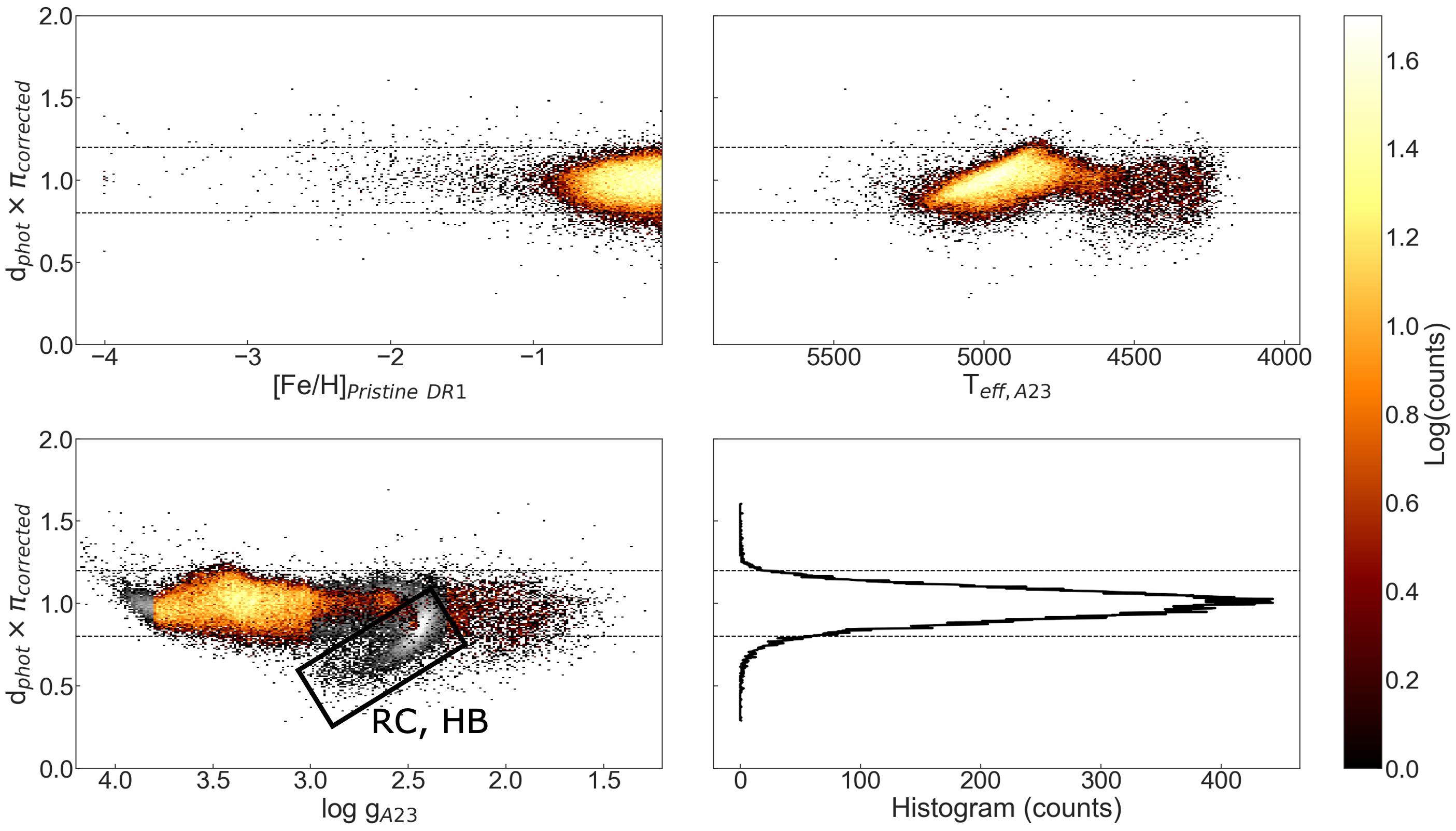}
\centering
Pristine-\textit{Gaia} synthetic (PGS) giants
  \includegraphics[width=0.5\textwidth]{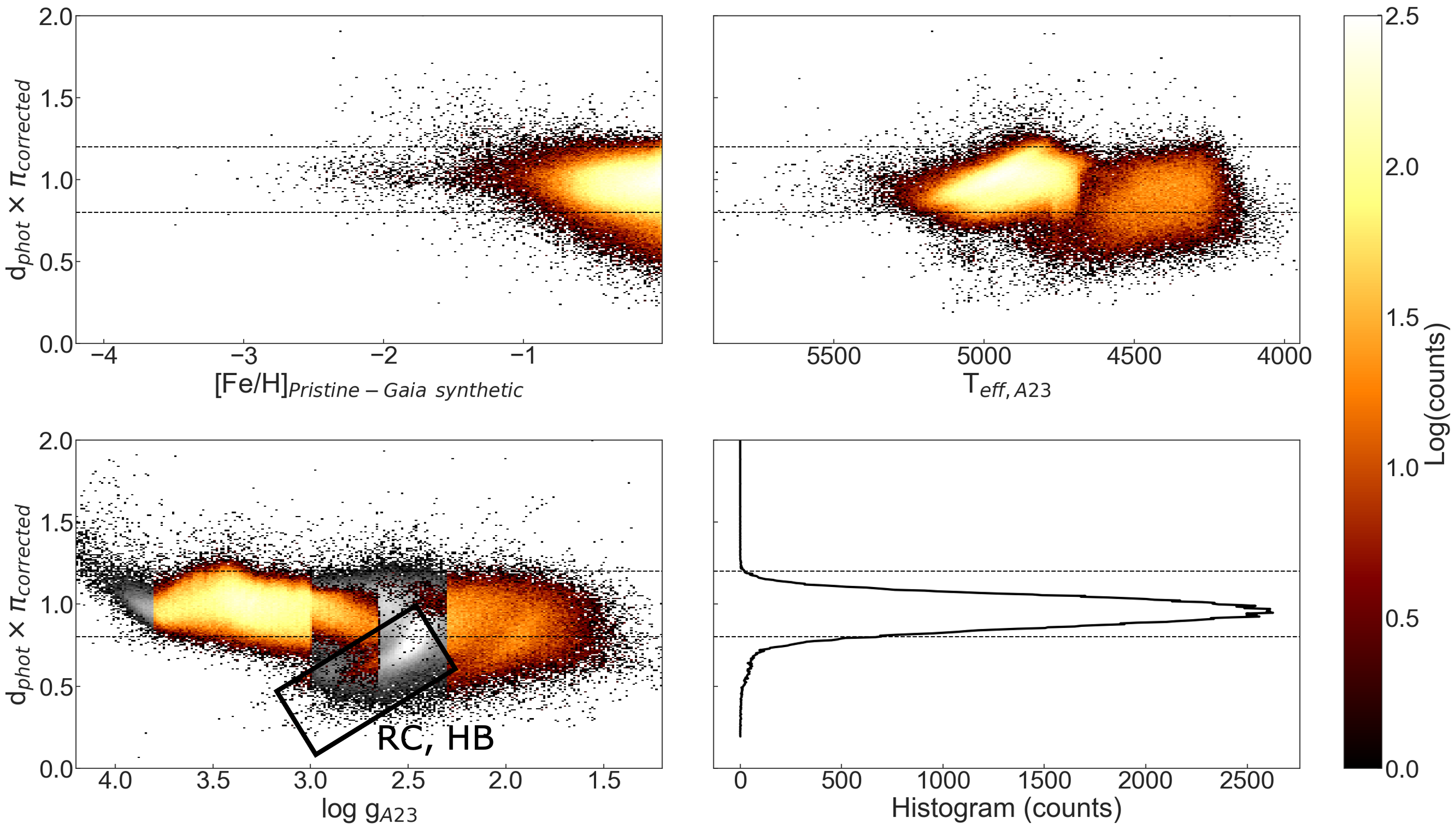}
  \caption{Photometric distance calculated in this work times the offset corrected parallax based on \citet{2021lindegreen} (which is supposed to be 1.0 in the ideal case) versus the input parameters to calculate these photometric distances, including photometric metallicities (top left) from PDR1 (top panels) and PGS catalogues (bottom panels), effective temperature (top right), and surface gravity (bottom left) from \citetalias{2023andrae} \texttt{XGBoost} catalogue. The input parameters are colour-coded by the log density and horizontal histogram of distance times parallax (bottom right). Note that the density distributions are in log scale, so most of the stars are within $\sim$10\% systematic uncertainties. The grey overdensities in the surface gravity plots are bad distance mismatches caused due to red clump stars and (colder) horizontal branch stars.}
  \label{fig:dist-comp}
\end{figure}

We constructed a 4D interpolation grid assuming a fixed stellar age of 10 Gyr, but varying metallicity, effective temperature, surface gravity, and absolute \textit{Gaia} G magnitude. This choice is justified by tests showing that even varying the age by a factor of two affects inferred distances by only $\sim$10\%, consistent with findings from \citet{2020bonaca}. For effective temperature and surface gravity, we crossmatch our final RGB samples from the PDR1 and PGS input catalogues with the \citet[][hereafter \citetalias{2023andrae}]{2023andrae} catalogue, which provides \texttt{teff\_xgboost} and \texttt{logg\_xgboost} estimates for 175 million stars, inferred from \textit{Gaia} XP spectra using the \texttt{XGBoost} algorithm \citep{2016chen}. This results in 408,524 PDR1-giants and 6,098,246 PGS-giants (see methods flowchart in Figure \ref{fig:flowchart}).

We find that using temperature-based isochrone matching leads to photometric distance estimates that agree better with high-quality parallax-based distances (\(f \leq 0.1\)) than those derived from \textit{Gaia} BP–RP colours. This is expected, as temperature incorporates the full XP spectrum, whereas BP–RP compresses this into a single colour. We avoid using \texttt{mh\_xgboost} metallicities from \citetalias{2023andrae}, as they are less reliable at low metallicities unless the star is both bright and has a good parallax (\(f \leq 0.04\)), as shown by the vetted RGB sample in Table 2 of \citetalias{2023andrae}. Although temperature and surface gravity from this source also degrade with poor parallaxes, they are more robust than the metallicities.

The goal of this work, however, is to demonstrate that RGB stars can be reliably selected using only photometry and parallax, without requiring spectroscopic parameters, distances, or radial velocities. Atmospheric parameters from \citetalias{2023andrae} are used solely for photometric distance inference and validation. This pipeline is applicable to any photometric catalogue with \textit{Gaia} parallaxes for RGB selection.

For each star, we generate an interpolated isochrone from the BaSTI model for an age of 10 Gyr and the star’s photometric metallicity from Pristine (CaHK-based for PDR1, or XP-based model for PGS). The isochrone spans a grid of effective temperatures, surface gravities, and \textit{Gaia} G absolute magnitudes. We place each star in the \(T_{\mathrm{eff}} - \log g\) space and find the closest point on the spline-interpolated isochrone, applying a weight on log \textit{g} that varies with evolutionary stage: 2500 near the turn-off and 500 near the RGB tip. This weighting minimises the difference between inferred photometric distances and parallax-based distances for the \(f \leq 0.1\) subset and reduces sensitivity to input parameters. The corresponding absolute \textit{Gaia} G magnitude is then used with the extinction-corrected apparent G magnitude to compute the photometric distance.

Extinction corrections for all \textit{Gaia} magnitudes are adopted from the input Pristine catalogues \citepalias{2023martin}. The inferred photometric distances show good agreement with inverted parallax distances (by design) and with the Starhorse and \citet{2021bailer-jones} photogeometric distances (see Appendix \ref{E}). In the PGS catalogue, the scatter in the distance comparison increases slightly with distance, as noted in Appendix B, indicating a mild distance-dependent uncertainty in the photometric distance estimation, which is expected in any outer halo catalogues using RGB stars.

\begin{figure}
  \includegraphics[width=\columnwidth]{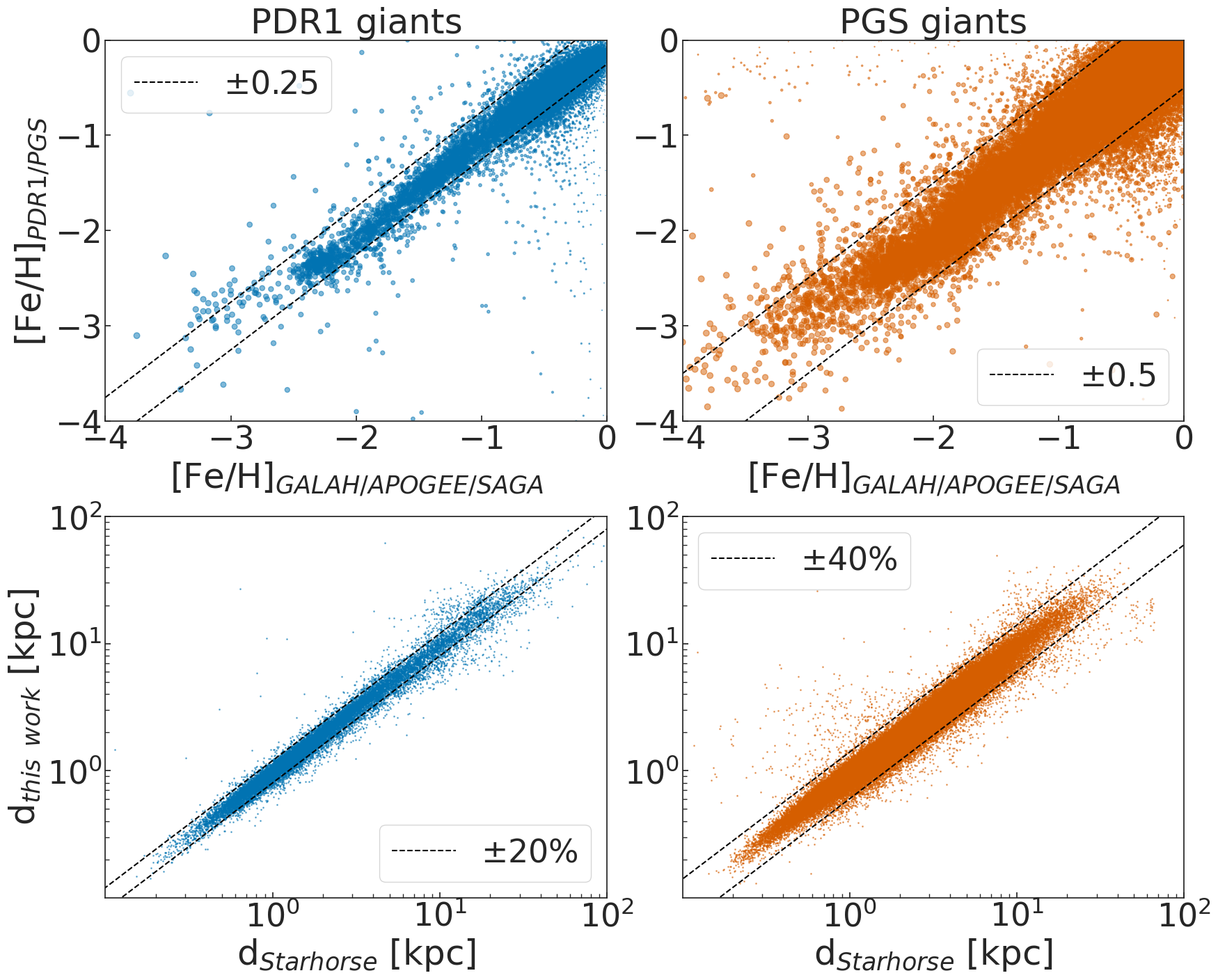}
  \caption{Validation of our photometric metallicities and distances with GALAH DR3 and APOGEE DR17 surveys and the high-resolution spectroscopic metallicities from the SAGA database of VMP stars (top) and Starhorse \textit{Gaia} DR3 (Bayesian isochrone-fitting code) distances for different spectroscopic surveys such as APOGEE, GALAH, SDSS SEGUE, LAMOST MRS, and \textit{Gaia} RVS surveys (bottom) for the PDR1 (left) and PGS (right) giants catalogue constructed in this work.}
  \label{fig:sh}
\end{figure}

We calculated systematic uncertainties on the inferred photometric distances instead of measurement uncertainties because we find that the measurement errors are negligible compared to the total dispersion inferred from the good parallax ($f$ $\leq$ 0.1) subset. This is because we do not have measurement uncertainties on effective temperatures or surface gravities from \citetalias{2023andrae} parameters that are inferred using machine learning. Available measurement uncertainties only depend on the measurement uncertainties on the photometric metallicities from the PDR1 and PGS input catalogues. These are, in turn, dependent on uncertainties on colour, and magnitudes that are relatively well-measured compared to other uncertainties.
Therefore, we stick to estimating only systematic uncertainties on the inferred photometric distances. For this, we use the 100 nearest neighbours of each star in the good parallax ($f$ $\leq$ 0.1) subsample, based on its input effective temperature, surface gravity and photometric metallicities (all of which are scaled between a range of 0 to 1, to make sure they have the same weights). We infer the dispersion between inverted-parallax distances and our inferred photometric distances for these 100 nearest neighbours as the systematic uncertainties on our photometric distances. 

We compared the photometric distances calculated in this work with the inverted-parallax distances against the input parameters used to calculate the photometric distances for both PDR1 and PGS-giants catalogue in Figure \ref{fig:dist-comp}. We show distance times parallax (ideally 1.0) versus photometric metallicities from PDR1/PGS catalogues from \citetalias{2023martin}, effective temperature and surface gravity from \citetalias{2023andrae}. The final panel shows a histogram of distance times parallax and it peaks at 0.99 and 0.95, with a 1$\sigma$ dispersion of 0.09 and 0.1 for PDR1 and PGS-giants respectively. From the Figure \ref{fig:dist-comp} top panel, we can see that the photometric distances agree very well with the inverted-parallax distances with no visible trend with the photometric metallicities, that is the main parameter used in our science cases in the next section. We see the same with PGS-giants in the bottom panel of Figure \ref{fig:dist-comp}, in terms of reliability of inferred photometric distances. However, the dispersion is higher for PGS-giants, mainly because of photometric metallicities that are based on lower S/N CaHK magnitudes inferred from \textit{Gaia} XP spectra, and the fact that the distance calculation pipeline is very much dependent on the reliablity of accurate metallicities. In addition to removing turn-off interlopers by using a log \textit{g}<3.5 cut (\texttt{logg\_xgboost}<3.5), we also use an additional quality cut to get the final catalogue of PDR1 and PGS-giants with reliable metallicities and distances used for the science cases in the results section. This is due to the red clump (RC) stars at the metal-rich end and (colder) horizontal branch (HB) stars at the metal-poor end that does not get a log \textit{g} from \citetalias{2023andrae} that agrees well with parallax-based inferences. This can be seen by the cloud of stars in the black box with underestimated distances shown in the bottom left panels for PDR1 and PGS-giants compared to parallax-based distances, shown in Figure \ref{fig:dist-comp}. To remove stars with underestimated distances in this region, we use a distance quality cut of 7\% or below within the log \textit{g} range of 2.3 and 3.0 (!((\texttt{phot\_dist\_errs}>0.07) \& ((\texttt{logg\_xgboost}>2.3) \& (\texttt{logg\_xgboost}<3.0)))). With all these quality cuts, we end up with 180,314 PDR1 and 2,420,898 PGS-giants, that have reliable metallicities and distances.
A small part of the Pristine survey sample ($\sim$1\%) can be selected as RGB stars using our method and are not part of the PDR1 catalogue, which we do not use in the rest of this paper.

We validated our photometric metallicities with GALAH DR3, APOGEE DR17, and SAGA database of VMP stars' high-resolution spectroscopic metallicities \citep{2021buder,2017majewski,2008suda} and our photometric distances with distances inferred by the Starhorse Bayesian-isochrone inferred distances \citep{2023queiroz} as shown in Figure \ref{fig:sh}. 
We chose GALAH DR3, APOGEE DR17, and SAGA (even though the crossmatch numbers are lower than when using low-resolution spectroscopic surveys and they have a bias towards brighter stars) because of its reliability across the metallicity scale used down to the EMP end. 
No particular trends within APOGEE, GALAH, or SAGA were seen.
To have a full understanding of the reliability of the photometric metallcities from PDR1 and PGS catalogues down to [Fe/H]$\sim$-4.0, we refer the readers to \citetalias{2023martin} and \citet{2024viswanathanb}. 
Based on these works, it is also important to keep in mind that in the VMP and EMP regime, the success rates of finding true VMP and EMP giants are 97\% and 38\% respectively, which makes these EMP giants good candidates and not always `true' EMPs. 
The validation of metallicities presented here is only to ensure that there are no offsets caused in the photometric metallicities due to the various selection made in the parent Pristine catalogues to select the RGB stars, and not to validate the metallicities themselves, as these are performed thoroughly in the data release paper \citepalias{2023martin}.
In the top panels of Figure \ref{fig:sh}, we see a comparison of metallicities from PDR1 (left) and PGS (right) giants with GALAH DR3 metallicities. 
We removed those stars with \texttt{flag\_sp}==0 or \texttt{flag\_fe\_h}==0 from the GALAH DR3 sample, and place a quality cut of \texttt{FE\_H\_FLAG}==0 on the APOGEE DR17 sample.
For the SAGA database of VMP stars, we use a 5.0 arcsec crossmatch radius as recommended by the SAGA catalogue makers due to lower precision in RA, Dec from older spectroscopic follow-ups as opposed to the 1.0 arcsec crossmatch radius used in the rest of this work.
We see good agreement within 0.25 dex and 0.5 dex for PDR1 and PGS-giants respectively. 
For the PGS-giants, if we only use stars with good S/N in the CaHK narrow-band (\texttt{error on CaHK magnitudes, d\_CaHK}<0.02), the scatter is as low as 0.2 dex. 
This is because, the photometric [Fe/H] uncertainty is $\sim$0.1 in PDR1 catalogue versus $\sim$0.4 in PGS catalogue at fainter magnitudes (G$\sim$16). 
This is also the reason why we see higher dispersion for PGS distances compared to parallax in the bottom panels of Figure \ref{fig:dist-comp}. Therefore, we recommend using this cut in PGS-giants for science cases where reliable distances and very reliable metallicities are a necessity. 

In the bottom panels of Figure \ref{fig:sh}, we show a comparison of our inferred distances with Starhorse distances (with <10\% uncertainties) using spectroscopic survey parameters from APOGEE DR17, GALAH DR3, SDSS SEGUE DR12, LAMOST DR8 MRS, and \textit{Gaia} DR3 RVS. No particular trends with specific surveys are seen. We choose these five surveys, because they have higher resolution or probe the fainter end of our catalogue (in the case of SEGUE). We see that the distances are not offset with a $\lesssim$20\% and $\lesssim$40\% scatter out to 100 kpc for PDR1 and PGS catalogues of giants respectively. 
The scatter between our distances and Starhorse distances is larger for PGS-giants than PDR1-giants (and slightly biased towards underestimation than overestimation), due to metallicity uncertainties.    

\subsection{Caveats with the catalogues}\label{3.5}

In this subsection, we discuss the two main caveats with the PDR1 and/or PGS-giants catalogue constructed in this work.

\subsubsection{Distances of red clump and horizontal branch stars}\label{3.5.1}

\begin{figure}
  \includegraphics[width=0.5\textwidth]{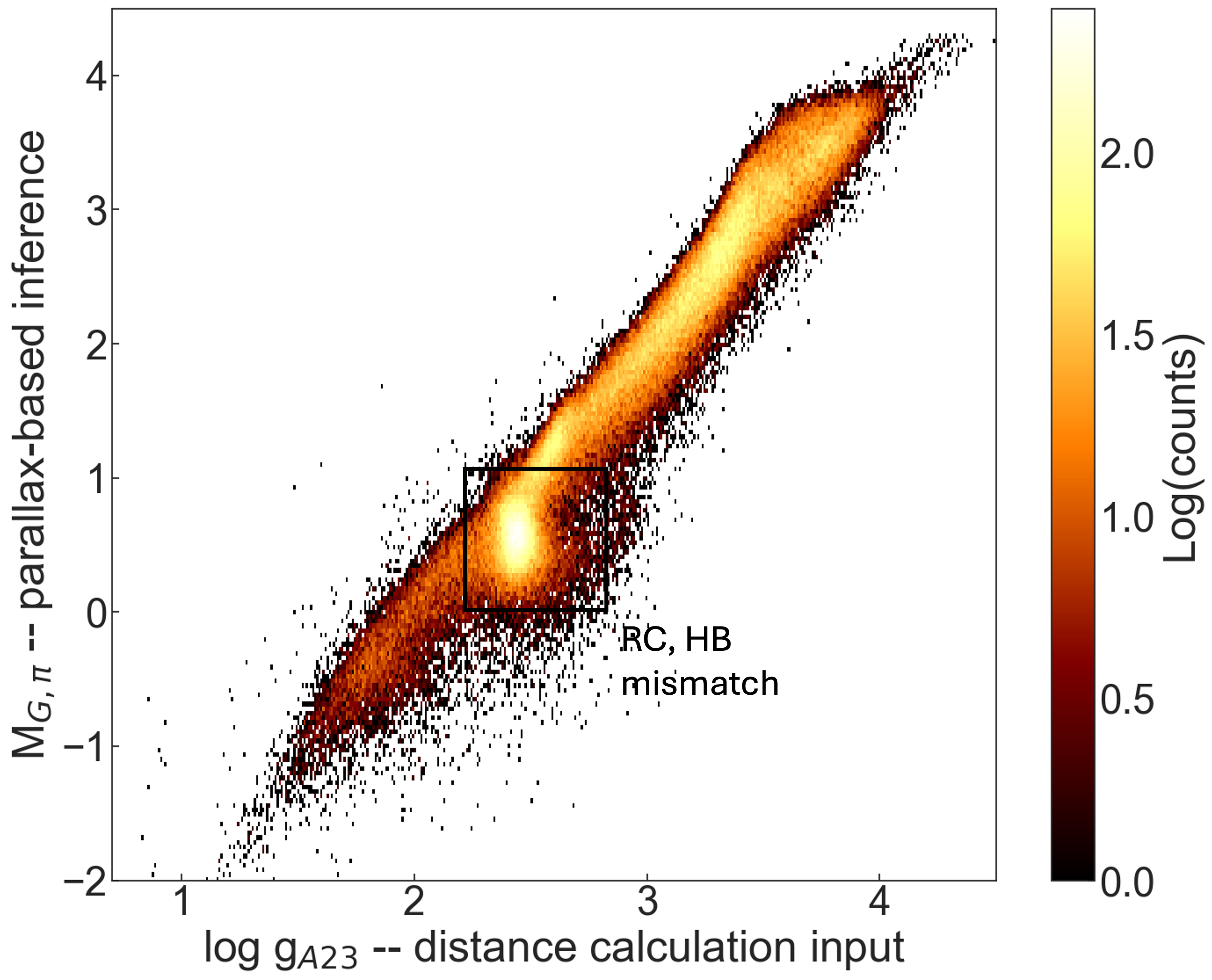}
  \caption{Absolute G magnitude calculated using distances inferred from inverted parallax versus log \textit{g} from the \citetalias{2023andrae} \texttt{XGBoost} catalogue. We used log \textit{g} as an input to calculate the photometric distances, whereas absolute magnitudes, which trace surface gravity, were inferred from parallax. One can clearly see the 1:1 trend mismatched at log \textit{g} that corresponds to red clump (metal-rich) and horizontal branch (metal-poor) stars.}
  \label{fig:rc}
\end{figure}

From the bottom left panels for PDR1 and PGS-giants in Figure \ref{fig:dist-comp}, we can see that the distances have a clear trend towards underestimation (of up to 50\%) for log \textit{g} values between 2.3 and 3.0. This region of log \textit{g} overlaps with where we find RC stars in the metal rich end and colder HB stars in the metal-poor end. RC stars are abundant stars that were once similar to the Sun and have since evolved into red giants, now sustained by helium fusion in their cores. Regardless of their specific age or composition, all RC stars achieve roughly the same absolute magnitude luminosity. 
Red clump stars are core helium-burning giants, valuable as standard candles due to their consistent luminosity and well-defined position in the H-R diagram.
The Pristine survey model assigns photometric metallicities for F, G, K stars between 0.5 and 1.5 in the \textit{Gaia} BP-RP colour range. This region inevitably overlaps with a few HB stars.
HB stars originate from low-mass stars that have finished their main-sequence lifetimes and experienced a helium flash at the conclusion of their red-giant phase. 
Consequently, HB stars are very old objects, making them useful markers in studies of the Galactic structure and formation history. 
However, our isochrone fitting code does not explicitly model RC and HB evolutionary phases. Instead, these phases are indirectly accounted for by matching each star to the nearest point on the isochrone in the \( T_{\mathrm{eff}} \)–log \textit{g} space. 
However, the log \textit{g} comes from the \citetalias{2023andrae} catalogue using \textit{Gaia} XP spectra that does not linearly scale with the absolute magnitudes calculated using parallaxes for the good parallax ($f$ $\leq$ 0.05) subset as illustrated in Figure \ref{fig:rc} where we show the mismatch in RC, and HB stars with the black box. Therefore, we use a quality cut in distances within these surface gravities to ensure that we end up with distances that are fully reliable (as discussed in previous subsection).

\begin{figure*}
\centering
\sidecaption
  \includegraphics[width=12cm]{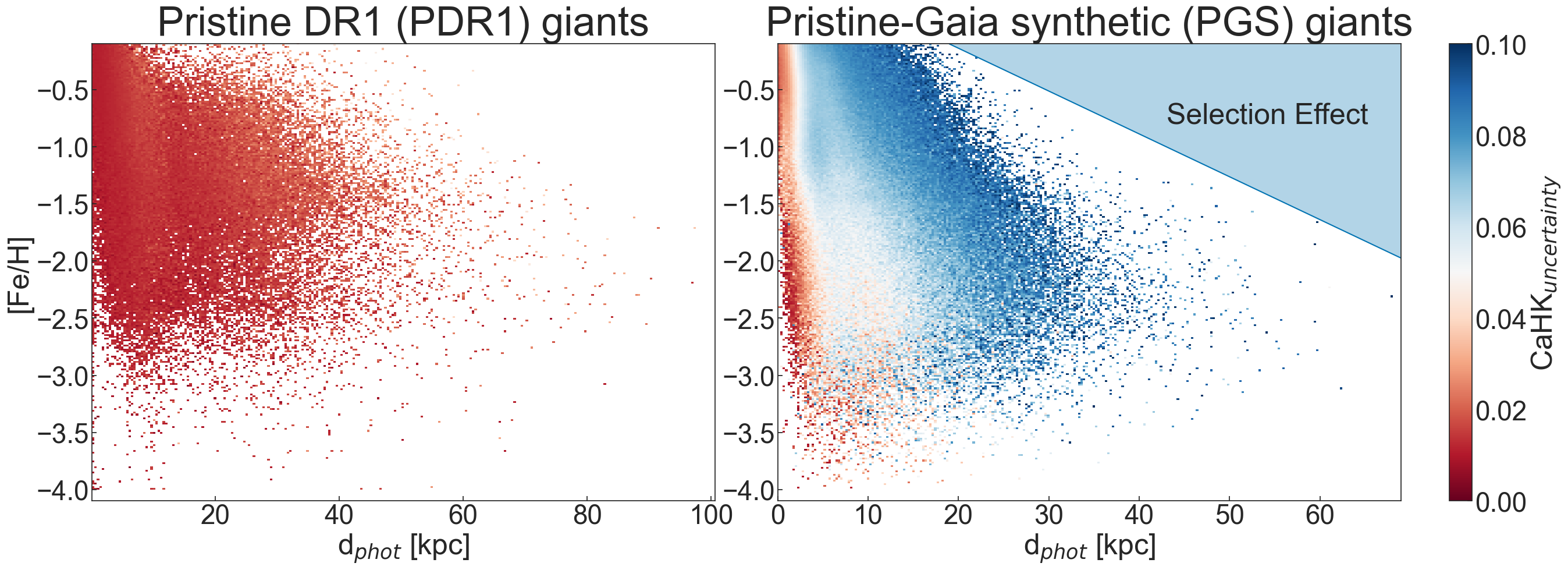}
  \caption{Metallicities versus heliocentric distances for PDR1 (left) and PGS (right) catalogues of giants presented in this work colour-coded by CaHK narrow-band magnitude uncertainties. The distance-metallicity selection effect caused by low S/N on the CaHK narrow-band from \textit{Gaia} XP spectra is highlighted by the blue filled area on the right.}
  \label{fig:sel-effect}
\end{figure*}

\subsubsection{Distance-metallicity selection effect on \textit{Gaia} XP}\label{3.5.2}

In Figure \ref{fig:sel-effect}, we show the photometric metallicities versus heliocentric distances for the PDR1 (left) and PGS (right) catalogues of giants, colour-coded by their mean uncertainty in the CaHK magnitudes used to calculate the photometric metallicities. The Pristine survey model assigns photometric metallicities for each star based on its CaHK narrow-band magnitudes in comparison with \textit{Gaia} BP-RP broad-band magnitudes. For this purpose, the model requires the CaHK uncertainties to be less than 0.1, which is the upper limit of our colour-coding in Figure \ref{fig:sel-effect}. The \textit{Gaia} XP spectra are magnitude limited down to G $\sim17.6$, with \textit{Gaia}'s scanning law limitations imprinted (see \citealt{2021riello} for more details about \textit{Gaia}'s scanning law effect). With the selection of the PGS-giants catalogue (which is also magnitude limited at 17.6), we are pushing the limits of the S/N in the CaHK region that is required to calculate photometric metallicities reliably. This problem is almost negligible in the PDR1-giants catalogue, because the Pristine survey goes much fainter than \textit{Gaia} XP spectra (down to G$\sim$21), with very high S/N in the CaHK narrow-band compared to \textit{Gaia} XP spectra.

The effect of CaHK uncertainty is visible clearly in Figure \ref{fig:sel-effect} for the PGS-giants. We see that the CaHK uncertainty increases clearly at larger distances. It is important to note that the CaHK magnitudes are brighter for metal-poor stars than for the metal-rich stars with respect to the broad-band magnitudes, which allows us to pick metal-poor stars more efficiently amidst the more metal-rich populations of our Galaxy. Given the magnitude limits and CaHK uncertainty limits of the PGS-giants catalogue, at larger distances, we only see metal-poor stars that are relatively brighter than the metal-rich stars. Therefore, the filled blue region in the right panel of Figure \ref{fig:sel-effect} is empty due to this distance-metallicity selection effect, and not due to physical conditions in the Galactic outer halo. This means that the PGS-giants cannot be used to study the metallicity variations at different distances in the Galactic halo. Given the low-to-no bias in distance versus metallicity in the PDR1-giants (apart from the small effects due to the colour boundaries corrected by weighing the metallicity bins in section \ref{4.2}), we can use this catalogue to study the metallicity structure of the Galactic halo out to large distances and down to the lowest metallicities, thereby probing deep, and far into the Galactic halo's earliest evolutionary times. Due to the distance-metallicity selection effects in the PGS-giants catalogue, we see almost no metal-rich stars in the outer halo, which means we can study the outer Galactic halo's oldest stars with a much cleaner sample of VMP stars than has been possible so far. We investigate some of these science cases in the next section(s).

\begin{figure*}
\center
    \includegraphics[width=0.24\textwidth]{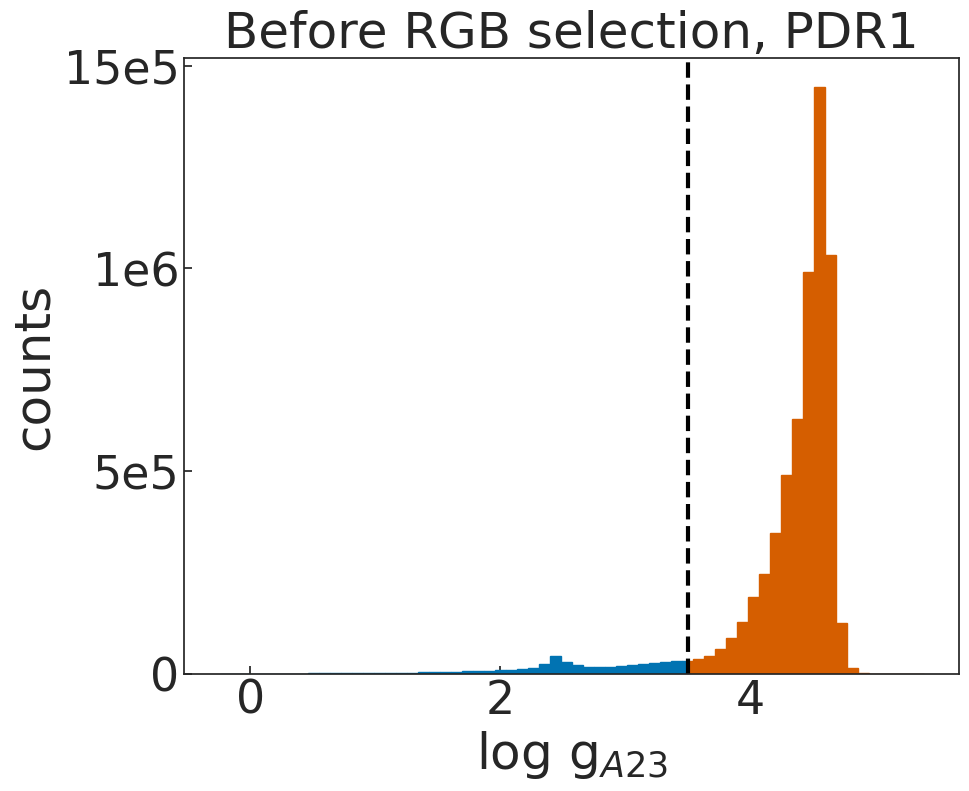}    \includegraphics[width=0.24\textwidth]{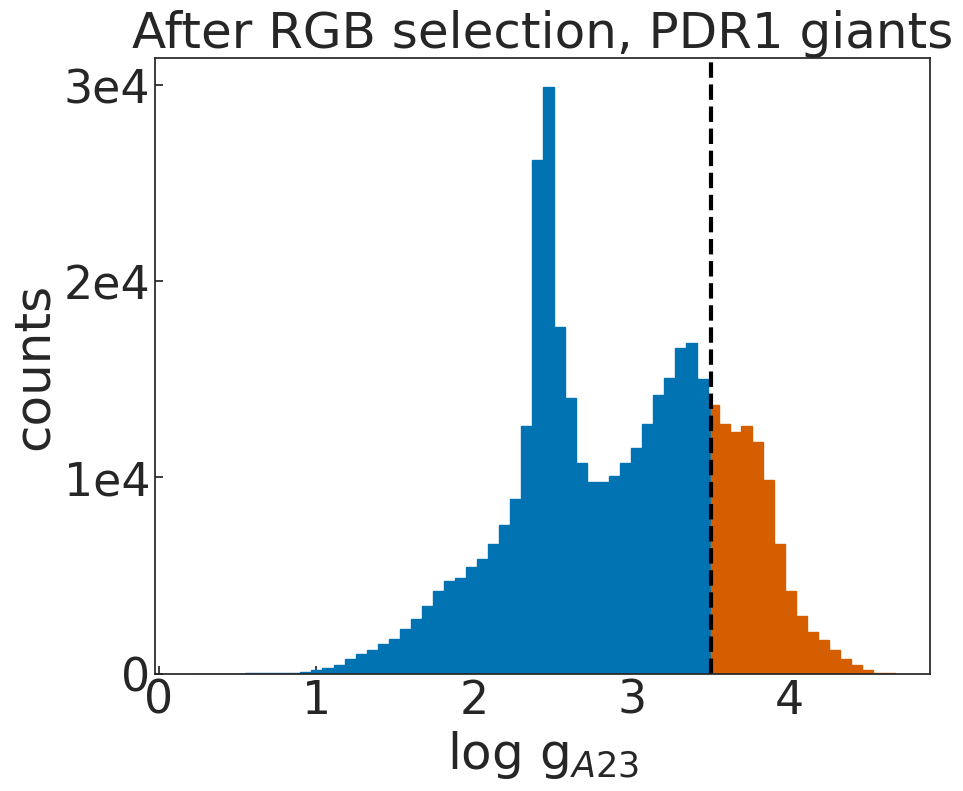}
    \hspace{4mm}
    \includegraphics[width=0.24\textwidth]{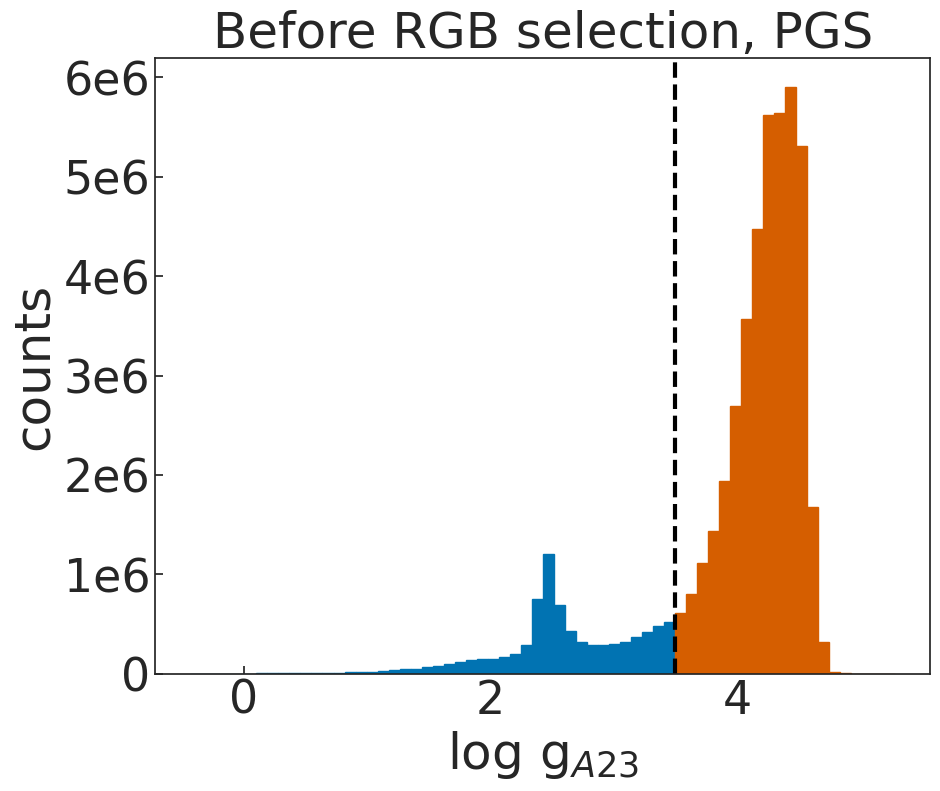}    \includegraphics[width=0.24\textwidth]{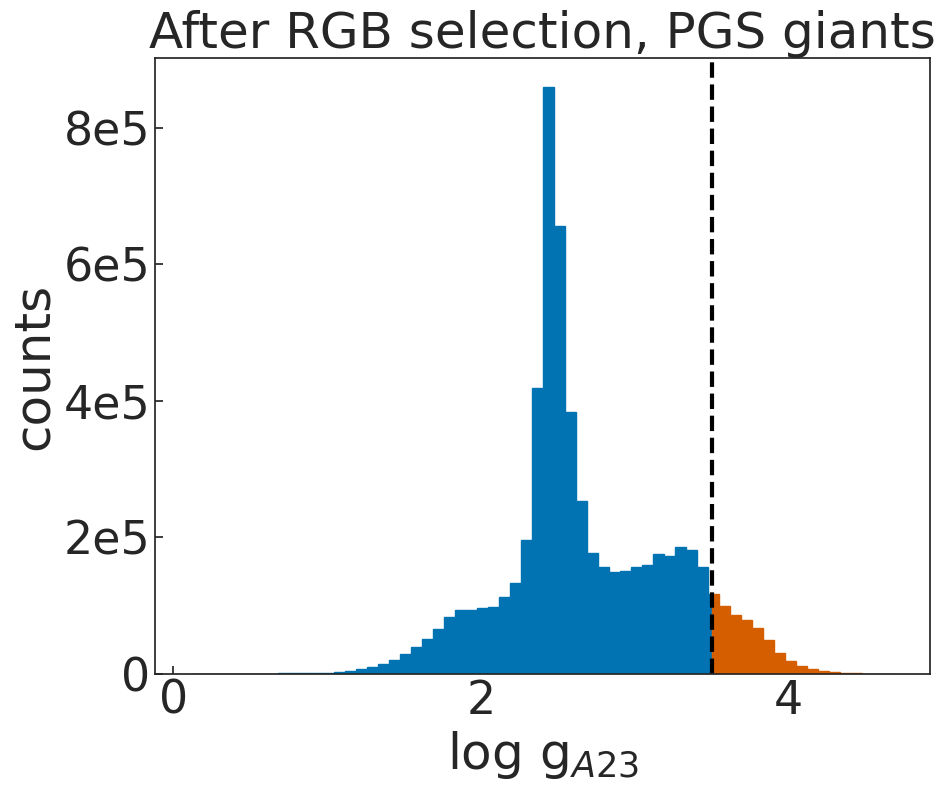}
\caption{Distribution of the \citetalias{2023andrae} surface gravities before and after the catalogue pipeline (described in Figure \ref{fig:flowchart}) is applied on the PDR1 (top) and PGS (bottom) input catalogues. The vertical line at log \textit{g}=3.5 shows the separation used to validate the selected sample of giants. In all the panels, the overdensity of stars at log \textit{g}$\sim$2.5 from RC stars are clearly visible.} 
\label{fig:loggdistributionandrae} 
\end{figure*}

\subsection{RGB selection validated using \citetalias{2023andrae} parameters}\label{5.1}

As a final check on the purity and completeness of our catalogues of giants, we use log \textit{g} values from the \citetalias{2023andrae} catalogue that is based on the \textit{Gaia} XP spectra and parallax. In Figure \ref{fig:loggdistributionandrae}, we show the log \textit{g} distribution from the \citetalias{2023andrae} catalogue before and after our RGB selection pipeline (see the Figure \ref{fig:flowchart} flowchart for a summary of the pipeline) has been applied to the PDR1 (top) and PGS (bottom) input catalogues.  
This figure shows the efficiency of our selection given the small number of stars that fall below log \textit{g} = 3.5, given the very low signal from RGB stars before the pipeline was applied (this is because dwarfs are 100 times more numerous than giants).
The final purity and completeness calculated based on the \citetalias{2023andrae} log \textit{g} is 78\% and 76\% respectively for the PDR1-giants and 92\% and 82\% respectively for the PGS-giants. These numbers are much higher than what we inferred based on the training sample. These numbers show that our RGB selection performs very well, achieving both high purity and completeness.
From a small subset of good parallax ($f$=0.05) stars that are misclassified with \texttt{logg\_xgboost}>3.5 in our RGB catalogues or unselected with \texttt{logg\_xgboost}<3.5 and not in our RGB catalogues, we see that most (90\%) of these stars are near the sub-giant branch and/or main sequence turn-off part of the CMD.

We only use the atmospheric parameters from \citetalias{2023andrae} for calculating photometric distances and validating our RGB star selection. 
Our goal is to demonstrate the effectiveness of selecting RGB stars using only parallax and photometry, without depending on atmospheric parameters, distances, or radial velocities. 
This approach is more widely applicable, as it allows for the reliable identification of RGB stars across any photometric catalogue that includes Gaia parallaxes.
The validation using log \textit{g} from \citetalias{2023andrae} showcase the power of our RGB selection.

\section{Results}\label{4}

In this section, we summarise the metallicity and distance properties of the catalogues of giants, and discuss the construction of 6D phase-space samples of PDR1/PGS-giants, Sagittarius (Sgr) stream members in the catalogues, calculation of phase-space information and integrals-of-motion (IOM) and how we can view the different accretion events in the metallicity view of the IOM space. We finally discuss the outer-halo substructures in our catalogues of giants. 

\subsection{Description of the catalogues}\label{4.1}

\begin{figure}
  \includegraphics[width=0.5\textwidth]{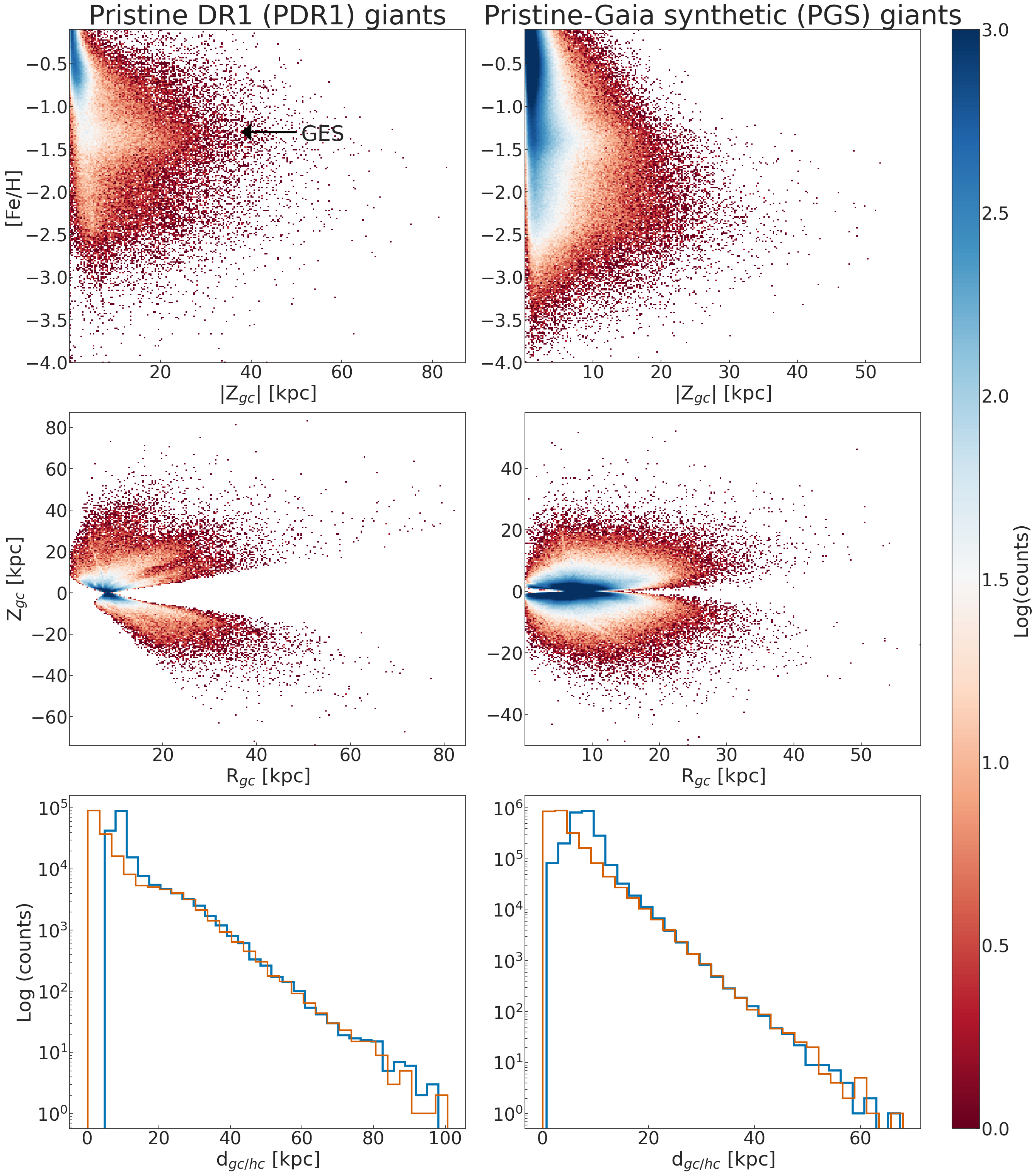}
  \caption{Density distribution of photometric metallicities versus absolute scale height (Z) (top), density distribution of giant stars in R versus Z frame (middle), and histogram of heliocentric (orange) and galactocentric (blue) distances probed by the giants catalogue (bottom) for PDR1 (left) and PGS (right) input catalogues. Note that the ranges of the x- and y-axes are different for PDR1- and PGS-giants in the middle and bottom panels}
  \label{fig:fehvsd}
\end{figure}

\begin{figure}
  \includegraphics[width=\columnwidth]{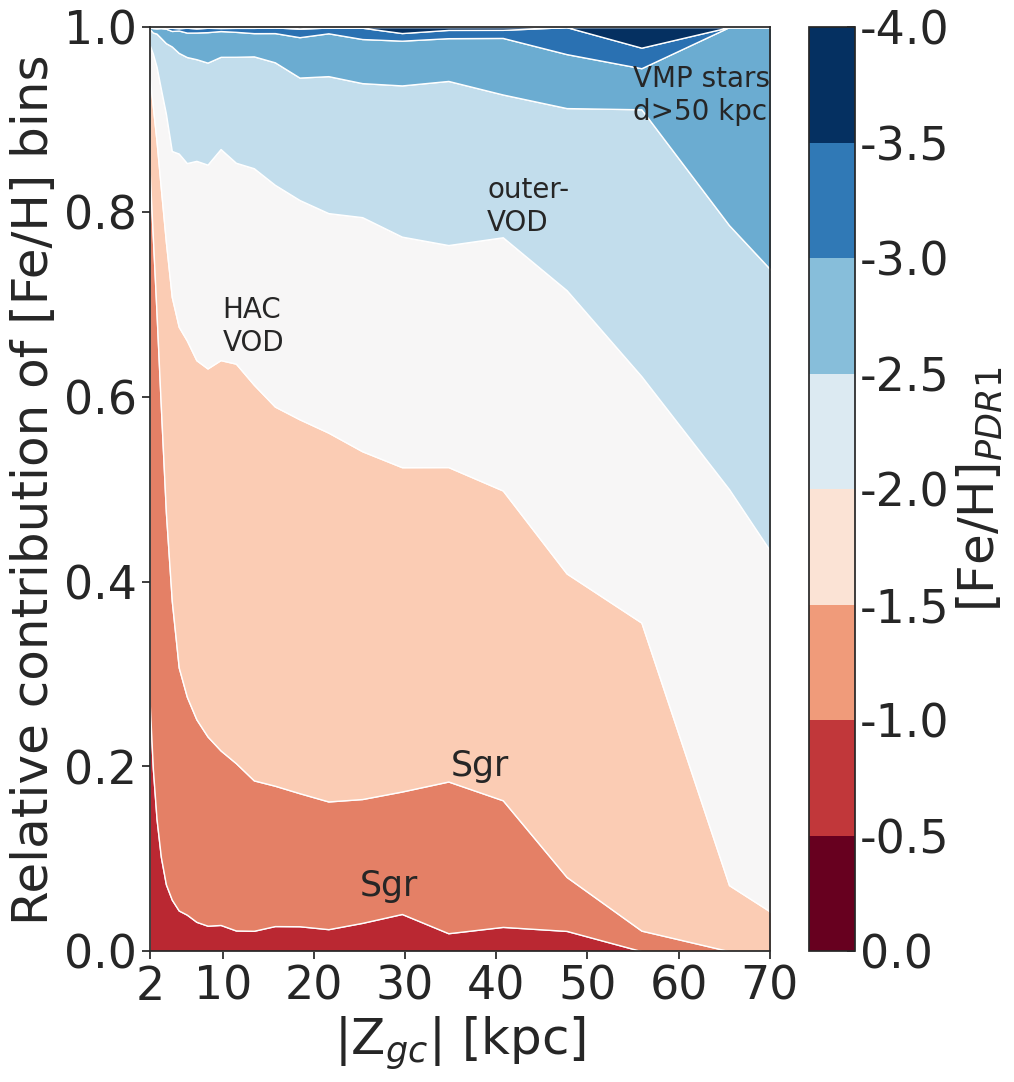}
  \caption{Metallicity structure of the halo as a function of height from the mid-plane (Z) for eight metallicity bins between -4 and 0 for the PDR1-giants catalogue. The fractional contribution of each metallicity bin to the population at a certain distance has been calculated. Stars below a scale height of 2 kpc have been cut away to avoid disc contamination.}
  \label{fig:relativez}
\end{figure}

We present a red giants branch catalogue using the Pristine survey and/or the \textit{Gaia} XP-based metallicities and photometric isochrone-fitted distances in this work. 
The PDR1-giants sample consists of 180,314 RGB stars and probes heliocentric distances up to 100.65 kpc (with mean uncertainties down to 12\% with a maximum of 40\% dependent on the quality of the input parameters, especially at the faint end). 
The PGS-giants sample consists of 2,420,898 RGB stars that probe heliocentric distances up to 68.03 kpc (with mean uncertainties down to 12\% with a maximum of 57\% dependent on the quality of the input parameters, especially in the faint end).  
The final purity and completeness based on the Pristine survey training sample is 90\% and 58\% respectively. 
Both the purity and completeness vary as a function of metallicities, surface gravities, and effective temperature, due to the colour cuts used to select them.
From the training sample, we see that the purity and completeness decreases as a function of metallicities (by $\sim$20\% between -4 and 0 dex), decreases as a function of temperature (by $\sim$40\% and $\sim$20\% between 4000 and 6000 K), and the completeness decreases as a function of surface gravity (by $\sim$20\% between 0.5 and 3.5), if we use log \textit{g}<3.5 as the pure sample of giants. 
However, it is important to note that the log \textit{g}<3.5 is not the purest and most complete selection of giants and also has a dependence on the metallicity.
It is important to note that the training sample is not necessarily representative of our input catalogues, and it is quite incomplete and has much fewer stars in the VMP end.
Some of these effects are corrected for when we measure the metallicity distribution function using model isochrones in subsection \ref{4.2}.

The mean metallicity uncertainties are 0.08 and 0.19 dex down to [Fe/H]$\sim$-4 in metallicity for the PDR1 and PGS catalogues respectively. 
The mean metallicity uncertainties increases up to 0.11 dex and 0.28 dex for VMP stars in the PDR1 and PGS catalogues respectively.
With such reliable metallicities and distances, we can study the metallicity distributions and (chemical and dynamical) substructures than make up the outer Galactic halo.

To calculate the Cartesian positions for all the RGB stars, a distance of 8.2 kpc between the Sun and the Milky Way's centre is assumed \citep{2018gravity}.
In the top panels of Figure \ref{fig:fehvsd}, we show metallicity versus absolute height above the Galactic plane for PDR1 and PGS-giants. We can see that the PGS-giants have a fairly smooth distribution with metal-rich stars in the inner halo and metal-poor stars in the outer halo. However, this view is biased due to the quality cut on CaHK uncertainties used in the making of the PGS input metallicity catalogue, as discussed in subsection \ref{3.5.2}. As a consequence of this, we see only metal-poor stars in the outer halo of PGS-giants, and do not trace the reality of the metallicity distribution of the outer halo. However, this sample can be used to study metal-poor substructures in the outer halo due to the low-to-no contamination from metal-rich substructures (that are usually large in number). We refrain from using the PGS catalogue for anything that involves studying the metallicity distribution or metallicity-distribution-dependent science cases. 

In the top left panel of Figure \ref{fig:fehvsd}, we show the metallicity versus distance space for the PDR1-giants, which is more representative of the Galactic halo out to $\sim$100 kpc. We can see a prominent overdensity of stars around the metallicity of $\sim$-1.3 out to about 30 kpc which we associate with the last major merger event, \textit{Gaia}-Enceladus-Sausage \citep[GES,][]{2018belokurov,2018helmi}.
This is reminiscent of the strongly radial orbits clustered at -1.0 to -1.6 in spectroscopic metallicities from the H3 survey out to large distances \citep{2019conroya,2019conroyb}. For a small subset of our sample with radial velocities, we find these stars to have high eccentricities and probe the radial regions of energy, E and vertical angular momentum, L$_z$ (see Figure \ref{fig:iom} and subsection \ref{4.3.3}). The small subset of these stars that overlap with the APOGEE high-resolution spectroscopic survey, have lower [$\alpha$/Fe], reminiscent of a dwarf galaxy stellar population that merged with the Milky Way, similar to the GES event. These checks allow us to conclude that this prominent peak in [Fe/H] versus |Z| space most likely belongs to the GES merger.
We also see other substructures and the distribution is not as smooth, indicating that the Galactic halo is made of stellar populations from several different merger events.
In Figure \ref{fig:fehvsd} middle panels, we see the on-sky distribution of the stars in galactocentric radius versus height above the plane for the PDR1 and PGS-giants. We can see the uneven footprint and northern coverage of the Pristine survey for the PDR1-giants on the left panel while the PGS-giants on the right panel are all-sky. The PDR1-giants probe further out to slightly larger distances ($\sim$100 kpc) than the PGS-giants ($\sim$60 kpc) due to the lower S/N of CaHK narrow-band magnitudes for the Gaia XP-based PGS-giants. This is, in turn, due to the relative brightness limit of \textit{Gaia} XP spectra of the PGS-giants sample. 
On the right panel, we see a strong selection function at lower scale height due to dust extinction cut in the disc plane. To see the metallicity distribution in a spatial view along Cartesian coordinates, we refer the reader to Appendix \ref{B}.
In the bottom panel of Figure \ref{fig:fehvsd}, we see the 1D-histogram of galactocentric and heliocentric distance probed by PDR1 (left) and PGS (right) giants. There is a steep decrease in the number of stars at larger distances, mostly due to the negative power law slope of about 4.0 in the halo \citep{2018hernitschek,2018deason,2018thomas,2019starkenburg}, but also due to selection functions in the underlying surveys and methods used to select the RGB stars. The small bump at about 20-40 kpc could be due to GES apocentre pile-ups \citep{2022perottoni} or the Sagittarius stream \citep{2020ibata}. 

\subsection{The bias corrected metallicity structure of the halo}\label{4.2}

Because the PDR1 catalogue contains both reliable distances and metallicities without major selection effects, we can present the metallicity distribution functions (MDFs) of the halo as a function of distance. 
Disc stars are removed to create a halo sample by using a cut of $|Z| < 3$ kpc\footnote{We opt to not use $|Z| < 2$ kpc as in the previous section to make sure we only measure the halo MDF. Since the MDFs themselves do not have any distance information in them, disentangling thick disc contributions to the halo MDF would be difficult. Therefore, the stricter |Z|<3 kpc cut is preferred.}.
We used six heliocentric distance bins: 3--4 kpc, 4--6 kpc, 6--10 kpc, 10--22 kpc, 22--45 kpc, and 45--101 kpc. 
These ranges ensure that both the numbers of stars in each bin and the distance range spanned by one bin change smoothly.

\begin{figure}
  \includegraphics{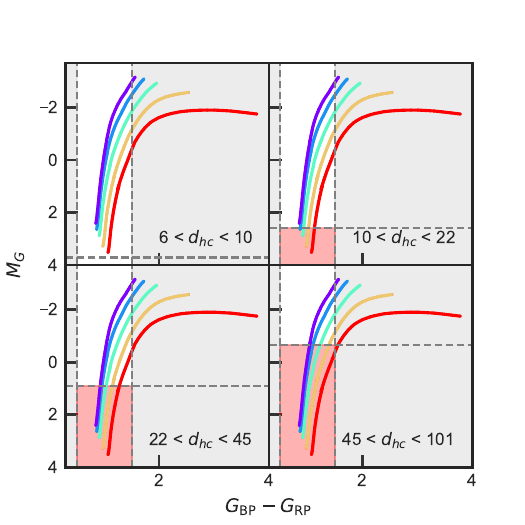}
  \caption{
  Five PARSEC RGB isochrones with metallicities at --0.25, --0.75, --1.25, --1.75, and --2.25 dex (i.e. centred on the metallicity bins we used for our weights) going from red to purple. 
  The colour range of $0.5 < G_\text{BP} - G_\text{RP} < 1.5$ is in dashed grey lines and the area that is not probed because of this colour cut is shown in grey.
  As metallicity increases, the colour cut means that we probe a smaller portion of the RGB, leading to an undersampling of metal-rich stars.
  We show four panels corresponding to the four most distant bins, given in kpc, where Malmquist bias reduces the range in absolute magnitude probed by each distance bin (where the probed area is seen in white and the area lost due to Malmquist bias is seen in red, the upper limit of which is simply the absolute magnitude of the upper limit of each distance bin).
  Only the three most distant bins are affected.
  Note how the most metal-rich isochrone in the 45--101 kpc bin does not enter into the white region at all.
  The combined effect of the colour and magnitude cut is that the metal-poor stars reach the brightest absolute magnitudes. Therefore, the further into the halo we probe, the smaller the fraction of metal-rich stars.
  }
  \label{fig:isochrones}
\end{figure}

We have two sources of bias in our MDFs. The first one is introduced because of our colour cut, and the second one is introduced when binning the sample in distance due to our magnitude cut.
We consider the former source of bias first.
The colour range of $0.5 < \text{G}_{BP,0} - \text{G}_{RP,0} < 1.5$ is the same for all stars, no matter their metallicity, but because the tip of the RGB becomes redder with increasing metallicity (see Figure \ref{fig:isochrones} that shows the probed colour range for a set of PARSEC isochrones with varying metallicities), we probed a smaller fraction of the metal-rich stars than the metal-poor stars. 
This led to an undersampling of metal-rich stars, which biases the shapes of our MDFs.
Now we consider the source of bias due to magnitude.
This bias arises because of our magnitude cut where we removed stars with $G > 17.6$, meaning that our MDFs are affected by a Malmquist bias.
Again, looking at Figure \ref{fig:isochrones}, the $0.5 < \text{G}_{BP,0} - \text{G}_{RP,0} < 1.5$ cut means that the brighter the absolute magnitudes that are probed, the fewer metal-rich stars that are included in the distant bin.
This again leads to an undersampling of metal-rich stars, but an undersampling that increases with distance, as can be seen in the increasing size of the red region with distance. 
The figure shows the three distance bins that are affected by this Malmquist bias.
We bias-corrected both of these effects by introducing weights to our MDFs for different metallicity ranges.
On top of this, we also have the underlying Gaia's scanning pattern selection effect on-sky \citep{2021riello}, which is the S/N needed for a star to have Gaia XP information in DR3. 
This is a strong function of the location on the sky, but does not affect the MDF as strongly. 

The weights are computed using PARSEC simulated stellar populations with \citet{2001kroupa,2002kroupa} canonical two-part-power law initial mass function (IMF), corrected for unresolved binaries, as they contain labels for evolutionary stage, so that we can select only RGB stars.
Using BaSTI isochrones requires manual removal of subgiant branch stars, which we aim to avoid. On the other hand, using MIST isochrones would require introducing an additional assumption—an $\alpha$-enhancement offset—which we also prefer not to impose. Instead, we adopt a method similar to that of \citet{2020youakim}, who corrected metallicity biases introduced by colour cuts in main-sequence turn-off stars. We generate five simulated stellar populations using PARSEC isochrones, each representing an equidistant metallicity bin and with a total simulated mass of 100,000 M$_\odot$. Rather than aiming for absolute population corrections—which would require knowledge of the total stellar mass within the catalogue's footprint (for local correction) or of the entire Galactic halo (for global extrapolation)—we focus on the shape of the MDFs. Therefore, relative weights are sufficient for our purposes, and we only present normalised histograms. To apply selection effects consistently, we impose the same cuts on the simulated populations as on the observational catalogue. First, we apply a colour cut of $0.5 < G_\text{BP} - G_\text{RP} < 1.5$, matching the Pristine survey's criteria. Since the simulated stars lack distances and apparent magnitudes, we replicate the parallax-based CaMD and magnitude cuts (as described in subsections \ref{3.1} and \ref{3.2}) by selecting stars with the RGB flag \texttt{label == 3}, which removes subgiant branch stars from the simulated data. As a result of this RGB-only selection, we also skip the colour–metallicity cut from subsection \ref{3.3}, which is designed to remove subgiants. The final simulated catalogue thus represents a population composed entirely of RGB stars.
Because the purity and completeness of our RGB catalogues are quite high, we can assume that this is a good approximation of our catalogue.

The colour bias weights are computed by taking the total mass in each metallicity bin after applying the \texttt{label == 3} cut to it, and divide that by the total mass after the \texttt{label == 3} and the colour cut has been applied.
The reference weight, which we divide all weights by, is taken as the $-1.5<\mathrm{[Fe/H]}<-1.0$ metallicity bin weight.
The resulting weights are seen in Table \ref{tab:colourbiasweights}.
Multiplying these times the MDFs will undo the undersampling of metal-rich stars that occurs because of our colour cut.

\begin{table}
	\centering
	\caption{
        Bias-correction due to colour cut in the RGB selection.
                }
        \label{tab:colourbiasweights}
	\begin{tabular}{lc}
		\hline
		Metallicity bin & Weight \\
		\hline
		  $-0.5<\mathrm{[Fe/H]}$ & 1.03  \\
		$-1.0<\mathrm{[Fe/H]}<-0.5$ & 1.01 \\
		$-1.5<\mathrm{[Fe/H]}<-1.0$ & 1.00 \\
		$-2.0<\mathrm{[Fe/H]}<-1.5$ & 0.99\\
           \multicolumn{1}{r}{$\mathrm{[Fe/H]}<-2.0$}   & 0.99 \\
		\hline
	\end{tabular}
    \tablefoot{The weights calculated from PARSEC simulated stellar populations, to bias-correct the metallicity bias introduced by the $0.5 < \text{G}_{BP,0} - \text{G}_{RP,0} < 1.5$ cut.
    The corresponding PARSEC isochrones are seen in Figure \ref{fig:isochrones}.
    The total mass in metallicity bin $-1.5<\mathrm{[Fe/H]}<-1.0$ is used as the reference mass, as we are only interested in normalised MDFs and thus their shape only.
    These weights are multiplied to the MDFs to bias-correct them.}
\end{table}

\begin{figure}
  \includegraphics{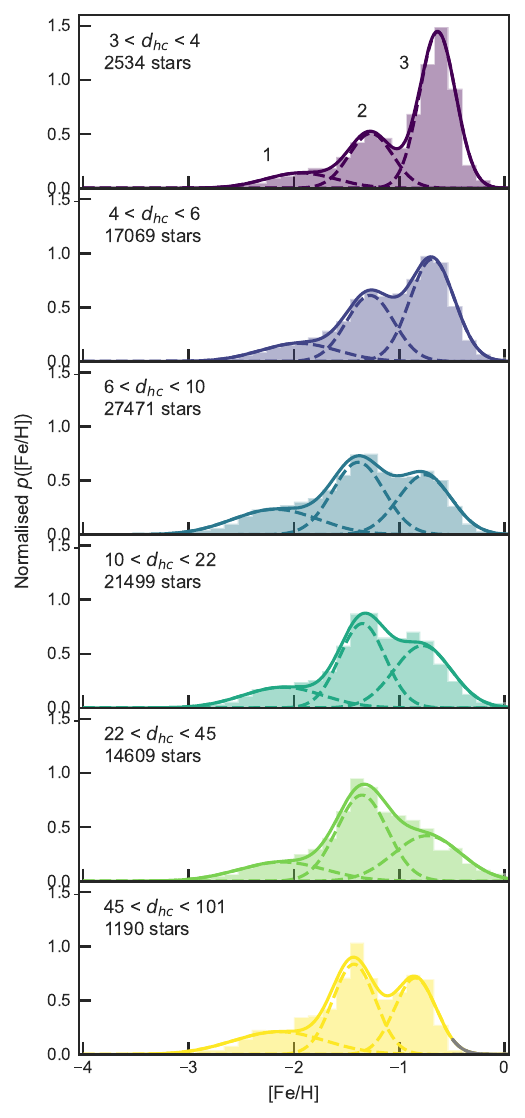}
  \caption{
  Metallicity distribution functions from PDR1-giants in six galactocentric distance bins and their corresponding GMMs (solid lines) and the three individual GMM components (dashed lines).
  The GMMs are shown next to each other in Figure \ref{fig:MDF2}.
  All distances are given in kpc.
  The individual GMM components are labelled as 1, 2 and 3 with decreasing metallicity in the top panel.
  The region [Fe/H] $> -0.5$ dex is greyed out for the most distant bin as we are not properly probing this region of the MDF, see Figure \ref{fig:isochrones}.
  }
  \label{fig:MDF1}
\end{figure}

We now move to computing the magnitude cut weights.
Only the three most distant bins are affected by Malmquist bias introduced by our magnitude cut, see Figure \ref{fig:isochrones}, and only they get weights assigned for this.
For each metallicity bin as in Table \ref{tab:colourbiasweights}, we compute the magnitude cut weights. For this, we divide the total mass for that metallicity bin by the mass for the same bin, but where the absolute magnitudes are brighter than the limiting magnitude for each distance bin.
Because we once again are only interested in the relative weights for a given distance bin, we normalise the weights within one distance bin by dividing all weights with the weight for the most metal-poor bin.
As we can see in Figure \ref{fig:isochrones}, the most metal-rich stellar population does not enter into the white region in the most distant bin, meaning that theoretically, we are not measuring the MDF of halo with heliocentric distances larger than 45 kpc where [Fe/H] $ > -0.5$.
This part of the MDF is greyed out in subsequent plots.
The weights are presented in Table \ref{tab:distancebiasweights} and we bias-correct the MDFs by multiplying them with these values. These weights are slightly overcorrecting the MDFs as we assume the largest distance at each distance bin to correct for the Malmquist bias and not the distribution of the distances itself in each bin, which is out of the scope of this work. 
There can also be an excess of metal-poor stars due to the inherent methodology of using CaHK narrow-band as a proxy for stellar metallicity. This is because metal-poor stars are brighter than metal-rich stars in CaHK magnitudes and therefore have a higher signal-to-noise. To correct for this, we need extensive modelling of the survey's photometry. However, we note that we used the brighter subsample of the survey, and thus, this effect should be very small, as seen in the left panel of Figure \ref{fig:sel-effect}.
The biases due to distance uncertainties and metallicity uncertainties in the faint end should also be small due to the large range of distances chosen in each bin, and the bin size chosen for the MDFs shown in Figure \ref{fig:MDF1}. 
Therefore, the simple bias-correcting technique presented in this work is a first step towards investigating the true view of the MDF of the outer halo.

\begin{table}
	\centering
	\caption{
    Bias-correction due to magnitude cut in the RGB selection.
                }
        \label{tab:distancebiasweights}
	\begin{tabular}{llc}
		\hline
		Distance bin [kpc] & Metallicity bin & Weight \\
		\hline
            \hline
		  $10<d_{hc}<22$ & $-0.5<\mathrm{[Fe/H]}$       & 2.07  \\
	                   	& $-1.0<\mathrm{[Fe/H]}<-0.5$ & 1.69  \\
	                   	& $-1.5<\mathrm{[Fe/H]}<-1.0$ & 1.27 \\
	                   	& $-2.0<\mathrm{[Fe/H]}<-1.5$ & 1.07 \\
            & \multicolumn{1}{r}{$\mathrm{[Fe/H]}<-2.0$}  & 1.00 \\
            \hline
		  $22<d_{hc}<45$ & $-0.5<\mathrm{[Fe/H]}$       & 3.91 \\
		                & $-1.0<\mathrm{[Fe/H]}<-0.5$ & 1.95 \\
                            & $-1.5<\mathrm{[Fe/H]}<-1.0$ & 1.46 \\
	                    & $-2.0<\mathrm{[Fe/H]}<-1.5$ & 1.20 \\
            & \multicolumn{1}{r}{$\mathrm{[Fe/H]}<-2.0$}  & 1.00 \\
            \hline
		  $45<d_{hc}<101$ & $-0.5<\mathrm{[Fe/H]}$      & - \\
		                  & $-1.0<\mathrm{[Fe/H]}<-0.5$ & 6.49 \\
		                  & $-1.5<\mathrm{[Fe/H]}<-1.0$ & 2.27\\
		                  & $-2.0<\mathrm{[Fe/H]}<-1.5$ & 1.32 \\
            & \multicolumn{1}{r}{$\mathrm{[Fe/H]}<-2.0$}  & 1.00 \\
		\hline
	\end{tabular}
    \tablefoot{The weights calculated from PARSEC simulated stellar populations, to bias-correct the metallicity bias introduced by binning the sample in distance.
        Because of our magnitude cut where we only keep stars with $G < 17.6$ mag, each distance bin probes a limited range in absolute magnitude, which because of the colour cut primarily affects the metal-rich stars, see Figure \ref{fig:isochrones}.
        Within each distance bin, the weights for each metallicity has been normalised to the weight in the $\mathrm{[Fe/H]}<-2.0$ bin.
        These weights are multiplied to the MDFs to bias-correct them.}
\end{table}

\begin{figure}
  \includegraphics{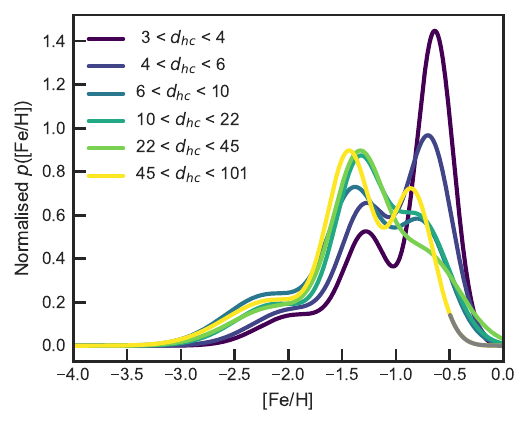}
  \caption{
  Best-fit GMMs to the MDFs seen in Figure \ref{fig:MDF1}.
  The region [Fe/H] $> -0.5$ dex is greyed out for the most distant bin as we are not properly probing this region of the MDF, see Figure \ref{fig:isochrones}.
  All distances are given in kpc.
  The means, standard deviations and weights of the different peaks can be seen in Table \ref{tab:MDFGMMcontributions}.
  }
  \label{fig:MDF2}
\end{figure}

After the weights in Tables \ref{tab:colourbiasweights} and \ref{tab:distancebiasweights} have been multiplied by the MDFs in each distance bin, we fit a Gaussian mixture model (GMM) decomposition to the MDFs. 
The number of components are chosen based on the lowest Bayesian Information Criteria (BIC), that ends up choosing three components as the optimal ones for all the different distance bins.
The MDFs and their corresponding GMMs, with each contributing component, are shown in Figure \ref{fig:MDF1} together with the amount of stars in each distance bin.
Figure \ref{fig:MDF2} illustrates the kernel density estimate (KDE) altogether per distance bin.
The GMM components and their means $\mu$, standard deviations $\sigma$ and component weights $\omega$ are shown in Table \ref{tab:MDFGMMcontributions}, both with and without the weights.

Both figures show that the metal-poor peak (coming from low-mass accretion events and the tail of more massive accretion events) marked as 1 getting stronger with distance, but peaks in the 6--10 kpc bin.
This is also seen in Table \ref{tab:MDFGMMcontributions}, where $\omega_1$ has the largest value in that bin, meaning that it has the most contribution from the metal-poor peak.
$\mu_1$ is also the most negative for that bin.
However, it has the largest dispersion $\sigma_1$ in the 45--101 kpc bin.
The variation in the mean between the 6-10 kpc and other more distant bins is very small, compared with the measured metallicity uncertainties. 
Therefore, it is safe to assume that this bin stays roughly constant past 6-10 kpc.
The medium metallicity peak (coming from more massive accretion events), peak 2, decreases in metallicity with distance ($\mu_2$), increases in strength with distance ($\omega_2$), but its dispersion ($\sigma_2$) roughly stays constant. This peak has its contribution mostly from the last major merger, GES.
The metal-rich peak (coming from `hot' thick disc stars, i.e. thick disc stars on halo-like orbits), peak 3, is most pronounced in the closest bin.
Its metallicity ($\mu_3$) also decreases with distance, its dispersion ($\sigma_3$) increases until the most distant bin, and its strength mostly decreases with distance ($\omega_3$).
This shows that the hot thick disc stars populate mostly the inner Galactic halo (d<10 kpc).
In the outermost halo (d>22 kpc), the metal-rich peak could also contain the disrupting Sagittarius dwarf galaxy stream \citep{2020ibata}, even though we only have a small number of stars that we associate with the stream as seen in upcoming subsections.
We do not attempt to remove the stream specifically to perform the MDF analysis as the number of members we find is very low (N<400), and the lack of 6D phase space information at larger distances makes the stream member removal less reliable and creates more selection effects. 
The effect with distance on the unweighted values is more continuous, which shows the need for our bias-correcting method using weights.
As we move further into the halo, the contribution to the stellar population from accreted dwarf galaxies increase \citep{2020naidu}, which explains the increase in metallicity dispersion with distance. 

We draw the conclusion that not only do the metal-poor components become stronger as distance increases, but each given component is also more metal-poor with distance.
It is also clear from the MDFs that the halo contains a metal-rich component, peak 3, that persists even at large distances (however, this might be a mix of `hot' thick disc and Sagittarius stream at large distances in the last distance bin).
The halo is known to contain a red and blue colour-magnitude diagram (CMD) population \citep{gaia2018}.
The blue sequence comes from the GES merger, where the mass ratio between the red and blue populations indicates that the GES was massive enough to perturb in situ MW stars in the old thick disc to halo-like kinematics \citep{2019gallart}. 
These kinematically heated stars can most clearly be seen at metallicities spanning the range --0.7 to --0.2 dex \citep{2020belokurov}.
The canonical thick (and thin) disc stars in our sample must be removed mostly by our $|Z| < 3$ kpc cut.
This coincides with the range of $\mu_3$ we measure at closer distances, and these splashed stars are likely the reason that we have such pronounced metal-rich peaks in our MDFs.

The decrease of metallicity with distance in the halo has been seen previously in both simulations \citep{2017astarkenburg} and observations \citep{dietz2020, Liu2022}. 
There are claims that this negative metallicity gradient with distance might be due to selection effects as other authors have observed a lack of this gradient out to 100 kpc \citep{2019conroyb}.
The MDFs presented in this work have been bias-corrected and should provide a much cleaner representation of the underlying metallicity structure of the halo.
We still clearly see a metallicity gradient with distance, but not as pronounced as it would be without the bias-correcting weights. This underscores the importance of our bias-correcting methods, while also highlighting the difference in the metallicity distribution of the Galactic halo (that is still present after accounting for the selection biases) at different distances out to 100 kpc.

\begin{table*}
	\centering
	\caption{
        Mean $\mu$, standard deviation $\sigma$, and component weights, $\omega$, of the three GMM components for the MDFs seen in Figure \ref{fig:MDF1}.
        }
        \label{tab:MDFGMMcontributions}
	\begin{tabular}{lccccccccc}
		\hline
		Distance bin [kpc] & $\mu_1$ & $\mu_2$ & $\mu_3$ & $\sigma_1$ & $\sigma_2$ & $\sigma_3$ & $\omega_1$ & $\omega_2$ & $\omega_3$  \\
		\hline
		\hline
		  $3<d_{hc}<4$ & --1.94 & --1.27 & --0.64 & 0.33 & 0.21 & 0.17 & 0.12 & 0.26 & 0.62 \\
		              & \textit{--1.94} & \textit{--1.27} & \textit{--0.64} & \textit{0.33} & \textit{0.21} & \textit{0.17} & \textit{0.12} & \textit{0.26} & \textit{0.62} \\
		\hline
		$4<d_{hc}<6$ & --1.96 & --1.28 & --0.69 & 0.37 & 0.23 &0.21 &0.15 &0.34 &0.50 \\
		            & \textit{--1.96} & \textit{--1.28} & \textit{--0.69} & \textit{0.37} & \textit{0.22} & \textit{0.21} & \textit{0.16} & \textit{0.35} & \textit{0.50} \\
		\hline
		$6<d_{hc}<10$ & --2.17 & --1.38 & --0.75 & 0.41 & 0.24 & 0.25 & 0.24 & 0.41 & 0.34  \\
		               & \textit{--2.17} & \textit{--1.38} & \textit{--0.76} & \textit{0.41} & \textit{0.24} & \textit{0.25} & \textit{0.24} & \textit{0.41} & \textit{0.34} \\
		\hline
		$10<d_{hc}<22$ & --2.11 & --1.35 & --0.78 & 0.38 & 0.22 & 0.27 & 0.18 & 0.42 & 0.40  \\
		              & \textit{--2.13} & \textit{--1.37} & \textit{--0.82} & \textit{0.38} & \textit{0.22} & \textit{0.28} & \textit{0.23} & \textit{0.45} & \textit{0.32} \\
		\hline
            $22<d_{hc}<45$ & --2.12 & --1.35 & --0.74 & 0.41 & 0.24 & 0.32 & 0.18 & 0.48 & 0.34 \\
                            & \textit{--2.17} & \textit{--1.40} & \textit{--0.86} & \textit{0.39} & \textit{0.23} & \textit{0.31} & \textit{0.25} & \textit{0.48} & \textit{0.27} \\
		\hline
            $45<d_{hc}<101$ & --2.14 & --1.43 & --0.85 & 0.44 & 0.20 & 0.20 & 0.23 & 0.42 & 0.35 \\
                            & \textit{--2.22} & \textit{--1.51} & \textit{--1.01} & \textit{0.43} & \textit{0.26} & \textit{0.38} & \textit{0.36} & \textit{0.45} & \textit{0.18} \\
		\hline
	\end{tabular}
    \tablefoot{The values in italics shows $\mu$, $\sigma$ and $\omega$ when no weights have been applied.
    '1' refers to the metal-poor GMM component, '2' to the middle component, and '3' to the metal-rich component.
    The total GMMs are shown in Figure \ref{fig:MDF2}.}
\end{table*}

\subsection{Dynamical view of metallicity substructures}\label{4.3}

In this section, we create the subsample of 6D positions and velocities of our PDR1 and PGS-giants catalogue and highlight the many different substructures seen in the outer halo down to lower metallicities using this 6D information.

\begin{figure}
  \includegraphics[width=\columnwidth]{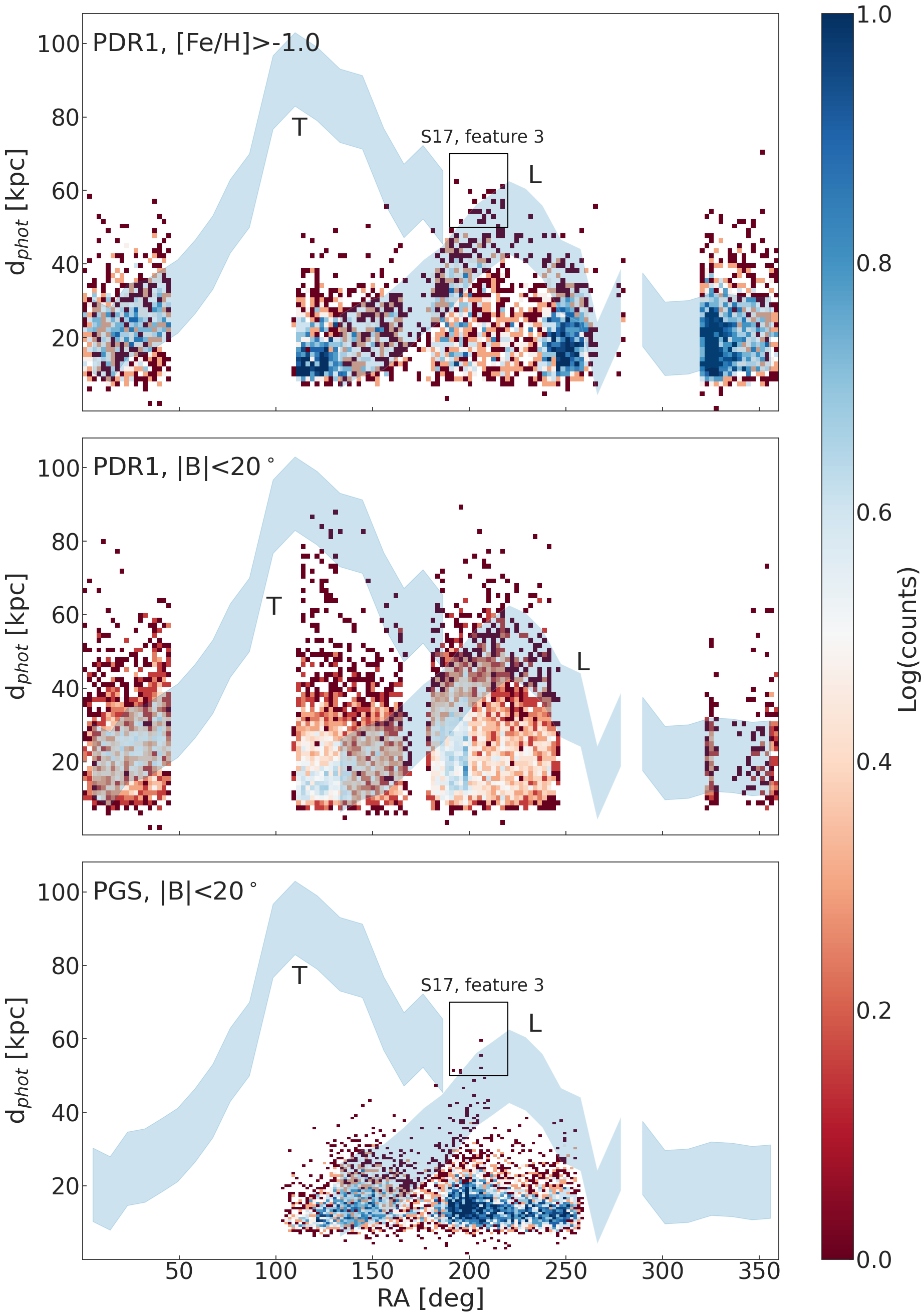}
  \caption{Sagittarius stream in our RGB catalogues. Photometric distance versus RA in the PDR1-giants catalogue at higher metallicities ([Fe/H]>-1.0, top), close to the Sagittarius stream plane (|B|<20$^\circ$, middle), and the latter for the PGS-giants catalogue (bottom). All panels have Sagittarius stream tracks adapted from \citet{2017hernitschek}. Note the spur feature (associated with feature 3 from \citet{2017sesar}). Leading and trailing arms are respectively labelled 'L' and 'T'. The stream members go further out and are more prominent with PDR1-giants than PGS-giants mostly due to the distance-metallicity selection effect in the latter catalogue.}
  \label{fig:sag}
\end{figure}

\begin{figure*}
\centering
  \includegraphics[width=0.5\textwidth]{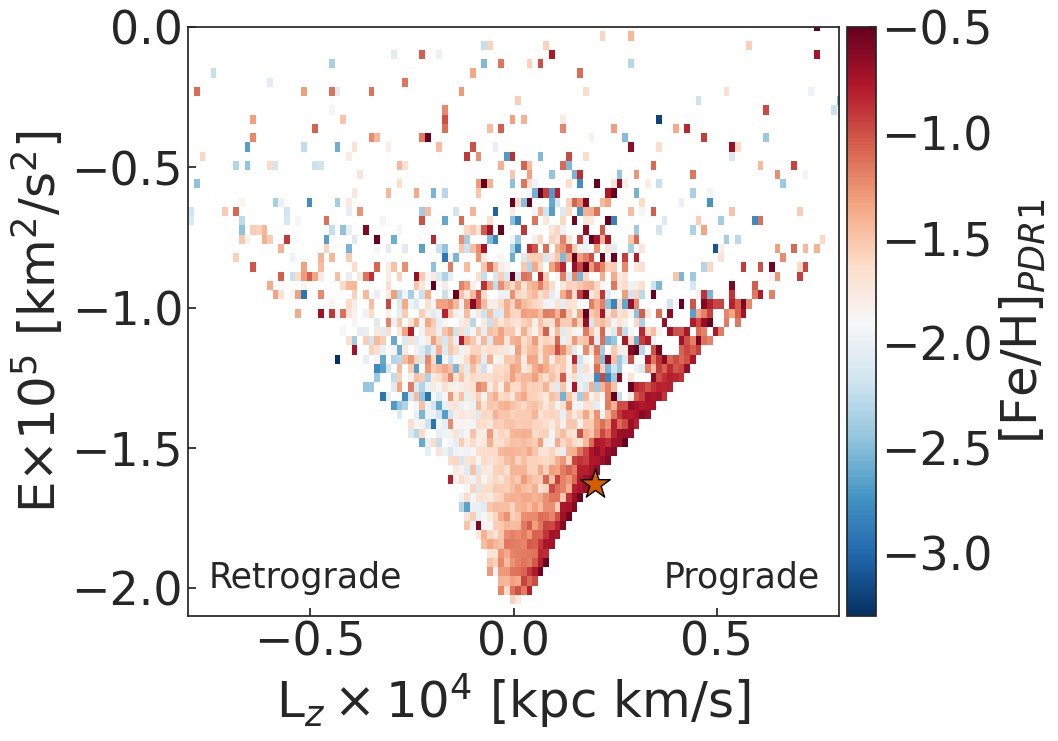}\\
  \includegraphics[width=\textwidth]{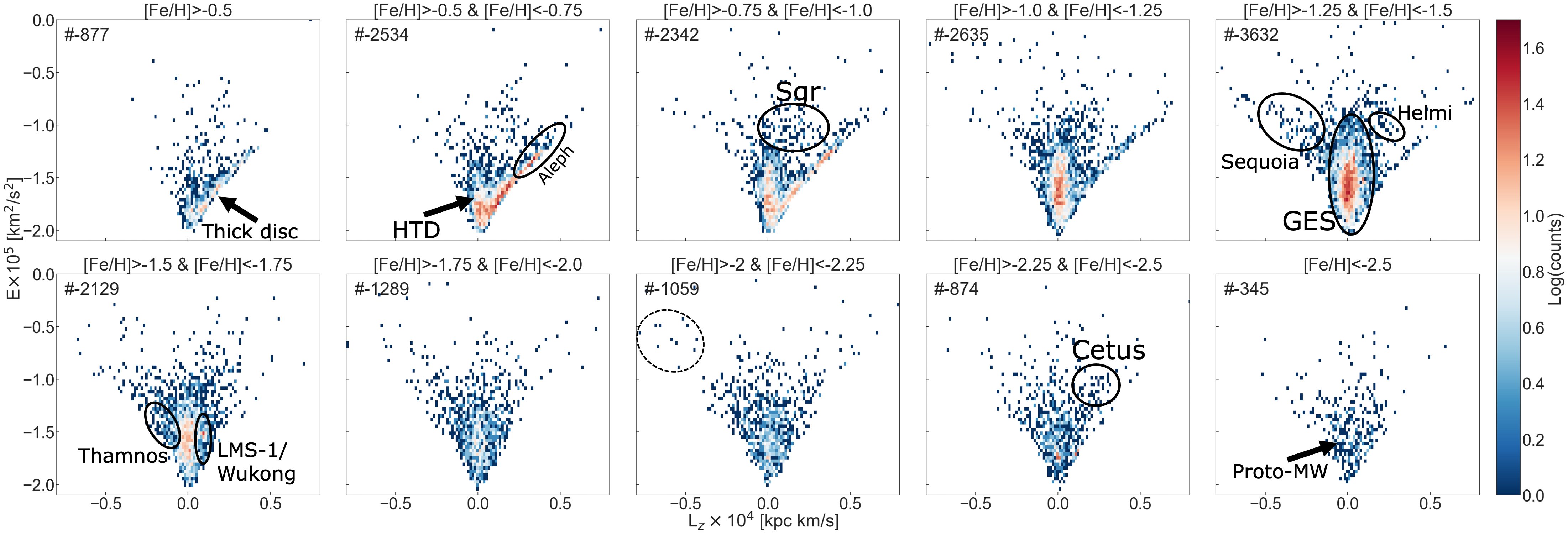}
  \caption{Metallicity view of the IOM space for the PDR1-giants catalogue. The energy versus angular momentum in the z-direction is colour-coded by the PDR1 metallicities, with the Sun's E-L$_z$ shown as a red star (top centre). The energy versus z-angular momentum at different slices in metallicities is colour-coded by their density distributions with different substructures highlighted for PDR1 6D giants (bottom panels). The following acronyms are used in this figure: HTD, hot thick disc; Sgr, Sagittarius; GES, Gaia-Enceladus-Sausage; LMS-1, low-mass stream-1; MW,  Milky Way.}
  \label{fig:iom}
\end{figure*}

\subsubsection{Radial velocities from spectroscopic surveys}\label{4.3.1}

To derive 6D phase-space information for a subset of our sample, we cross-matched our PDR1/PGS-giants with catalogs from the SDSS DR12 Sloan Extension for Galactic Understanding and Exploration \citep[SEGUE,][]{2000york}, the Large Sky Area Multi-Object Fiber Spectroscopic Telescope Medium and Low Resolution Surveys \citep[LAMOST MRS, \& LRS DR7,][]{2006zhao}, the RAdial Velocity Experiment \citep[RAVE DR6,][]{2006steinmetz}, the Galactic Archaeology with HERMES spectroscopic survey \citep[GALAH DR3,][]{2021buder}, the APO Galactic Evolution Experiment \citep[APOGEE DR17,][]{2017majewski}, Southern Stellar Stream Spectroscopic Survey \citep[S5 DR1,][]{2019li}, and \textit{Gaia} Radial Velocity Spectrometer \citep[\textit{Gaia} RVS DR3,][]{2023gaia}. These surveys complement each other in probing lower to higher latitudes, brighter to fainter stars, and northern and southern hemispheres. Even though the combination of all of these surveys along with the Pristine survey and/or the \textit{Gaia} XP sample gives raise to a complex selection function, we try to extract as much information as possible from the literature for our giants catalogue, and refrain from modelling the selection function, given the simple science cases shown in this work. The radial velocities are corrected for their zero point offsets with each other using \textit{Gaia} RVS radial velocities as the zero point \citep[similar to what has been done in][]{2022ruizlara}. In this work, we refer to stars with 6D information as `PDR1/PGS 6D giants' and the full catalogues as `PDR1/PGS-giants' implying 5D information without line-of-sight velocities. The PDR1 6D giants go out to $\sim$65 kpc and PGS 6D giants go out to $\sim$45 kpc in heliocentric distances. 

\subsubsection{Sagittarius stream}\label{4.3.2}

In Figure \ref{fig:sag}, we show the inferred photometric distances versus right ascension (RA) for PDR1-giants (top two panels) and PGS-giants (bottom panel), with the Sagittarius stream track from \citet{2017hernitschek} overlaid to guide the eye through the Sagittarius member stars in our sample. All the panels have $\pi$<0.05 mas (\texttt{parallax}<0.05, distance>20 kpc) to remove nearby field giants (same as removing stars d<10 kpc using our inferred photometric distances). In the top panel, we show the metal-rich stars ([Fe/H]<-1.0) to pick out the structures we see in Figure \ref{fig:relativez}. We can see the leading arm traced out to $\sim$60 kpc along with the spur feature 3 reported by \citet{2017sesar}. 
It is important to note that the spur feature seen in the top and bottom panels of Figure \ref{fig:sag} is at the same distance probed by their RR Lyrae counterpart, showing the reliability of our distances. This spur feature close to the apocentre of the leading arm is selected with the Sagittarius stream coordinate absolute latitude cut within 20$^\circ$, but it remains up to 9$^\circ$, with few members going down to 5$^\circ$, consistent with what is seen in the literature with standard candle tracers. This confirms the association of the spur feature with the stream itself and the existence of such apocentre lumps is seen in most Sagittarius simulations for the disruption of a Sagittarius dwarf Spheroidal (dSph)-like galaxy \citep{2017sesar}. 
However, the nearby distances are still quite noisy due to field star contamination. This does not change with the parallax cut that removes nearby stars. In the middle panel, we show all stars within 20$^\circ$ of the Sagittarius stream coordinate latitude (using the \citealt{2021vasilievsag} coordinate conversion). From the bulk of stars from GES that are more metal poor than Sagittarius, we do not see the Sagittarius signal clearly in this plot. However, we trace the trailing arm out to the apocentre more cleanly (with a small offset in distance that matches well within the distance uncertainties) in this view. Adding a metallicity cut on top of the stream latitude cut improves the selection at nearby distances but we do not see the trailing arm at larger distances anymore. In the bottom panel, we show the same for PGS-giants and we clearly trace the leading arm and the spur feature 3 in this catalogue. We do not see Sagittarius as much in PGS-giants, mostly due to the distance-metallicity selection effect due to which we do not see any metal-rich stars at higher distances and the Sagittarius stream is relatively metal-rich compared to the bulk of halo stars at large distances. A similar analysis using PDR1/PGS 6D giants is discussed in detail in Appendix \ref{C}. This results in a sparser, more nearby, but cleaner selection of Sagittarius stream members due to the availability of 6D information for a full kinematic selection. 
The cleanest selection of Sagittarius in both the catalogues of giants are using 6D phase space information where available and if not, using a metal-rich cut for PDR1-giants or a latitude cut on PGS-giants as can be seen in the top and bottom panels of Figure \ref{fig:sag}. 
However, we know that the Sagittarius streams have a clear metallicity gradient, which will impact the MDF in lower metallicities as well \citep{2024cunningham}.

\subsubsection{Metallicity view of integrals-of-motion space}\label{4.3.3}

Substructures from small or large dwarf galaxies accreted onto the Milky Way have a distinct chemistry to the field halo stars (in situ). They will have similar metallicities with a relatively smaller metallicity dispersion and spread in chemical abundances as seen in merger events \citep{2012leaman} such as GES ([Fe/H] = --1.18$\pm$0.3), Helmi streams ([Fe/H] = --1.28$\pm$0.19), Sequoia ([Fe/H] = --1.59$\pm$0.25), Thamnos ([Fe/H] = --1.9$\pm$0.41), LMS-1/Wukong ([Fe/H] = --1.58$\pm$0.23), Sagittarius ([Fe/H] = --1.0$\pm$0.2), and Cetus ([Fe/H] = --2.17$\pm$0.20, [Fe/H] = $\sim$--2.0 in some other works \citealt{2022thomas,2022yuan}) \citep[see e.g.][for more details on these values]{2022malhan}. 
It is important to note that even though the reported metallicity dispersions of these accreted dwarf galaxies are approximated to be Gaussians, in reality these galaxies have a long tail towards the metal-poor end that is more difficult to measure and constrain reliably \citep{2012leaman}.
These dispersion measurements in chemical abundances works well for substructures with [Fe/H]>--2.5.
For lower metallicities, its harder to separate accretion events from the general halo using chemical abundances \citep[see e.g.][]{2024sestito}.
However, in this study, a large fraction of our stars have [Fe/H] $>$ --2.5, which makes it easier to separate accretion events by combining chemical and dynamical information, such as by binning in metallicity and examining their distribution in IOM space, as shown in subsection \ref{4.3.3}.

This implies that different accretion events contributing to various regions of the Galactic halo—particularly their still-unmixed debris—can manifest as over- or underdensities in the relative number of stars at specific metallicities within narrow distance ranges. Studying the relative contribution of different metallicity bins to different distances in the Galaxy also allow us to understand the metallicity structure of the halo. Figure \ref{fig:relativez} shows the relative fraction of stars in 8 different metallicity bins along the absolute height above the disc plane. We remove stars with |Z|<2 kpc to avoid thin/thick disc contamination. Fractions are computed for 40 bins in distances, spaced evenly on a logarithmic scale between 2 and 90 kpc. From Figure \ref{fig:iom}, we can see the relatively metal rich stars ([Fe/H]>-1.0) coming from Sagittarius stream \citep{2024cunningham} at large Z. On the intermediate metallicities (-1.0<[Fe/H]<-1.5), we see the apocentre pile-ups of GES merger, such as Hercules-Aquila Cloud (HAC), and Virgo Overdensity (VOD) at smaller Z (<30 kpc) and Outer Virgo Overdensity (OVO) and outer HAC at larger Z (>30 kpc) \citep[see e.g.][]{2007belokurov,2007newberg,2017sesar}. At scale heights larger than 40 kpc, the contribution from VMP stars increases very steeply. We associate about 40\% of the total halo to VMP stars and 20\% to EMP stars in the outermost halo (d>65 kpc). These numbers demonstrate the power of our RGB catalogue in probing further out to 100 kpc, down to the most metal-poor stars in the Galaxy. 

Merger debris from different accretion events that made up the Milky Way halo in the distant past are clustered in the IOM space \citep{2000helmi}. Here, we use two typical quantities as IOMs: the angular momentum in the z-direction (L$_z$), and the total energy (E). L$_z$ is truly conserved in an axisymmetric potential, while varying slowly in a triaxial potential, maintaining a certain degree of clustering for stars on similar orbits from the same accretion event, though it is not fully conserved. The total energy, E, was computed as 

\begin{equation}
  E = \frac{1}{2} v^2 + \Phi(r),  
\end{equation}

where $\Phi(r)$ is the Galactic gravitational potential at the star's location. For this analysis, we used the same potential as in \citet{2022lovdal}: a Miyamoto-Nagai disc, Hernquist bulge, and Navarro-Frenk-White halo with parameters (a$_d$, b$_d$) = (6.5, 0.26) kpc, M$_d$ = 9.3 $\times$ 10$^{10}$ M$_\odot$ for the disc, c$_b$ = 0.7 kpc, M$_b$ = 3.0 $\times$ 10$^{10}$ M$_\odot$ for the bulge, and r$_s$ = 21.5 kpc, c$_h$ = 12 kpc, and M$_{halo}$ = 10$^{12}$ M$_\odot$ for the halo. We used a low renormalised unit weight error (\texttt{ruwe} < 1.4) to use stars with good quality astrometry and remove potential binaries. We assume V$_{LSR}$ = 232 km/s, a distance of 8.2 kpc between the Sun and the Galactic centre \citep{2017mcmillan}, and (U$_\odot$, V$_\odot$, W$_\odot$) = (11.1, 12.24, 7.25) km/s for the peculiar motion of the Sun \citep{2010schonrich}.

We show the IOM space (energy versus angular momentum in z-direction) colour-coded by the mean metallicities for the PDR1 6D giants on the top panel of Figure \ref{fig:iom}. We choose PDR1 over PGS 6D giants due to the larger distances probed, the fact that we are looking for phase-mixed structures (not coherent ones) which reduces the impact of the Pristine survey footprint, the higher quality of the photometric metallicities, and the low-to-no metallicity biases. 
In the bottom panels of Figure \ref{fig:iom}, we show the IOM space in ten different bins of metallicities indicating the different structures/accretion events identified in the literature \citep{2019koppelman,2019myeong,2020yuan,2020naidu,2022lovdal,2022ruizlara,2022malhan,2022thomas,2022yuan,2023horta,2023dodd}. The bins are chosen wide enough to see most of one dwarf accretion event in one bin, given their metallicity dispersion scales. All these panels and subpanels are plotted for an absolute scale height greater than 3 kpc cut to remove disc stars (|Z|>3 kpc). This inevitably also removes foreground and/or background stars from a spherical halo distribution, but the effects of this should be minimal given that the halo completeness matters less than the purity for studying phase mixed halo substructures.

In the first panel, for [Fe/H] > -0.5, we can clearly see prograde thick disc stars still left over in our sample. In the next bin (-0.75 < [Fe/H] < -0.5), we see the hot thick disc/Splash stars that are thick disc stars kicked up to halo-like orbits, likely resulting from the heating of the primordial high-$\alpha$ thick disc due to early mergers. We also find the 'Aleph' structure in this bin, a highly circular structure that is significantly enriched ([Fe/H] = -0.5, [$\alpha$/Fe] = 0.2), and may be associated with the enigmatic globular cluster Palomar 1. Its origin, whether in situ or ex situ, is still ambiguous.
From the next panels, we start seeing accretion events that made up the Milky Way halo. Between metallicities of -1.0 and -0.75, we see the now-disrupting Sagittarius stream. We are probing the lower energies of Sagittarius in this work, mostly due to the fact that the trailing arm (at higher energy) is not covered as well as the leading arm (at lower energy) in our PDR1 6D giants. Parts of the Sagittarius stream are also visible at lower metallicities, but fewer in number. It has a clear and distinct negative L$_y$, which we use to select 6D members of the Sagittarius stream in Appendix \ref{C} \citep[see][who also use the same criteria to select Sagittarius stream]{2023chandraa}. 
From [Fe/H] < -1.0, we already start seeing the last major merger, GES, and metal-poor end of Sgr stream, down to metallicities of -2.0. However, the density of GES stars at L$_z\sim$ 0 peaks in the metallicity bin -1.5<{Fe/H]<-1.25. We probe large distances and higher energies of this last major merger event, down to lower metallicities, for the first time.
The lower energies of GES contour does not necessarily have to belong to GES, but could belong to the old protogalaxy or the metal-poor tail of the hot thick disc. 
However, we need more chemical abundances to distinguish them. 
This is the same metallicity bin where we see the highly retrograde high energy Sequoia event and the Helmi streams, which is one of the first halo structures discovered through IOMs \citep{1999helminature}. In the more metal-poor bin (-1.75<[Fe/H]<-1.5), we see two more metal-poor structures namely, Thamnos, a low-mass ($\sim$2$\times$10$^6$M$_\odot$), retrograde structure deep in the potential well of the Galaxy, and LMS-1/Wukong that is still disrupting and is reported to be VMP in some studies \citep{2021malhan}. 
However, we find a strong density peak around this metallicity bin, which is about 0.5 dex more metal-rich than some literature studies \citep{2021malhan}, but similar in metallicities to the some other \citep{2020naidu}. 
The next two metallicity bins look cleaner with no indication of any significant substructures. We do see a group of highly retrograde (L$_z$>-0.5 kpc km/s) and high energy (E$\sim$-0.5 km$^2$/s$^2$) VMP stars ([Fe/H]<-2) in the metallicity bin -2.25<[Fe/H]<-2 separate from the rest of the distribution, the origin or reality of which needs a bigger sample of homogenously analysed 6D giants to be characterised out to large distances. In the future, with the upcoming WEAVE, 4MOST and DESI surveys \citep{2024weave,20194most,2023cooper} and our own high-resolution spectroscopic follow-up, we will have this 6D information and chemistry to understand the accretion history of our Milky Way at the VMP end. In the VMP bin -2.5<[Fe/H]<-2.25, we see the very prograde, still-disrupting Cetus stellar stream which is one the lowest metallicity structures from a dwarf galaxy accreted onto to the Milky Way. The final and most metal-poor bin still consists of 345 PDR1 6D giants, which is one of the largest collation of giants out to large distances and down to very low metallicities. This bin is almost free of substructures, but is slightly prograde and centrally concentrated, reminiscent of the proto-galaxy/poor-old-heart/Aurora population that are thought to be of in situ origin tracing the infant Milky Way stage \citep{2022belokurov,2022rix,2023belokurov,2024ardernarentsen}. 
Among these many accretion events, we do not recover some of them namely, Shiva, Shakti and Pontus \citep{2022malhanpontus,2024malhan}. 
This could be due to selection effects in the many different spectroscopic surveys and the Pristine survey itself. 
On the other hand, this could also be explained by the selection effect in the Gaia RVS radial velocities \citep{2022lane,2023dillamore} and ridges caused by bar resonances \citep{2024dillamore} for Shiva and Shakti events and Pontus being captured as the low energy part of the last major merger, GES \citep{2022amarante}. 
The same plots are made but colour-coded by galactocentric distances, to understand the region of the Galaxy probed by these accretion events (wherein the stars with lower energies have lower distances and higher energies have higher distances, as would be expected) and summarised in Appendix \ref{D}. In the top panel of Figure \ref{fig:iom}, we can already associate the many different substructures described above with their places in the IOM by eye as they are colour-coded by the mean metallicities. 
With a significant number of stars in each metallicity bin going further out into the halo, this is the first time we are able to associate all these substructures with their metallicity view of the IOM space.

\begin{figure*}
\sidecaption
  \includegraphics[width=12cm]{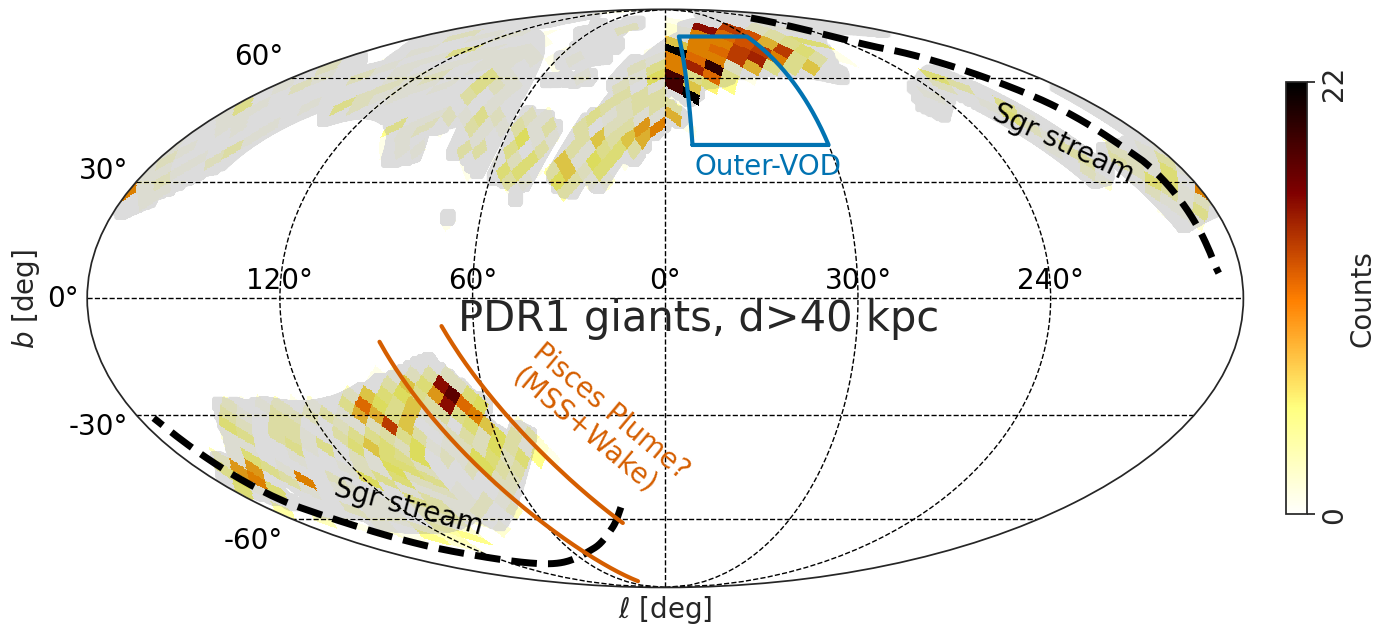}
    \caption{Outer halo of the Milky Way as viewed by PDR1-giants (top, Pristine footprint) and PGS-giants (bottom, \textit{Gaia}'s all-sky view) at heliocentric distances greater than 40 kpc. The colour map is a density distribution in Galactic coordinates at absolute latitudes greater than 20$^\circ$. We highlight the significant overdensities that dominate the outer halo and identify their most-likely progenitors. In the top panel, we also show the footprint of the entire Pristine survey in grey. Some of the pixels in the outer halo part of PDR1-giants are larger than the Pristine footprint due to the chosen healpix level.}
    \includegraphics[width=12cm]{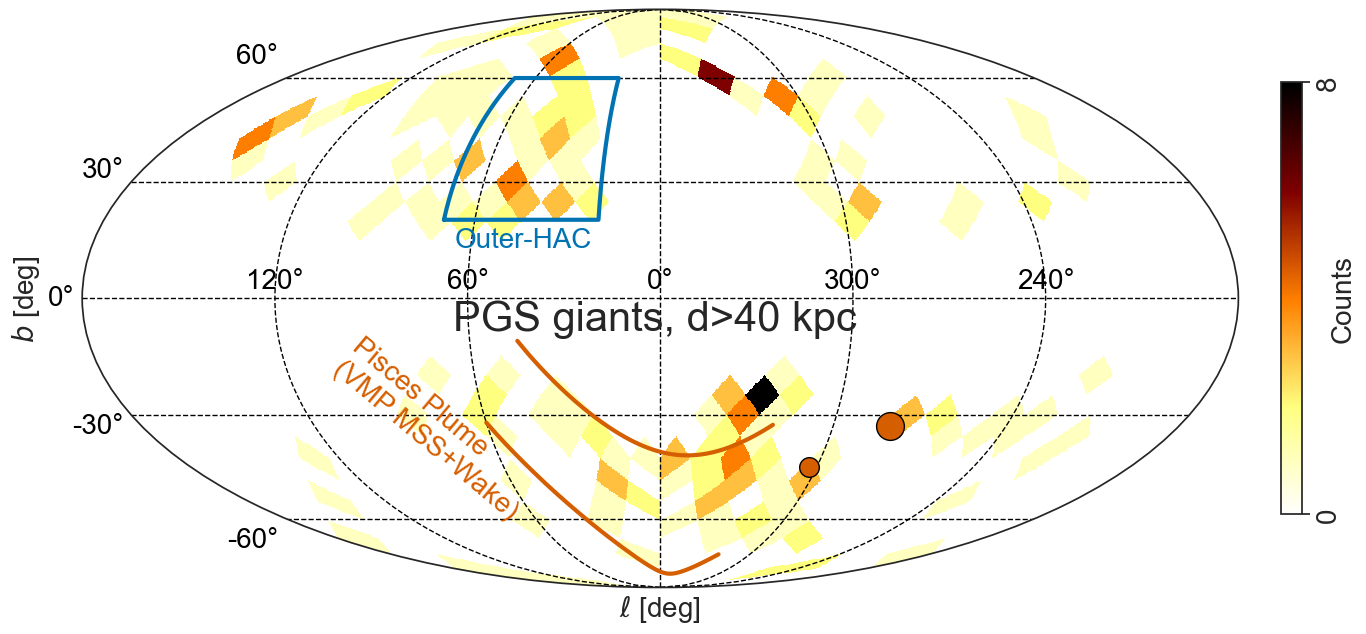}
  \label{fig:outerhalo}
\end{figure*}

\subsection{Outer halo metal-poor substructures}\label{4.4}

The outer halo is intriguing because it holds clues about the formation and evolution of the Milky Way, including the remnants of past mergers and accretion events and their turning points. Fully characterizing the global extent of structures found in local samples also requires looking further into the outer halo.
In Figure \ref{fig:outerhalo}, we show the on-sky distribution of the outer halo (d>40 kpc) colour-coded by the density of stars. We have $\sim$2000 PDR1-giants and $\sim$200 PGS-giants in the outer halo (that are all VMP due to the distance metallicity selection effect in PGS-giants), at distances greater than 40 kpc and Galactic latitudes higher than 20$^\circ$ to avoid extinction affecting our distance and metallicity estimates in the outer halo. We have almost no radial velocity \textit{from publicly available spectroscopic surveys} in this outer halo subsample. This is the largest collection of VMP stars out to such large distances, allowing us to study some of the earliest times of our galaxy's evolution. 

In the top panel of Figure \ref{fig:outerhalo}, we clearly see the Pristine survey footprint preventing us from looking at the all-sky distribution of outer halo substructures. However, the outer halo is full of many substructures, both dynamical and chemical. We clearly see overdensities near the Sagittarius stream tracks shown as black dashed lines. We see the highest density of stars around part of the region that overlaps significantly with the outer Virgo overdensity (outer-VOD) \citep{2017sesar}. This has been associated to the apocentric pileup of debris from the GES accretion event using a very complementary sample of outer halo giants by \citet{2023chandraa}.
The peak of the metallicity distribution for these stars is also close to the mean metallicity of GES. However, a larger number of 6D members would be needed to confirm this association.
Linking overdensities such as HAC and VOD to larger accretion events has also been explored by \citet{2021balbinot}.
We also see an overdensity of stars in the southern hemisphere around the same region as the Pisces Plume with the Magellanic wake overdensities but at higher longitudes. We guide the eye using the following track on the sky: $\ell$ = -60 + 0.2b + 0.01b$^2$ shifted by $\pm$10$^\circ$ (modified slightly from what was reported in \citealt{2023chandraa} for the Pisces Plume). However, this could also simply be explained by more stars being present in this region as it is getting closer to lower latitudes, where the stellar density is higher along with the \textit{Gaia}'s scanning law effect in the same region (see Figure 10 in \citetalias{2023martin}). A larger on-sky stretch of stars in this region would allow for a disentangling of these effects. 

In the bottom panel of Figure \ref{fig:outerhalo}, we see an all-sky view of the metal-poor Milky Way outer halo. Due to the distance-metallicity selection effect (see subsection \ref{3.5.2}), we are free of metal-rich stars in the outer halo, which allows us to study the outer halo's earliest evolution more easily. This is also why we do not see the Sagittarius stream which is one of the most prominent outer-halo metal-rich substructures. 
We see a clear overdensity of stars at the same region as HAC in the north \citep{2007belokurov}. However, these stars are at very large distances compared to the distances probed by HAC North and South (d<20 kpc) and these are all VMP. The origin of this overdensity is unclear and due to the unavailability of radial velocities, it is impossible to derive 6D-phase space information and orbital parameters. 
Therefore, we cannot associate it to GES or any other accretion events yet. We indicate this with a blue box and an outer-HAC label on Figure \ref{fig:outerhalo} bottom panel. 

\begin{figure}
  \includegraphics[width=\columnwidth]{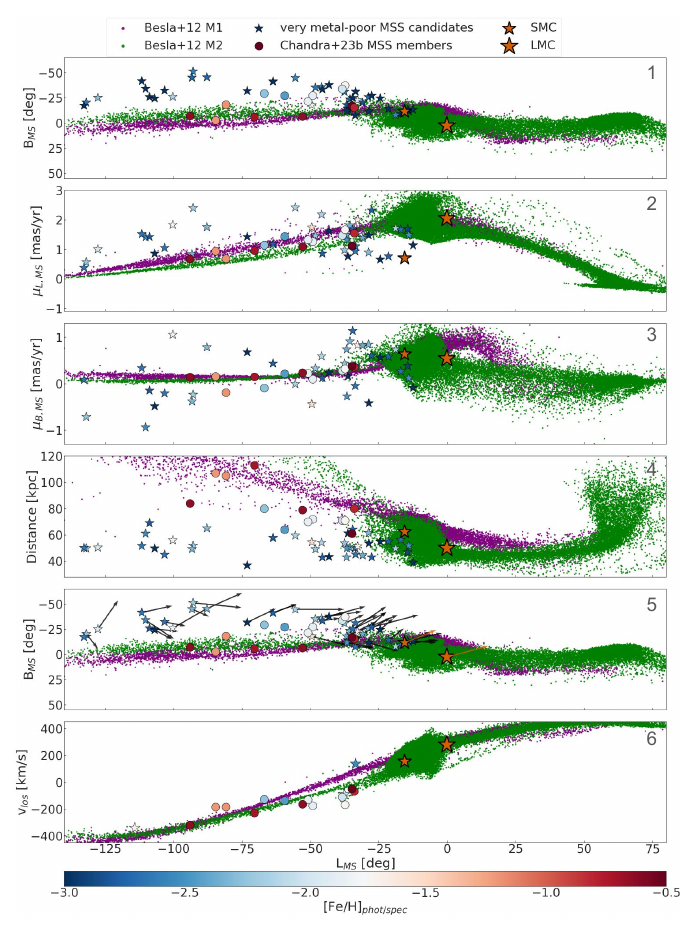}
      \caption{Our VMP MSS member candidates projected along the Magellanic stream coordinates (panel 1), proper motions along and transverse to the stream (panels 2 and 3), heliocentric distance (panel 4), along projected stream coordinates with their transverse motion highlighted using velocity vectors (panel 5), and line-of-sight radial velocity (where available, from PGS 6D giants, panel 6). All the parameters are plotted against the Magellanic stream longitude. The projection includes particles from both tidal debris models in \citet{2012besla}, LMC, SMC parameters adapted from \citet{2006kallivayalilb,2006kallivayalila}, and the Magellanic stellar stream debris discovered by \citet{2023chandrab}.}
  \label{fig:mss}
\end{figure}

In this bottom panel of Figure \ref{fig:outerhalo} in the south, we see a clear overdensity that we associate with the Pisces Plume \citep{2019belokurov} which could be a mix of the dynamical wake due to LMC's infall and the stellar counterpart of the Magellanic stream (MSS). 
We show the LMC and SMC (infalling satellites of the Milky Way) as large and small orange circles in the same figure. This overdensity is almost fully co-incident with the infalling orbits of the Magellanic Clouds. 
The overdensity close just above the Pisces plume and closer to inner Galaxy maybe caused due to Gaia's scanning law and metal-poor end of the Sagittarius dwarf, but we need 6D members to confirm this.
We discuss the Pisces Plume substructure more in detail in the next subsection.

\subsubsection{The Pisces Plume in the outer halo}\label{4.4.1}

Using \textit{Gaia} DR2, \citet{2019belokurov} used all-sky RR Lyrae stars to uncover a plume-like elongation near the Pisces Overdensity (we only map the plume and not the overdensity, mostly due to \textit{Gaia}'s scanning law pattern that creates northern and southern caps of underdensities in these regions) extending to larger distances. This elongation, aligned with the direction of the gaseous MS, suggests a connection to the Clouds. 
Based on the kinematics and metallicities of a small subsample of BHB stars in the Pisces Plume of this 'Pisces Plume', they argued that it predominantly represents the dynamical friction wake imprinted on the Milky Way’s halo by the infall of the LMC, rather than being composed of stripped stars from the Clouds. \citet{2021conroy} used all-sky RGB stars to suggest that the southern overdensity and a northern counterpart correspond to the dynamical friction wake and the 'collective response' of the LMC’s infall, matching predictions from simulations \citep{2019camargo,2021camargo}. However, they found that the southern overdensity is twice as strong in the data as predicted by simulations, possibly indicating multiple populations within the Pisces Plume.

The origin of this Pisces Plume is uncertain; it is still unclear whether it contains debris stripped from the Magellanic Clouds, ram-pressure stripping \citep[the Magellanic stream,][]{2003putman}, or if it primarily represents the dynamical friction wake of the Large Magellanic Cloud \citep{2019camargo,2021conroy}.
It is important to note that almost all of our member stars in this region are VMP ([Fe/H]<-2.0, most of them are below -2.5).
The Magellanic Stream is an extensive gaseous structure gracefully encircling the Milky Way and spanning over 140 degrees of the southern sky. 
Despite decades of dedicated observations and simulation efforts, the precise origin of the Magellanic Stream remains elusive. 
Two major formation processes, tidal disruption and ram-pressure stripping, are competing explanations for its existence. 
To complicate matters further, its trailing arm is also the region on-sky that experiences the Large Magellanic Cloud's dynamical friction wake. 
The theorised stellar counterpart to the gaseous Magellanic Stream is the MSS, which was recently traced in the relatively metal-rich end by \citet{2023chandrab} using a \textit{Gaia} XP spectra giants catalogue that is complementary to ours.
The stellar stream provides strong constraints on the distance and kinematics of the gaseous Magellanic Stream, helping us understand the past orientation and interaction history between the Clouds and the Milky Way. 
By accurately characterizing the MSS's detailed chemical abundances, we can study the chemical evolution in the outskirts of the Clouds and the interaction of their haloes with the outer Galactic halo.

\subsubsection{The Magellanic stellar stream down to the very metal-poor end}\label{4.4.2}

We select all our MSS member candidates between the orange polynomial lines shifted by $\pm$20$^\circ$ as shown in the bottom panel of Figure \ref{fig:outerhalo} at distances larger than 25 kpc. We choose this value because at large distances, our method tends to underestimate distances more than it overestimates them (see Figure \ref{fig:sh} for the comparison with Starhorse distances). 
In Figure \ref{fig:mss}, we show various kinematic properties of our VMP MSS member candidates in star symbols along the Magellanic stream longitude L$_{MS}$ (coordinate conversion based on the \citet{2008nidever} stream axis definition).
We overlay the two Magellanic stream models from \citet{2012besla,2013besla} that was modelled to trace the gas for comparison.
Model 1 (M1) was designed to best match the velocity of the Magellanic stream (MS) and model 2 (M2) to match the kinematics of the Clouds themselves. 
The LMC and SMC are also shown as big and small orange stars in all the panels. Their positions and velocities are taken from \citet{2006kallivayalilb, 2006kallivayalila} and the LMC and SMC have a median metallicity of -0.5 and -1.0 respectively.
We also overlap \citet{2023chandrab} members as circles. These members and our member candidates are colour-coded by spectroscopic metalliccities where available and, if not, PGS photometric metallicities. 
We find 41 member candidates by association to the MSS's trailing arm in the south in proper motion and positions, 47 are VMP ([Fe/H]<-2), 32 are [Fe/H]<-2.5 and 9 are EMP ([Fe/H]<-3) stars out to 70 kpc. 
\citet{2023chandrab} confirmed 7 relatively metal-rich members but also serendipitously discovered 6 members that are relatively metal poor ([Fe/H]<-1.5).
Their metal-rich population is described as extended and stream-like, while the metal-poor population is more diffuse and clumpy.
In summary, we find the metal-poor population to also exhibit a stream-like elongated orientation. 

In Figure \ref{fig:mss} panel 1, we see the members in stream coordinates and our members lie in the same region occupied by the six metal-poor members from \citet{2023chandrab}, but more elongated. These are slightly offset from the models which are closely tracing the gas. This could be because the gas is tracing stars that are tidally disrupted from the disc regions rather than the outskirts of the Clouds. These VMP stars must be some of the oldest stars associated with the Clouds that got kicked away into a stream-like structure from the outskirts of the SMC (also maybe the LMC, even though it is relatively metal-rich) from the LMC-SMC interaction \citep[see][who trace stream-like SMC stars accreted onto the LMC halo at L$_{MS}$>0]{2019navarrete}. Panel 2 and 3 show the stream longitude versus proper motions in L$_{MS}$ and B$_{MS}$, respectively. 
The VMP MSS candidates match well with the broad direction of the model's proper motions, but have a larger range and dispersion than the \citet{2023chandrab} members, especially in the B$_{MS}$ direction. Panel 4 shows the heliocentric distances that are much closer than the \citet{2023chandrab} members, reminiscent of the cloud-associated debris from \citet{2020zaritsky}. These stars have relatively nearby distances compared to the metal-rich members. The metal-poor members from \citet{2023chandrab} have similar (relatively nearby) distances comparable to our MSS VMP candidates' distances, which could mean that the metal-poor members are closer than the metal-rich members. On the other hand, we caution that our distances tend to be biased towards closer distances. Thus, we need more reliable spectrophotometric distances to confirm this.
Panel 5 shows the same as panel 1, but with transverse velocity vectors, most of which point in the same direction as the Clouds themselves. 
However, the members at smaller L$_{MS}$ have a bit more random motion than the members close to the Clouds. 
Therefore, these members are less likely to be members of the MSS.
Panel 6 shows the line-of-sight velocities for three members in our selection that are in common with the PGS 6D giants subset. Two of these members are from the S5 (DR1) survey and one member is from LAMOST (DR7 MRS) survey. All of these members are VMP and one of the S5 members has a spectroscopic metallicity [Fe/H] = -2.55$\pm$0.07, which is already the most metal-poor member of the MSS discovered yet. All these 6D stars are remarkably consistent with the models in panel 6 with respect to the line-of-sight velocities that are expected to trace the gaseous MS.

In this work, we find 41 stars at the metal-poor end, in the outer halo that we tentatively associate with the Pisces Plume overdensity/the Magellanic stellar stream. 
Even with the kinematic parameters roughly aligning with the gaseous MS, its models and MSS members presented in \citet{2023chandrab}, there is a possibility that part of these stars could be associated with the dynamical wake due to the LMC's infall. 
In the work of \citet{2023chandrab}, from their Pisces plume members, they find that at least 7 out of 45 stars (or $\geq$15\%) in their Pisces Plume overdensity appear to be confidently identified as debris from the Clouds. 
In our work, from our member candidates of Pisces plume, to clearly understand the percentage contribution from the Clouds themselves and the halo response at the VMP end, we need full 6D phase space information for all these stars and detailed chemical abundances. 
We have an ongoing spectroscopic follow-up program with the Gemini GHOST instrument \citep{2024mcconnachie} for >10 stars in this region covering the full extent of the plume at the bright end as a pilot program. From a full chemodynamical analysis of the stars in this region through follow-up spectroscopy, we can uncover the origin story of these VMP stars around the Pisces Plume, assess the contribution from the Clouds' tidal debris versus the field halo's response to the LMC's infall, while also understanding the true origin of the metal-poor stars in the MSS. 


\begin{table*}
\caption{Description of the columns of the PDR1 and PGS catalogues of giants made publicly available in this work.}  
\label{catalogue}
\centering
\begin{tabular}{llll}
\hline \hline
 Column & Description & Unit & Type \\
 \hline
source\_id & \textit{Gaia} DR3 Source ID & unitless & longint \\
ra & \textit{Gaia} DR3 right ascension (RA) in ICRS (J2016) format & degrees & float \\
dec & \textit{Gaia} DR3 declination in ICRS (J2016) format & degrees & float \\
parallax & \textit{Gaia} DR3 offset-corrected parallax & mas & float\\
parallax\_error & Offset-corrected uncertainty on the \textit{Gaia} DR3 offset-corrected parallax & mas & float\\
pmra & \textit{Gaia} DR3 proper motion in the RA direction & mas/yr & float\\
pmra\_error & Uncertainty on the \textit{Gaia} DR3 proper motion in the RA direction & mas/yr & float\\
pmdec & \textit{Gaia} DR3 proper motion in the declination direction & mas/yr & float\\
pmdec\_error & Uncertainty on the \textit{Gaia} DR3 proper motion in the declination direction & mas/yr & float\\
G\_0 & de-reddened \textit{Gaia} G magnitude & unitless & float\\
BP\_0 & de-reddened \textit{Gaia} G$_{BP}$ magnitude & unitless & float\\
RP\_0 & de-reddened \textit{Gaia} G$_{RP}$ magnitude & unitless & float\\
phot\_dist & Photometric distance inferred in this work & kpc & float\\
phot\_dist\_unc & Uncertainty on the photometric distance inferred in this work & kpc & float\\
v\_los & Radial velocity crossmatched with spectroscopic surveys & km/s & float \\
v\_los\_unc & Uncertainty in the radial velocity crossmatched with spectroscopic surveys & km/s & float \\
FeH\_phot & Photometric metallicity derived in the PDR1/PGS catalogues & unitless & float \\
FeH\_phot\_unc & Measurement uncertainty on the photometric metallicity derived in the PDR1/PGS catalogues & unitless & float \\
\hline
\end{tabular}
\end{table*}

\section{Conclusions and outlook}\label{6}

In this work, we have constructed two large RGB catalogues using the publicly available Pristine data: PDR1 (based on CFHT CaHK photometry) and PGS (based on Gaia XP spectra). These catalogues contain reliable photometric distances and metallicities, probing down to [Fe/H] = --3.5 and out to distances of $\sim$100 kpc. Below, we state the technical and scientific highlights of the paper.

\subsection*{Technical highlights}
\begin{itemize}
    \item The RGB selection pipeline combines (i) a CaMD cut using good enough parallaxes (uncertainty $\le$ 50\%) to separate giants from dwarfs (see subsection \ref{3.1}, Figures \ref{fig:puritycompletenessfvalues}, \ref{fig:ACMDtrainingsamplegiantdwarfcut}), (ii) a magnitude cut (G < 17.6) to retain bright giants with poor parallaxes while removing dwarfs (subsection \ref{3.2}, Figure \ref{fig:app-gmag}), and (iii) a colour–metallicity cut using MIST isochrones to remove subgiants (subsection \ref{3.3}, Figure \ref{fig:colormetallicitycut}; pipeline summarised in Figure \ref{fig:flowchart}).

    \item Photometric distances are derived via isochrone interpolation using each star's [Fe/H], T$_\text{eff}$, and log \textit{g}. Uncertainties ($\sim$12\%) are estimated from offsets in a good-parallax control sample (subsection \ref{3.4}, Figures \ref{fig:dist-comp}, \ref{fig:sh}). Distances are validated using inverted parallaxes and StarHorse estimates (maximum scatter < 20–40\%). PDR1 metallicities are more precise (uncertainty $\sim$0.1 dex) than PGS ($\sim$0.4 dex at G$\sim$16) due to a higher S/N in the narrowband CaHK photometry.

    \item Caveats include a slight underestimation of distances for red clump and metal-poor HB stars, as these are not explicitly modelled in the isochrone fitting (subsection \ref{3.5.1}, Figure \ref{fig:rc}). PGS also shows a distance-metallicity selection bias due to stricter CaHK uncertainty thresholds, preferentially selecting metal-poor stars at larger distances (subsection \ref{3.5.2}, Figure \ref{fig:sel-effect}).

    \item After all the quality cuts (including removal of subgiants with log\textit{g}>3.5), we obtained 180,314 RGB stars in PDR1 and 2,420,898 in PGS, with reliable metallicities and distances (Figure \ref{fig:sh}).

    \item The final purity and completeness based on the surface gravities from \citetalias{2023andrae} are 78\% and 76\% for PDR1 and 92\% and 82\% for PGS (Figure \ref{fig:loggdistributionandrae}).
\end{itemize}

\subsection*{Scientific highlights}

\begin{itemize}
    \item The [Fe/H]–distance plane (Figure \ref{fig:fehvsd}) shows distinct halo substructures, including the GES merger ([Fe/H]$\sim$-1.4) out to $\sim$50 kpc. PDR1 gives a bias-corrected view of the outer halo; PGS provides an all-sky map of VMP halo stars. Relative contributions of different metallicities as a function of height above the plane reveal overdensities such as Sagittarius, HAC, and VOD and a steep increase in the VMP halo beyond 40 kpc (Figure \ref{fig:relativez}).

    \item Using bias-corrected metallicity distributions based on PARSEC simulated stellar populations, we modelled MDFs in six galactocentric distance bins. A three-component GMM fit shows that with increasing distance, the halo becomes more metal-poor and the dispersion increases (Figures \ref{fig:MDF1} and \ref{fig:MDF2}).

    \item We constructed a 6D sample (with radial velocities) of 111,305 PDR1 and 1,706,006 PGS-giants, probing to $\sim$70 and $\sim$50 kpc, respectively (Figure \ref{fig:iom}).

    \item Sagittarius stream members were identified using both 5D and 6D selections. In the absence of radial velocities, metal-rich ([Fe/H]>--1) stars from PDR1 and a latitude cut (|B|<20$^\circ$) on PGS stars work well to select the Sagittarius stream. With 6D, the cleanest selection is via a Ly<--0.25$\times$10$^4$ kpc km/s cut (Figures \ref{fig:sag}, \ref{fig:app-sag1}, \ref{fig:app-sag2}).

    \item Mapping the IOM space with 6D PDR1-giants and slicing in metallicity revealed known substructures, including Sagittarius, GES, Helmi streams, Sequoia, LMS-1, Cetus, and more. A new retrograde, high-energy, VMP outer halo component was also detected (Figure \ref{fig:iom}).

    \item In the outer halo (d > 40 kpc), we identified Sagittarius, outer-VOD (possibly GES debris), outer-HAC, and a prominent overdensity along the Pisces Plume, possibly tied to the Magellanic Clouds (Figure \ref{fig:outerhalo}).

    \item We identified 41 Magellanic Stream candidates among PGS VMP stars, three with 6D data that trace gaseous MS velocities. One has [Fe/H] = --2.55$\pm$0.07, the most metal-poor MSS candidate to date. These require follow-up chemodynamical confirmation (Figure \ref{fig:mss}).

    \item Both RGB catalogues are made publicly available in the format shown in Table \ref{catalogue}.
\end{itemize}

The PDR1 catalogue enables bias-corrected studies of the outer halo's MDF and accretion history, while the all-sky PGS catalogue maps the VMP halo. These RGB samples can probe the metallicity gradient, identify ancient dwarf galaxy debris, and trace faint halo substructures. With future chemodynamical studies from WEAVE \citep{2024weave}, 4MOST \citep{20194most}, DESI \citep{2023cooper,2024Koposov}, and our own Gemini/GHOST follow-up, we will unveil the complex assembly of the Galaxy’s outermost regions-which has the most exciting discovery space that has not been probed by most Galactic studies yet.

\section*{Data availability}

PDR1/PGS-catalogues of giants are available at the CDS via \href{https://cdsarc.cds.unistra.fr/viz-bin/cat/J/A+A/706/A195}{https://cdsarc.cds.unistra.fr/viz-bin/cat/J/A+A/706/A195}.

\begin{acknowledgements}
      AV thanks Ewoud Wempe, and Vedant Chandra for their helpful discussion on this work. AV also thanks Tomas Ruiz-Lara and Eduardo Balbinot for making the \textit{Gaia} DR3 catalogues and relevant survey cross-matches available to the 'Galactica' group members immediately after the \textit{Gaia} Data Release 3. AV and CN thank Gurtina Besla, and Himansh Rathore for sharing their simulations of tidal Magellanic stream debris from \citet{2012besla,2013besla}.
      AV gratefully acknowledges support from the Canadian Institute for Theoretical Astrophysics (CITA) through a CITA National Fellowship and the International Astronomical Union (IAU) and the Gruber Foundation through a IAU Gruber Fellowship.
      AB acknowledges the European Union's Erasmus+ Traineeship program, and the Edinburgh Doctoral College Scholarship (ECDS) that funded her contribution to this work.
      ES, and MM acknowledge funding through VIDI grant "Pushing Galactic Archaeology to its limits" (with project number VI.Vidi.193.093) which is funded by the Dutch Research Council (NWO).
      TM was supported by a Gliese Fellowship at the Zentrum f\"{u}r Astronomie, University of Heidelberg, Germany. 
      NFM gratefully acknowledge support from the French National Research Agency (ANR) funded project "Pristine" (ANR-18-CE31-0017) along with funding from the European Research Council (ERC) under the European Unions Horizon 2020 research and innovation programme (grant agreement No. 834148).
      FS and KAV thank the National Sciences and Engineering Research Council of Canada for funding through the Discovery Grants and CREATE programs.
      AAA acknowledges support from the Herchel Smith Fellowship at the University of Cambridge and a Fitzwilliam College research fellowship supported by the Isaac Newton Trust. GB acknowledges support from the Agencia Estatal de Investigaci\'on del Ministerio de Ciencia en Innovaci\'on (AEIMICIN) and the European Social Fund (ESF+) under grant PRE2021-100638. 
      GB, and GF acknowledges support from the Agencia Estatal de Investigaci\'on del Ministerio de Ciencia en Innovaci\'on (AEI-MICIN) and the European Regional Development Fund (ERDF) under grant number AYA2017-89076-P, the AEI under grant number CEX2019-000920-S and the AEI-MICIN under grant number PID2020-118778GBI00/10.13039/501100011033. Based on observations obtained with MegaPrime/MegaCam, a joint project of CFHT and CEA/DAPNIA, at the Canada-France-Hawaii Telescope (CFHT) which is operated by the National Research Council (NRC) of Canada, the Institut National des Sciences de l'Univers of the Centre National de la Recherche Scientifique of France, and the University of Hawaii. This work has made use of data from the European Space Agency (ESA) mission Gaia (https://www.cosmos.esa.int/gaia), processed by the Gaia Data Processing and Analysis Consortium (DPAC, https://www.cosmos.esa.int/web/gaia/dpac/consortium). Funding for the DPAC has been provided by national institutions, in particular the institutions participating in the Gaia Multilateral Agreement. 
      This research was supported by the International Space Science Institute (ISSI) in Bern, through ISSI International Team project 540 (The Early Milky Way). This research has been partially funded from a Spinoza award by NWO (SPI 78-411). AV also thanks the availability of the following packages and tools that made this work possible: \texttt{vaex} \citep{2018vaex}, \texttt{pandas} \citep{2022pandas}, \texttt{astropy} \citep{2022astropy}, \texttt{NumPy} \citep{2006numpy,2011numpy}, \texttt{SciPy} \citep{2001scipy}, \texttt{matplotlib} \citep{2007matplotlib}, \texttt{seaborn} \citep{2016seaborn}, \texttt{agama} \citep{2019vasiliev}, \texttt{gala} \citep{2017pricewhelan}, \texttt{galpy} \citep{2016jupyter}, \texttt{healpy} \citep{healpy}, \texttt{gaiadr3-zeropoint} \citep{2021lindegreen}, \texttt{jax} \citep{2019jax}, \texttt{scikit-learn} \citep{2011scikit-learn}, \texttt{JupyterLab} \citep{2016jupyter}, and \texttt{topcat} \citep{2018topcat}.
\end{acknowledgements}

\vspace{-0.95cm}
\bibliographystyle{aa}
\bibliography{references}

\begin{appendix}

\section{\textit{Gaia} G for dwarfs with good parallax}\label{A}

\begin{figure}[htbp!]
  \includegraphics[width=\columnwidth]{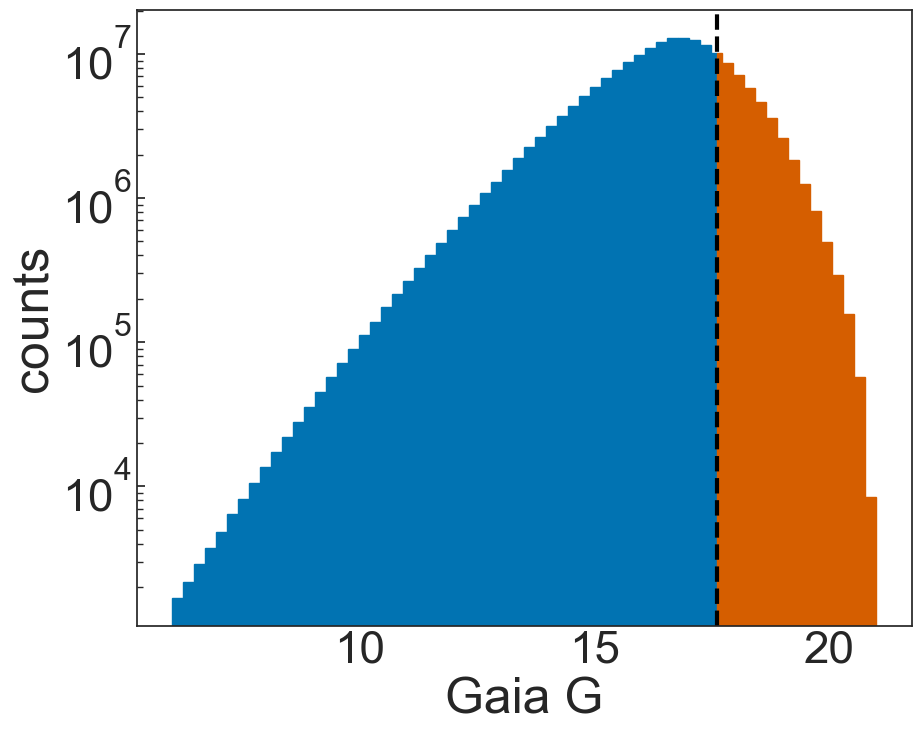}
  \caption{\textit{Gaia} G magnitude distribution of dwarfs with good parallax ($f$ $\leq$ 0.2) removed using CaMD cuts shown in Figure \ref{fig:ACMDtrainingsamplegiantdwarfcut} before applying the magnitude cut. We can see that our magnitude cut falls just before the distribution falls off due to incompleteness in magnitudes where dwarfs do not have good parallaxes ($f$ $\leq$ 0.2) (in orange). Selected stars' distribution used for the CaMD cut is shown in blue.}
  \label{fig:app-gmag}
\end{figure}

In Figure \ref{fig:app-gmag}, we show the \textit{Gaia} G apparent magnitude distribution for dwarfs that are discarded using the CaMD cut described in subsection \ref{3.1} and Figure \ref{fig:ACMDtrainingsamplegiantdwarfcut}. 
From the distribution in Figure \ref{fig:app-gmag}, we can see that at magnitude of about 17.3, the number of dwarfs with `good enough' parallax drops steeply, indicating that dwarfs with good parallax ($f$ $\leq$ 0.2) are not complete beyond this limit, i.e. not all dwarfs beyond this magnitude have good enough parallax that allow us to remove dwarfs and keep giants based on their `bad parallax'. We empirically choose the magnitude limit to select giants based on their bad parallax at G>17.6. This is very close to the G<17.3 cut suggested by Figure \ref{fig:app-gmag}, but slightly different as this is also approximately the limiting magnitude of the \textit{Gaia} XP spectra and in turn, the limiting magnitude of publicly available PDR1 catalogue, allowing us to select giants in these publicly available catalogues. 

\section{Comparison of inferred distances with \citet{2021bailer-jones} photogeometric distances}\label{E}

\begin{figure}[htbp!]
    \includegraphics[width=\columnwidth]{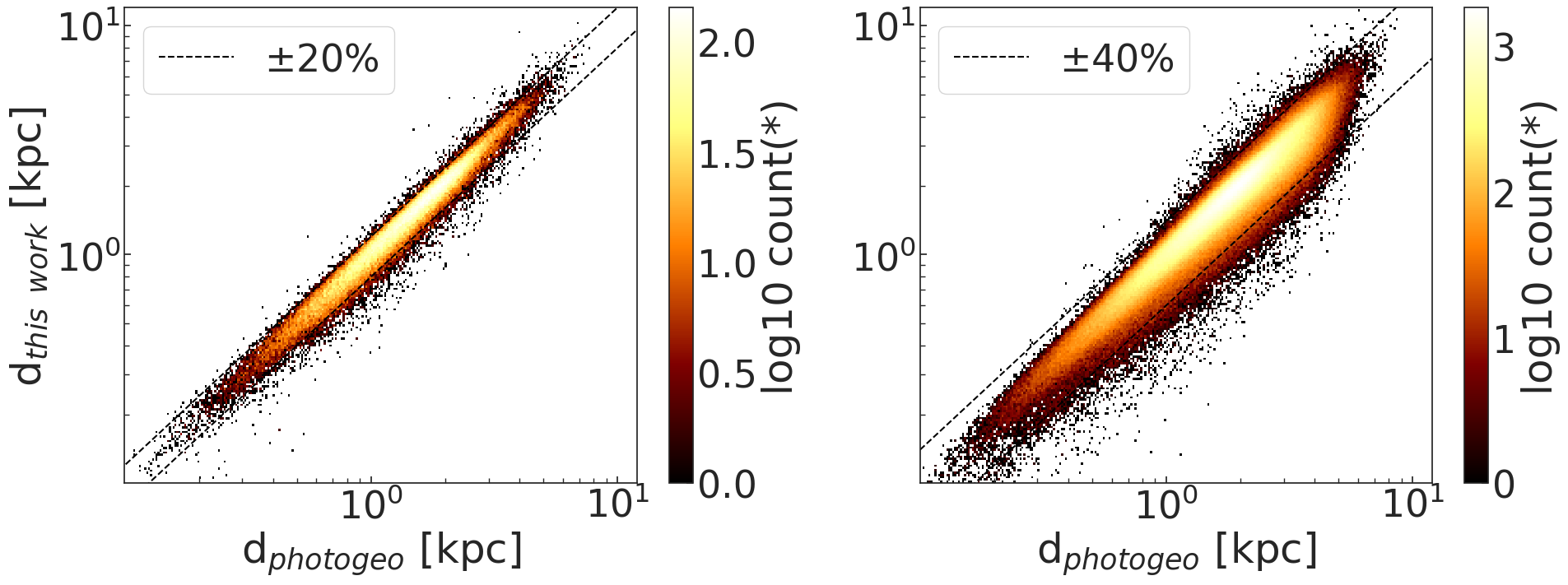}
    \caption{Comparison of our photometric distances with Gaia DR3 \citet{2021bailer-jones} photogeometric distances for PDR1 (left) and PGS (right) giants catalogue constructed in this work.}
    \label{fig:bailer}
\end{figure}

In Figure \ref{fig:bailer}, we show a comparison of our inferred photometric distances against \citet{2021bailer-jones} good quality photogeometric distances (<5\% uncertainties). 
The photogeometric distances utilise the G magnitude and BP-RP colour from Gaia DR3. 
These distance estimates incorporate direction-dependent priors based on a detailed model of the Galaxy’s 3D structure, taking into account the distribution, colours, and magnitudes of stars as observed by Gaia. 
This model also factors in interstellar extinction and the Gaia selection function. 
Tests with mock data, alongside validation against independent measurements and open clusters, indicate that these estimates remain reliable up to several kpc. 
However, for faint or distant stars, the prior often plays a significant role in the estimation, which is one of the main reasons we scale our inferred distances to be constrained well with the inverted-parallax measurements which is fully observational with no assumptions on the Galaxy, making it more reliable for VMP stars, as the priors on the Galaxy distribution do not always apply the same way for the most metal-poor stars.
From Figure \ref{fig:bailer}, we can see that both the PDR1 and PGS-giants are in good agreement with the photogeometric distances, with the PDR1 distances with much lower spread (<20\%) compared to PGS-giants ($\leq$40\%) due to the quality of photometric metallicities being way higher for PDR1-giants compared to PGS-giants \citepalias[see][]{2023martin}.

\section{Metallicity structure with spatial distributions}\label{B}

\begin{figure}[htbp!]
  \includegraphics[width=\columnwidth]{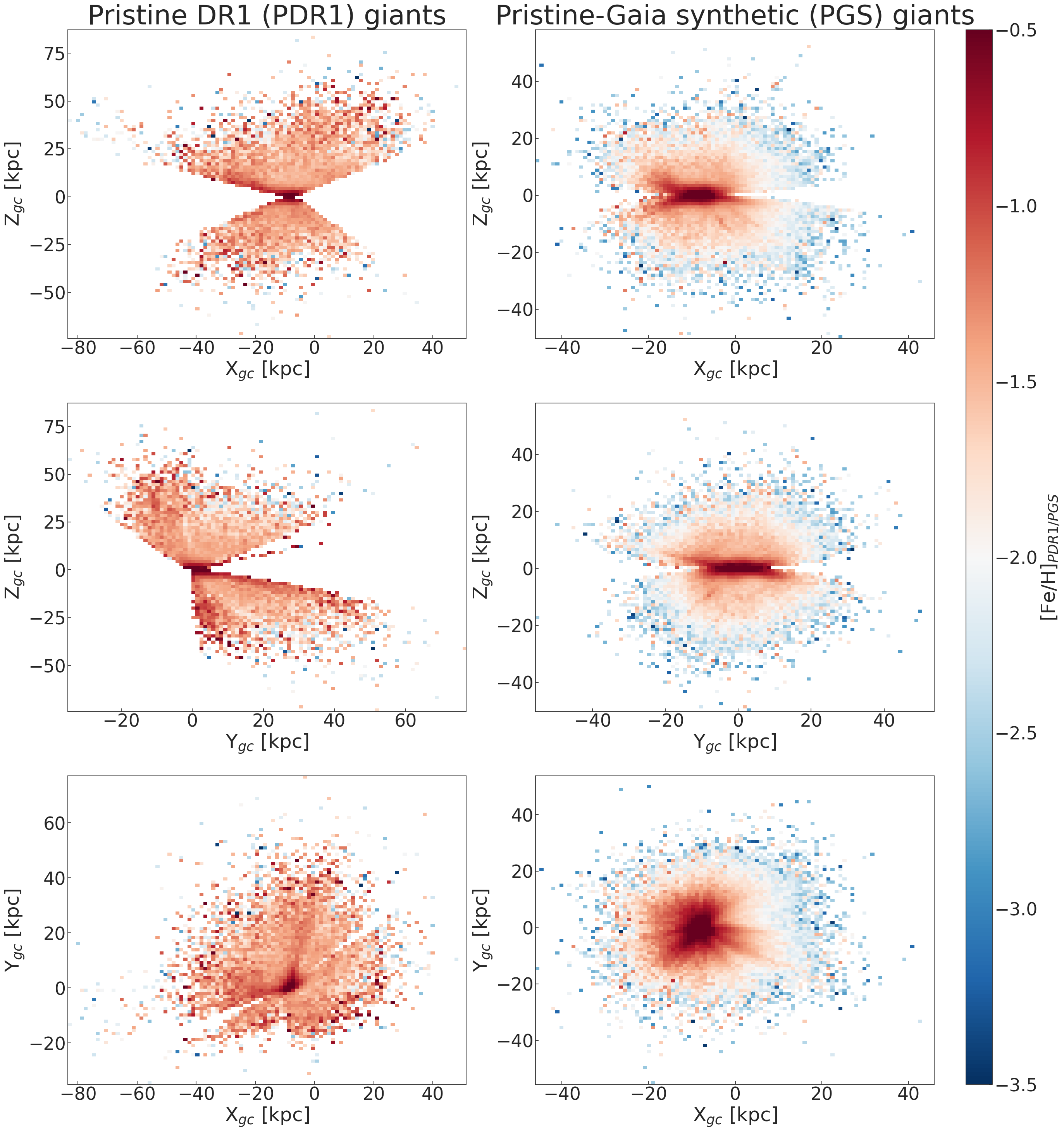}
  \caption{Spatial distribution of galactocentric cartesian Z versus X (top), Z versus Y (middle), and Y versus X (bottom), colour coded by mean metallicity for PDR1 (left) and PGS (right) giants.}
  \label{fig:app-met-cartesian}
\end{figure}

In Figure \ref{fig:app-met-cartesian}, we show the spatial distribution (in X-Z, Y-Z, and X-Y planes) of our PDR1 (left) and PGS-giants (right) colour-coded by their mean metallicities. The consequence of having the Pristine survey footprint is seen on left panels with PDR1-giants whereas the right panels based on PGS-giants cover the entire sky. On the right panels, we can see how the mean metallicities drop to lower metallicities at larger distances showing the power of the catalogue to look for metal-poor substructures in the outskirts of our Galxy, despite the distance-metallicity selection effect. On the left panels, we see a less biased view of mean metallicities across the spatial cartesian coordinates. We can also clearly see the large distances probed by both these catalogues of giants. 

\section{Sagittarius in 6D giants}\label{C}

\begin{figure}[htbp!]
  \includegraphics[width=\columnwidth]{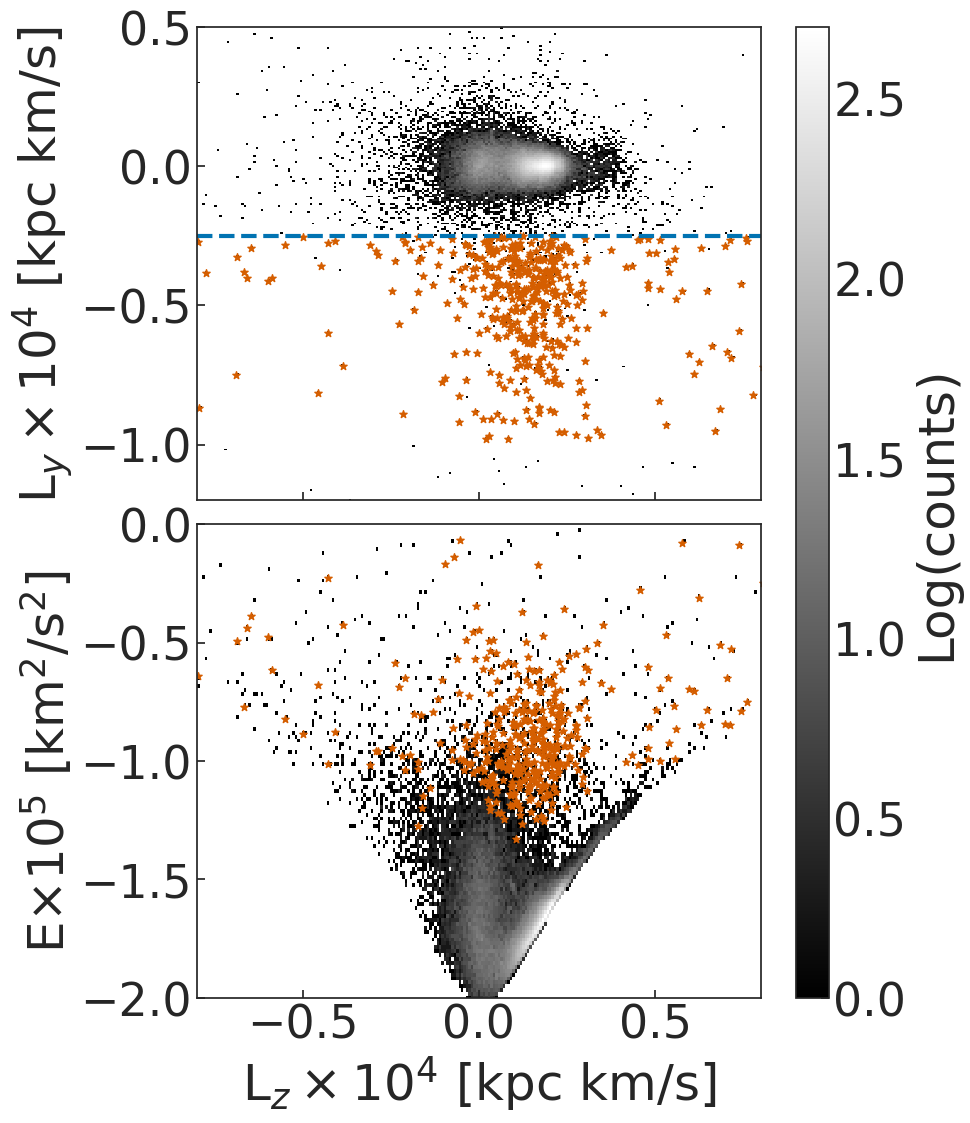}
  \caption{Angular momentum in the y-direction versus z-direction (top) and energy versus angular momentum in z-direction for PDR1 6D giants. Sagittarius stream is seen as a clear overdensity in negative y-direction angular momentum and is separated using a simple L$_y$ cut. These Sagittarius stream stars are shown in orange in both the panels. The same cut is used in PGS 6D giants to isolate Sagittarius stream members but the number of member stars are significantly lower.}
  \label{fig:app-sag1}
\end{figure}

In this subsection, we show how we select 6D Sagittarius stream members. Figure \ref{fig:app-sag1} shows the angular momentum in y-direction versus the z-direction in the top panel and energy E versus z-component angular momentum in the bottom panel for PDR1-giants which has more Sagittarius members than the PGS-giants. We select Sagittarius based on the hallmark negative L$_y$ values (less than -0.5$\times$10$^4$ kpc km/s) and show them in orange in both the panels. These orange stars fall nicely in the region of Sagittarius stream in the IOM space (also seen in Figure \ref{fig:iom}) with few highly prograde and retrograde contaminants. 

In Figure \ref{fig:app-sag2}, we show the distance versus RA view of the 6D stream members in PDR1 (top) and PGS (bottom) giants with the \citet{2017hernitschek} track for the Sagittarius stream overlaid in blue. 
We can see that the stream members are selected much more efficiently and with cleaner and clearer overdensities along the stream tracks (see Figure \ref{fig:sag} for a comparison with 5D selection) using this 6D giants selection.
We end up with 374 and 409 confident 6D stream members in PDR1 and PGS-giants sample.

\begin{figure}[htbp!]
  \includegraphics[width=\columnwidth]{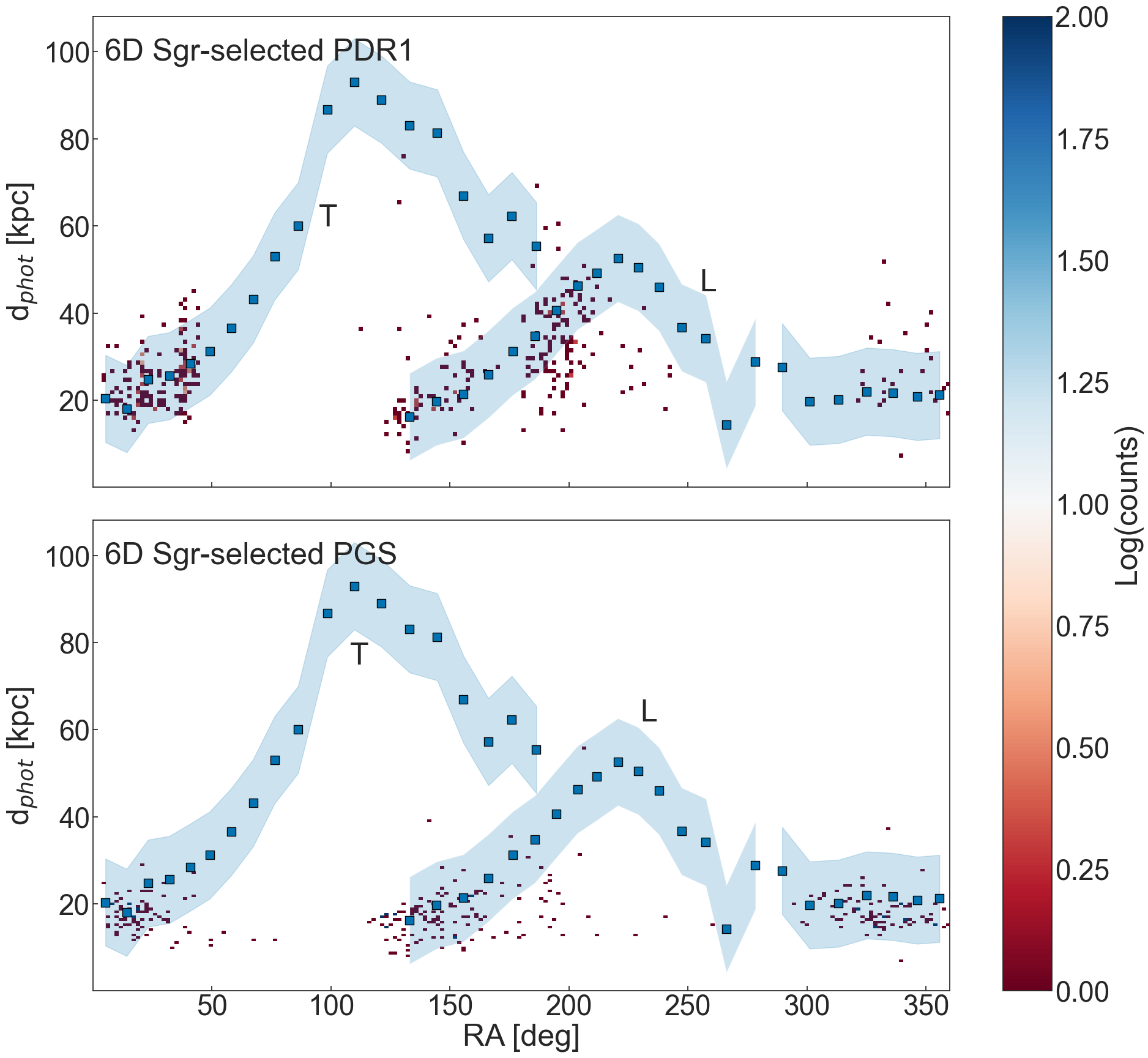}
  \caption{PDR1 (top) and PGS (bottom) 6D Sagittarius stream members in distance versus RA plane with Sagittarius stream tracks adapted from \citet{2017hernitschek}. Leading and trailing arms are indicated with a 'L' and 'T' respectively.}
  \label{fig:app-sag2}
\end{figure}

\section{IOM with distances} \label{D}

\begin{figure*}[htbp!]
  \includegraphics[width=\textwidth]{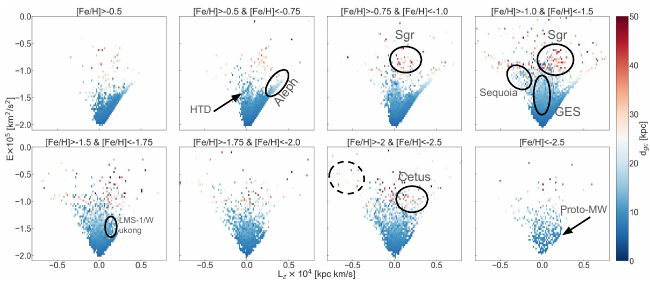}
  \caption{Energy versus z-angular momentum at different slices in metallicities colour-coded by their galactocentric distances for PDR1 6D giants with some known outer halo substructures annotated}.
  \label{fig:app-iom-dist}
\end{figure*}

In Figure \ref{fig:app-iom-dist}, we show the IOM space sliced in bins of photometric metallicities colour-coded by their mean galactocentric distances for PDR1 6D giants. From this figure, we can associate the different substructures and phase-mixed streams to the inner/outer halo. We see the Sagittarius stream clearly dominating the metal-rich end of the outer Galactic halo, and Cetus stream dominates the outer Galactic halo in the metal-poor end. We can see the range of distances probed by the GES merger event (up to about 40 kpc). We find LMS-1/Wukong substructure in the intermediate metallicities at intermediates distances as well. We also find other nearby halo substructures such as Thamnos and Helmi streams. We also see the retrograde intermediate metallicities at higher distances than just the solar neighbourhood; however, the Sequoia merger is still mostly in the nearby halo. 
In the VMP bin, -2.5<[Fe/H]<-2.0, we see distant stars in highly retrograde and higher energy orbits (shown with dashed circle). This substructure, if real, needs a more thorough chemodynamical investigation. In the most metal-poor bin, we see the stars being more centrally concentrated reminiscent of a proto-Galaxy state of the Milky Way, with oldest stars in the inner galaxy \citep{2010tumlinson,2018starkenburg,2018elbadry,2022belokurov,2022rix}, but also find V/EMP stars ([Fe/H]<-2.5) out to 50 kpc. 
This can be already seen from [Fe/H]<--1.5, wherein the more rotating stars are more centrally concentrated than in [FeH]>--1.5.
The general trend when colour-coded by mean distances is that the stars close to the inner galaxy have lower energies as they have sunken into the deep potential well of the Galaxy while the stars with large distances occupy the more higher energy orbits, as expected. 

\end{appendix}

\end{document}